\documentclass[a4paper,11pt]{article}

\usepackage{jheppub} 
\usepackage[utf8]{inputenc}
\usepackage{hyperref}
 \usepackage{hyperref}
 \hypersetup{
     colorlinks=true,
     linkcolor=blue,
     filecolor=blue,
     citecolor = blue,      
     urlcolor=blue,
     }
\usepackage{booktabs}
\usepackage[T1]{fontenc} 
\usepackage{siunitx} 
\usepackage{graphicx}
\usepackage{diagbox}
\usepackage{colortbl}     
\usepackage{array}
\usepackage{comment}
\usepackage{subcaption}
\usepackage[dvipsnames]{xcolor}
\usepackage{dcolumn}
\usepackage{bm}
\usepackage[normalem]{ulem}
\usepackage{braket} 
\usepackage{amsmath}
\usepackage{cancel}
\usepackage{slashed}
\usepackage{multicol}
\usepackage{multirow}
\usepackage{cleveref}
\usepackage{arydshln}
\usepackage{lscape}
\usepackage{pdflscape}
\usepackage{makecell}
\usepackage{graphicx}
\usepackage{mathrsfs}
\usepackage[export]{adjustbox}
\usepackage{tikz}
\usepackage[compat=1.1.0]{tikz-feynman}
\usepackage[compat=1.1.0]{tikz-feynhand}
\usepackage{simpler-wick}

\newcommand{\lzm}{\left(}
\newcommand{\dzm}{\right)}
\newcommand{\lzs}{\left[}
\newcommand{\dzs}{\right]}
\newcommand{\lzv}{\left\{}
\newcommand{\dzv}{\right\}}

\newcommand{\cL}{\mathcal{L}}
\newcommand{\cO}{\mathcal{O}}
\newcommand{\cR}{\mathcal{R}}
\newcommand{\cM}{{\mathcal M}}

\newcommand{\cB}{{\mathcal B}}
\newcommand{\cD}{{\mathcal D}}
\newcommand{\cQ}{{\mathcal Q}}

\newcommand{\cC}{{\mathcal C}}
\newcommand{\cS}{{\mathcal S}}

\newcommand{\cX}{{\mathcal X}}

\newcommand{\cN}{{\mathcal N}}

\newcommand{\mev}{\mathrm{MeV}}
\newcommand{\gev}{\mathrm{GeV}}
\newcommand{\tev}{\mathrm{TeV}}

\newcommand{\hermc}{\text{h.c.}}
\newcommand{\ud}[2]{\phantom{}^{#1}\phantom{}_{#2}}
 
\newcommand{\U}{\mathrm{U}}
\newcommand{\SU}{\mathrm{SU}}

\newcommand{\Tr}{\mathop{\mathrm{Tr}}}
\newcommand{\eminus}{\vcenter{\hbox{\scalebox{0.6}[1]{$ - $}}}}	

\newcommand{\rep}[1]{\mathbf{#1}}
\newcommand{\repbar}[1]{\overline{\mathbf{#1}}}
\newcommand{\sscript}[1]{{\scriptscriptstyle \mathrm{#1}}}
\newcommand{\mel}[3]{%
  \left\langle #1 \middle| #2 \middle| #3 \right\rangle}

\newcommand{\dbtwo}{|\Delta B|=2}

\definecolor{deepskyblue}{rgb}{0.0, 0.75, 1.0}
\definecolor{aqua}{rgb}{0.0, 1.0, 1.0}
\definecolor{bronze}{rgb}{0.8, 0.5, 0.2}
\definecolor{electricyellow}{rgb}{1.0, 1.0, 0.0}
\definecolor{goldenyellow}{rgb}{1.0, 0.87, 0.0}
\definecolor{glaucous}{rgb}{0.38, 0.51, 0.71}

\colorlet{blueRef}{blue!80!black}
\colorlet{CodeColor}{red!70!black} 

\title{EFT Pathways to $\bm{|\Delta B| =2}$: Chiral Constructions and Phenomenology}

\author[a]{Arnau Bas i Beneito,}
\author[a]{Ajdin Palavri\'c,}
\author[b,c,d]{Andrea Sainaghi}

\affiliation[a]{Institut de F\'isica Corpuscular (IFIC), Consejo Superior de Investigaciones Cient\'ificas (CSIC) and Universitat de Val\`encia (UV), 46980 Val\`encia, Spain}
\affiliation[b]{Physik-Institut, Universit\"at Z\"urich, Winterthurerstrasse 190, CH–8057 Z\"urich, Switzerland}
\affiliation[c]{Istituto Nazionale di Fisica Nucleare, Sezione di Padova, 35131 Padova, Italy}
\affiliation[d]{Dipartimento di Fisica e Astronomia ”Galileo Galilei”, Universit\`a di Padova, 35131 Padova, Italy}

\emailAdd{arnau.bas@ific.uv.es}
\emailAdd{ajdin.palavric@ific.uv.es}
\emailAdd{andrea.sainaghi@phd.unipd.it}

\begin{document}

\abstract
{We develop a systematic effective field theory framework for studying $|\Delta B|=2$ interactions across energy scales. Using chiral symmetry, we construct the complete and non-redundant set of operators governing these interactions at low energies and establish their connection to the corresponding operators in the Standard Model effective field theory, as well as to their realizations in baryon chiral perturbation theory. The framework is then applied to the phenomenology of baryon–antibaryon oscillations and dinucleon decay. While oscillations probe only a limited subset of operator structures, dinucleon decay is sensitive to a significantly broader class, including transitions that are otherwise inaccessible. In addition, we identify previously unexplored dinucleon decay channels, which can probe these unconstrained regions of parameter space. More generally, this formalism makes explicit the complementarity of different probes and provides a consistent way to trace baryon-number-violating effects from their ultraviolet origin to low-energy hadronic observables, thereby providing a basis for systematic studies of ultraviolet models generating such interactions.
} 
\maketitle

\clearpage
\section{Introduction} \label{sec:intro}

Baryon number violation (BNV) is one of the most compelling signals of physics beyond the Standard Model (SM). At the fundamental level, baryon number is only an accidental global symmetry of the renormalizable SM Lagrangian, and therefore there is no deep principle enforcing its exact conservation. BNV is additionally motivated by the observed baryon asymmetry of the Universe. As first pointed out by Sakharov, its dynamical generation requires baryon-number violation, together with $C$ and $CP$ violation and a departure from thermal equilibrium~\cite{Sakharov:1967dj}. This makes BNV observables a useful probe of the ultraviolet (UV) structure of particle physics.

The best-known class of BNV observables is associated with $|\Delta B|=1$ transitions, most notably proton decay~\cite{Abbott:1980zj,Nath:2006ut,Langacker:1980js,Claudson:1981gh,Weinberg:1980bf,Weinberg:1979sa} (see also Ref.~\cite{Beneito:2023xbk} for a recent EFT treatment of BNV nucleon decay), which has long served as a paradigmatic prediction of Grand Unified Theories (GUTs)~\cite{Georgi:1974sy,Fritzsch:1974nn,Pati:1973uk}. However, $|\Delta B|=2$ processes provide a complementary and, in several respects, qualitatively distinct probe of baryon-number violation. Their most iconic realization is neutron--antineutron oscillations~\cite{Kuzmin:1970nx,Caswell:1982qs,Rao:1982gt,Rao:1983sd,Mohapatra:1980qe,Kuo:1980ew,Mohapatra:1980de,Chang:1980ey,Basecq:1983hi,Zwirner:1983dgv,Barbieri:1985ty,Mohapatra:1986bd,Kabir:1983qx,Arun:2022eqs}, which have a long theoretical and experimental history and have received renewed attention in recent years in connection with dedicated experimental efforts~\cite{Addazi:2020nlz}, baryogenesis~\cite{Babu:2012vc,Fridell:2021gag,Babu:2006xc,Babu:2008rq,Babu:2013yca,Gu:2011ff,Grojean:2018fus,Mohapatra:2021aig,Aitken:2017wie,Dev:2015uca,Bowes:1996ew}, and low-scale UV completions~\cite{Babu:2001qr,Dorsner:2016wpm,Berezhiani:2018xsx,Dorsner:2025pwe,Dorsner:2025epy}. More generally, $|\Delta B|=2$ interactions can induce oscillations of other neutral baryons, dinucleon decay processes inside nuclei~\cite{Goity:1994dq,Aitken:2017wie,Beneito:2025ond}, and a broader class of exotic hadronic transitions~\cite{He:2021sbl,Hao:2019qjb}. A broader overview of possible BNV processes and their current experimental status can be found in Ref.~\cite{Heeck:2019kgr}, while recent reviews and white papers on BNV processes can be found in Refs.~\cite{Broussard:2025opd,Dev:2022jbf,Phillips:2014fgb}.

From the effective-field-theory (EFT) viewpoint, $|\Delta B|=2$ interactions first arise through dimension-nine six-quark operators below the electroweak scale. These operators provide the natural bridge between UV completions and hadronic observables, and have therefore been studied in several complementary frameworks, including model-independent analyses, simplified-model constructions, and explicit UV completions~\cite{Arnold:2012sd,Baldes:2011mh,Beneito:2025ond,Heeck:2026dmh,Chen:2022gjd,Gardner:2018azu,Nieves:1981tv,Bowes:1996xy}, as well as their renormalization-group evolution in the context of neutron--antineutron oscillations~\cite{Ozer:1982qh,Buchoff:2015qwa,ThomasArun:2025rgx}. Nevertheless, turning this general description into a systematic low-energy framework remains nontrivial. In the hadronic regime, the relevant degrees of freedom are no longer quarks and gluons, but baryons and mesons, and the effects of the six-quark operators must be reorganized according to the approximate chiral symmetry of QCD.

This observation motivates the main goal of the present work: to develop a systematic EFT framework for $|\Delta B|=2$ interactions, bridging quark-level operators and hadronic observables. The natural organizing principle is the approximate chiral symmetry of three-light-flavor QCD, namely $\SU(3)_R \times \SU(3)_L$, which governs the low-energy realization of these interactions and provides the appropriate language for matching onto baryon chiral perturbation theory (B$\chi$PT)~\cite{Weinberg:1978kz,Scherer:2002tk,Gasser:1984gg,Jenkins:1990jv,Bernard:1995dp,Pich:1995bw,Gasser:1983yg,Gasser:1987rb}. Building on earlier two-flavor analyses in the $\SU(2)_R \times \SU(2)_L$ limit, particularly for neutron--antineutron oscillations~\cite{Rinaldi:2018osy,Oosterhof:2019dlo,Rao:1982gt,Buchoff:2015qwa,Rinaldi:2019thf,Bijnens:2017xrz}, we extend the chiral classification to the full three-flavor case and construct the complete non-redundant set of dimension-nine operators relevant for $|\Delta B|=2$ processes. This extension is motivated both formally, since it allows one to determine systematically which operator structures are independent, redundant, or absent once color, flavor, and Lorentz symmetries are imposed, and phenomenologically, since an $\SU(3)_R \times \SU(3)_L$ treatment is required to describe not only nucleon observables but also strange processes such as $\Lambda-\bar\Lambda$ oscillations and kaon final states in dinucleon decays. Lastly, a further class of $|\Delta B|=2$ processes is provided by dinucleon decays into dilepton final states. However, since these violate both $|\Delta B|=2$ and $|\Delta L|=2$, they can only arise from operators of dimension twelve or higher (see Refs.~\cite{He:2021mrt,Girmohanta:2019cjm,Girmohanta:2020eav,Caswell:1982qs,Helset:2021plg,Grossman:2018rdg,Bramante:2014uda,Bryman:2014tta,Feinberg:1978sd,Arnellos:1982nt} for representative studies) and will not be considered in this work. The construction of a complete, non-redundant operator basis for $|\Delta B|=2$ interactions provides the foundation for relating these operators to the corresponding dimension-nine Standard Model EFT (SMEFT)~\cite{Grzadkowski:2010es,Buchmuller:1985jz,Brivio:2017vri,Henning:2014wua,Liao:2020jmn} structures after electroweak symmetry breaking, thereby enabling a direct link to their possible UV origins. It also enables their realization in B$\chi$PT in terms of baryon and meson degrees of freedom.

{\color{black}In addition, this construction provides a starting point for the phenomenological study of $|\Delta B|=2$ observables, in particular baryon--antibaryon oscillations, especially $n-\bar n$ and $\Lambda-\bar\Lambda$, as well as dinucleon decay.} We place special emphasis on the latter because, once the B$\chi$PT expansion is taken into account, it probes a broader set of operator structures than baryon oscillations, including meson-emission channels arising from direct two-baryon operators as well as from contact interactions with one or two mesons, including derivative couplings. As a result, dinucleon decay is highly complementary to oscillation searches: the same Wilson coefficients that govern neutral-baryon oscillations can also contribute to nuclear decay amplitudes. As we show, this complementarity is phenomenologically significant, yielding {\color{black}{order-of-magnitude bounds}} on $n-\bar n$ transitions comparable to those from direct oscillation searches and substantially stronger indirect limits on $\Lambda-\bar\Lambda$ transitions.

Taken together, the framework developed in this work provides a systematic link between the various effective descriptions of $|\Delta B|=2$ physics, from UV completions to low-energy six-quark operators and their hadronic realization. This connection is necessary to compare constraints consistently, propagate flavor assumptions across scales, and interpret the sensitivity of different experiments within a common framework. In this sense, the present work establishes an EFT pathway from UV dynamics to the hadronic phenomenology accessible to low-energy experiments.

The paper is organized as follows. In Section~\ref{sec:chiral_framework_sec2}, we construct the complete non-redundant basis of $|\Delta B|=2$ six-quark operators classified according to their transformation properties under $\SU(3)_R \times \SU(3)_L$, and establish the connection to SMEFT by determining how the relevant dimension-nine SMEFT operators project onto the chiral basis. In Section~\ref{sec:matching_BxPT}, we match the chiral operators onto B$\chi$PT and derive the corresponding hadronic realizations. In Section~\ref{sec:pheno_applications}, we apply this framework to phenomenology, first discussing baryon--antibaryon oscillations and then turning to dinucleon decay, for which we develop a systematic B$\chi$PT treatment and extract the resulting constraints. In addition, we provide a supplementary \texttt{Mathematica} notebook to facilitate the use of the results derived in this section for the study of the relevant $|\Delta B|=2$ phenomenology. We conclude in Section~\ref{sec:conclusions} by emphasizing the main results of our analysis. Additional technical details and derivations
are collected in the Appendices.

\section{Chiral Framework for \texorpdfstring{$\bm{\dbtwo}$}{x} Operators}
\label{sec:chiral_framework_sec2}

In this section, we first review the relevant aspects of chiral symmetry and subsequently develop the corresponding operator construction within the chiral framework.

\subsection{Overview of Chiral Symmetry}
\label{sec:chiral_sym_overview}
The low-energy dynamics of the light-quark sector is governed by the approximate chiral symmetry of QCD. We consider the QCD Lagrangian restricted to the three lightest quark flavors
\begin{equation}\label{eq:light_quarks_SU3}
	q=\begin{bmatrix}u&d&s\end{bmatrix}^\intercal\,,
\end{equation}
which, in terms of chiral components, can be written as
\begin{equation}
\mathcal{L}_\sscript{QCD}=\bar q_L i\gamma^\mu D_\mu q_L+\bar q_R i\gamma^\mu D_\mu q_R-\left(\bar q_L M q_R +\hermc\right)\,.
\end{equation}
Here, the covariant derivative and quark mass matrix are given by
\begin{equation}
    D_\mu=\partial_\mu-ig_s A_\mu^a T^a\,,
    \qquad 
    M=\begin{bmatrix}
        m_u & & \\ & m_d & \\ & & m_s
    \end{bmatrix}\,.
\end{equation}
In the chiral limit $M \to 0$, the QCD Lagrangian exhibits an enhanced global flavor symmetry
\begin{equation}
	G_\chi=\SU(3)_R \times \SU(3)_L \times \U(1)_B\,,
\end{equation}
where the factor $\U(1)_B$ corresponds to baryon number conservation, under which each quark carries charge $1/3$.\footnote{More precisely, in the absence of quark masses the classical QCD Lagrangian is invariant under $\U(1)_V \times \U(1)_A$, corresponding to vector and axial transformations. The vector subgroup $\U(1)_V$ is identified with baryon number, while the axial symmetry $\U(1)_A$ is broken at the quantum level by the chiral anomaly and therefore does not constitute a symmetry of the full theory.} The left- and right-handed quark fields transform independently under $G_\chi$ as
\begin{equation}\label{eq:quark_ch_reps}
	q_R\sim(\rep3_R,\rep1_L)_{1/3}\,,
	\qquad
	q_L\sim(\rep1_R,\rep3_L)_{1/3}\,,
\end{equation}
or, equivalently,
\begin{equation}
	q_R\to e^{i\alpha_B/3} R\,q_R\,,
	\qquad
	q_L\to e^{i\alpha_B/3} L\,q_L\,,
	\qquad
	R,L\in\SU(3)_{R,L}\,,
\end{equation}
where $\alpha_B$ denotes the parameter of the $\U(1)_B$ transformation. In addition, the theory is also invariant under parity, which acts on the quark fields as
\begin{equation}
	q_{R,L}(t,\vec x) \xrightarrow{P} \gamma^0 q_{L,R}(t,-\vec x)\,.
\end{equation}
In the QCD vacuum, the chiral symmetry is spontaneously broken to the diagonal subgroup
\begin{equation}
	\SU(3)_R \times \SU(3)_L\to \SU(3)_V\,,
\end{equation}
as a consequence of the formation of a non-vanishing quark condensate
\begin{equation}
	\bra{0}\bar q_L^iq_R^j\ket{0}\sim \delta^{ij}\Lambda_\chi^3\,,
\end{equation}
which is invariant only under vector transformations acting identically on left- and right-handed fields. This condensate therefore aligns the vacuum in a way that preserves the vector subgroup $\SU(3)_V$, while breaking the axial generators of $\SU(3)_R \times \SU(3)_L$.

This pattern of spontaneous symmetry breaking provides the natural group-theoretic framework for organizing quark-level operators according to their transformation properties under $\SU(3)_R \times \SU(3)_L$. In particular, it enables a systematic classification of $|\Delta B| = 2$ operators in terms of irreducible chiral representations, which will be constructed in what follows.

Lastly, it should be emphasized that the above discussion is formulated at the quark level and therefore does not yet capture the appropriate degrees of freedom below the QCD confinement scale. In the low-energy regime, the dynamics is instead described in terms of hadronic fields, such as mesons and baryons, within the framework of baryon chiral perturbation theory (B$\chi$PT). The matching to this effective description will be performed in Section~\ref{sec:matching_BxPT}, where the chiral structures identified here will be realized at the hadronic level.

\subsection{Construction of Chiral Representations}
\label{sec:chiral_ops_reps_constr}
We now turn to the construction of the chiral irreducible representations relevant for $|\Delta B| = 2$ interactions. At energies below the electroweak scale, these transitions are described by dimension-nine operators composed of six quark fields. Restricting our attention to the light-quark sector defined in Eq.~\eqref{eq:light_quarks_SU3}, we consider operators of the form~\cite{Buchoff:2015qwa,Rinaldi:2018osy,Rinaldi:2019thf,Chang:1980ey,Kuo:1980ew,Rao:1982gt,Rao:1983sd,Caswell:1982qs}
\begin{equation}\label{eq:chiral_ops_defs}
    \begin{alignedat}{4}
    &\cO_{\chi_1\chi_2\chi_3}^{prstvw}
    &&=(q^{\alpha p} P_{\chi_1} q^{\beta r})(q^{\gamma s}  P_{\chi_2} q^{\rho t})(q^{\sigma v} P_{\chi_3} q^{\tau w})T^{AAS}_{[\alpha\beta][\gamma\rho]\{\sigma\tau\}}\,,
    \\[3pt]
    &\widetilde\cO_{\chi_1\chi_2\chi_3}^{prstvw}
    &&=(q^{\alpha p} P_{\chi_1} q^{\beta r})(q^{\gamma s}  P_{\chi_2} q^{\rho t})(q^{\sigma v} P_{\chi_3} q^{\tau w})T^{SSS}_{\{\alpha\beta\}\{\gamma\rho\}\{\sigma\tau\}}\,.
\end{alignedat}
\end{equation}
In the above expressions, $p,r,s,t,v,w$ denote light-quark flavor indices, while $\alpha,\beta,\gamma,\rho,\sigma,\tau$ are color indices. The projectors $P_{\chi_i}$, with $\chi_i = L,R$, select definite chiralities of the quark fields. Furthermore, throughout this work, we employ a two-component spinor notation, in which the operators are written in terms of left- and right-handed Weyl fields.\footnote{Our notation follows the two-component spinor formalism reviewed in Ref.~\cite{Dreiner:2008tw}. However, for definiteness and ease of reading, we keep the chiral labels $L$ and $R$ explicit throughout. These fields are to be identified with the left- and right-handed Weyl components of a four-component Dirac spinor, and later we will make the connection with the corresponding four-component notation more explicit. In particular, bilinears of the form $(q P_\chi q)$ correspond to charge-conjugated contractions in four-component notation, involving the charge-conjugation matrix $C$, which take the form $(\bar q_\chi^c q_\chi)=(q^\intercal C P_\chi q)$.}

The tensors $T^{AAS}$ and $T^{SSS}$ encode the possible contractions of color indices, where square and curly brackets denote antisymmetrization and symmetrization, respectively. Their explicit form is given by
\begin{equation}\label{eq:ch_color_tensors}
    \begin{alignedat}{4}
        &T^{SSS}_{\{\alpha\beta\}\{\gamma\rho\}\{\sigma\tau\}}&&=\varepsilon_{\alpha\gamma\sigma}\varepsilon_{\beta\rho\tau}+\varepsilon_{\beta\gamma\sigma}\varepsilon_{\alpha\rho\tau}+\varepsilon_{\alpha\rho\sigma}\varepsilon_{\beta\gamma\tau}+\varepsilon_{\alpha\gamma\tau}\varepsilon_{\beta\rho\sigma}\,,
        \\[2pt]
    	&T^{AAS}_{[\alpha\beta][\gamma\rho]\{\sigma\tau\}}&&=\varepsilon_{\alpha\beta\sigma}\varepsilon_{\gamma\rho\tau}+\varepsilon_{\alpha\beta\tau}\varepsilon_{\gamma\rho\sigma}\,.
    \end{alignedat}
\end{equation}
Additional details, including the identities satisfied by these tensors, are provided in Appendix~\ref{app:color_tensors}. For the operators in Eq.~\eqref{eq:chiral_ops_defs}, expressing the color structure in terms of the tensors $T^{AAS}$, and $T^{SSS}$ proves particularly convenient for the analysis of the transformation properties under $\SU(3)_R \times \SU(3)_L$ allowing for a direct and systematic extraction of the corresponding chiral irreducible representations. Finally, we note that the operators defined above constitute a subset of the general $|\Delta B| = 2$ operator basis in the Low-Energy Effective Field Theory (LEFT), restricted to the three lightest quark flavors. Their origin within a more general EFT framework, in particular through matching from the SMEFT, will be discussed in Section~\ref{sec:projection_SMEFT_ch_bas}.

Building on the symmetry structure introduced in Section~\ref{sec:chiral_sym_overview}, we now proceed to determine the transformation properties of the operators in Eq.~\eqref{eq:chiral_ops_defs} under $\SU(3)_R \times \SU(3)_L$. As outlined in Eq.~\eqref{eq:quark_ch_reps}, the quark fields transform in the fundamental representations of the chiral symmetry, which provides the starting point for the construction of the corresponding irreducible representations. The basic building blocks of this construction are given by quark bilinears of definite chirality, which we define as
\begin{equation}\label{eq:sixq_ident_bilinears}
	[\cQ_\chi^{\alpha \beta}]^{pr}\equiv(q_\chi^{\alpha p} q^{\beta r}_\chi)\,.
\end{equation}
By construction, the bilinear $[\cQ_\chi^{\alpha \beta}]^{pr}$ is symmetric under the simultaneous exchange of color and flavor indices. This follows from the properties of two-component Weyl spinors, for which contractions of the form $(q_\chi^{\alpha p} q_\chi^{\beta r})$ are symmetric under interchange of the fermion fields. In terms of these building blocks, the operators introduced in Eq.~\eqref{eq:chiral_ops_defs} can be written in a compact form as products of three bilinears, for instance
\begin{equation}
	\cO_{\chi_1\chi_2\chi_3}^{prstvw}
	=[\cQ_{\chi_1}^{\alpha\beta}]^{pr}[\cQ_{\chi_2}^{\gamma\rho}]^{st}[\cQ_{\chi_3}^{\sigma\tau}]^{vw}T^{AAS}_{[\alpha\beta][\gamma\rho]\{\sigma\tau\}}\,,
\end{equation}
and analogously for $\widetilde\cO$.

Since the quark fields transform in the fundamental representation of the chiral group, the $[\cQ_\chi^{\alpha \beta}]^{pr}$ bilinear transforms in the product representation $\rep3_\chi\otimes\rep3_\chi=\repbar3_\chi\oplus\rep6_\chi$. The corresponding irreducible components can be isolated by decomposing the bilinear into its antisymmetric ($\repbar3_\chi$) and symmetric ($\rep6_\chi$) components as\footnote{Throughout this work, we denote symmetrization over indices by parentheses, $2\,\psi_{(a|}\chi_{|b)}=\psi_a\chi_b+\psi_b\chi_a$, and antisymmetrization by square brackets $2\,\psi_{[a\lvert}\chi_{\lvert b]}=\psi_a\chi_b-\psi_b\chi_a$.}
\begin{equation}\label{eq:def2Qs}
	\begin{alignedat}{8}
		&[\cD_{\chi}^{\alpha\beta}]_a&&\equiv[\cX_{\repbar3_\chi}\lvert_a]_{ij}[\cQ_\chi^{\alpha\beta}]^{ij}&&\sim\repbar3_\chi\,,
		&&\quad\qquad
		&&[\cX_{\repbar3_\chi}\lvert_a]_{ij}&&=\frac{1}{2}\varepsilon_{aij}\,,
		\\[3pt]
		&[\cD_\chi^{\alpha\beta}]^{ab}&&\equiv[\cX_{\rep6_\chi}\lvert^{ab}]_{ij}[\cQ_\chi^{\alpha\beta}]^{ij}&&\sim\rep6_\chi\,,
		&&\quad\qquad
		&&[\cX_{\rep6_\chi}\lvert^{ab}]_{ij}&&=\delta\ud{a}{(i|}\delta\ud{b}{|j)}\,,
	\end{alignedat} 
\end{equation}
where $\cX_{\cR_\chi}$ denotes the chiral tensor, $i,j$ denote flavor indices of the light quarks, while $a,b$ label the components of the corresponding irreducible representations. 

These components serve as the fundamental building blocks for the construction of the chiral irreducible representations associated with the $|\Delta B| = 2$ six-quark operators. In particular, the operator structures can be obtained by forming tensor products of the bilinears and decomposing them into irreducible representations of $\SU(3)_R \times \SU(3)_L$. This procedure allows for a systematic identification of the relevant chiral irreps and the extraction of the corresponding chiral tensors encoding their transformation properties. The complete set of relevant irreducible representations, along with the tensor products from which they are extracted and the associated chiral tensors, is summarized in Table~\ref{tab:chiral_reps_tensors_overview}.

\begin{table}[t]
    \centering
\scalebox{0.84}{
\begin{tabular}{c@{\hspace{0.5cm}}c@{\hspace{0.5cm}}>{\raggedright\arraybackslash}p{2.5cm}@{\hspace{1.cm}}c}
\toprule
\multirow{1}{*}{\textbf{Irrep}}
&\multirow{1}{*}{\textbf{Decomposition}}
&\multirow{1}{*}{\textbf{Tensor}}
&\multirow{1}{*}{\textbf{Definition}}
\\
\midrule
\addlinespace[0.2cm]
$\repbar3_\chi$&--&$[\cX_{\repbar3_\chi}\lvert_a]_{ij}$&$\frac{1}{2}\varepsilon_{aij}$
\\[0.2cm]
$\rep6_\chi$&--&$[\cX_{\rep6_\chi}\lvert^{ab}]_{ij}$&$\delta\ud{a}{(i|}\delta\ud{b}{|j)}$
\\[0.2cm]
\noalign{\vskip 0.0cm}
\cdashline{1-4}[.4pt/2pt]
\noalign{\vskip 0.2cm}
$\repbar6_\chi$&$\repbar3_\chi\otimes\repbar3_\chi$&$[\cX_{\repbar6_\chi}\lvert_{ab}]_{ijk\ell}$&$\frac{1}{4}\varepsilon_{(a|ij}\varepsilon_{|b)k\ell}$
\\[0.2cm]
$\rep3_\chi$&$\rep6_\chi\otimes\repbar{3}_\chi$&$[\cX_{\rep3_\chi}\lvert^a]_{ijk\ell}$&$\frac{1}{4}\delta\ud{a}{(i|}\varepsilon_{|j)k\ell}$
\\[0.2cm]
$\rep{15}_\chi$&$\rep6_\chi\otimes\repbar3_\chi$&$[\cX_{\rep{15}_\chi}\lvert\ud{ab}{c}]_{ijk\ell}$
&$\frac{1}{2}\delta\ud{a}{(i|}\delta\ud{b}{|j)}\varepsilon_{ck\ell}-\frac{1}{8}\lzm \delta\ud{a}{c}\delta\ud{b}{(i|}\varepsilon_{|j)k\ell}+a\leftrightarrow b  \dzm$
\\[0.2cm]
$\rep{15}_\chi'$&$\rep6_\chi\otimes\rep6_\chi$&$[\cX_{\rep{15}'_\chi}\lvert^{abcd}]_{ijk\ell}$
&$\frac{1}{24}\sum_{\sigma\in S_4} \delta\ud{a}{\sigma(i)}\delta\ud{b}{\sigma(j)}\delta\ud{c}{\sigma(k)}\delta\ud{d}{\sigma(\ell)}$
\\[0.2cm]
\noalign{\vskip 0.0cm}
\cdashline{1-4}[.4pt/2pt]
\noalign{\vskip 0.2cm}
\multirow{2}{*}{$\rep{27}_\chi$}&\multirow{2}{*}{$\rep6_\chi\otimes\repbar3_\chi\otimes\repbar3_\chi$}&\multirow{2}{*}{$[\cX_{\rep{27}_\chi}\lvert\ud{ab}{cd}]_{ijk\ell mn}$}
&$\frac{1}{4}\delta\ud{a}{(i|}\delta\ud{b}{|j)}\varepsilon_{(c|k\ell}\varepsilon_{|d)mn}+\frac{1}{40}\delta\ud{(a|}{c}\delta\ud{|b)}{d}\delta\ud{e}{(i|}\delta\ud{f}{|j)}\varepsilon_{(e|k\ell}\varepsilon_{|f)mn}$
\\[0.2cm]
&&&$-\frac{1}{20}\lzs  \lzm \delta\ud{a}{c}\delta\ud{b}{(i|}\delta\ud{e}{|j)}\varepsilon_{(d|k\ell}\varepsilon_{|e)mn}+c\leftrightarrow d \dzm+a\leftrightarrow b  \dzs$
\\[0.35cm]
$\rep{28}_\chi$&$\rep6_\chi\otimes\rep6_\chi\otimes\rep6_\chi$&$[\cX_{\rep{28}_\chi}\lvert^{abcdef}]_{ijk\ell mn}$
&$\frac{1}{720}\sum_{\sigma\in S_6}\delta\ud{a}{\sigma(i)}\delta\ud{b}{\sigma(j)}\delta\ud{c}{\sigma(k)}\delta\ud{d}{\sigma(\ell)}\delta\ud{e}{\sigma(m)}\delta\ud{f}{\sigma(n)}$
\\[0.2cm]
$\rep1_\chi$&$\rep6_\chi\otimes\repbar3_\chi\otimes\repbar3_\chi$&$[\cX_{\rep1_\chi}]_{ijk\ell mn}$&$\frac{1}{24}\varepsilon_{(i|k\ell}\varepsilon_{|j)mn}$
\\[0.2cm]
\bottomrule
\end{tabular}
}
    \caption{Overview of the relevant chiral irreducible representations, together with the tensor-product decompositions from which they arise and the corresponding chiral tensors. The dotted lines separate groups of irreducible representations according to the number of same-chirality flavor indices, corresponding to structures involving two, four, and six quarks of identical chirality. The table lists a non-redundant set of representations relevant for the $|\Delta B| = 2$ operators considered in this work. The indices $a,b,c,d,e,f$ label the components of the chiral irreducible representations, while $i,j,k,\ell,m,n$ denote flavor indices of the light quarks. Further details on the derivation of these tensors, as well as the treatment of alternative (redundant) constructions, are provided in Appendix~\ref{app:constr_chiral_irreps}.}
    \label{tab:chiral_reps_tensors_overview}
\end{table}

The main conclusion that can be drawn from Table~\ref{tab:chiral_reps_tensors_overview} is that the symmetry and index structure of the chiral tensors already largely determine the form of the corresponding operator structures. In particular, the symmetry properties of the flavor indices must be matched by the corresponding color contractions within each bilinear, such that the overall operator does not vanish. Based on these considerations, two observations can be made.
\begin{itemize}
    \item Starting with the four chiral irreducible representations whose tensors involve four light-quark indices, we notice that $\rep3_\chi$ and $\rep{15}_\chi$ representations feature one symmetric and one antisymmetric bilinear. In contrast, the chiral tensor corresponding to the $\repbar6_\chi$ representation is antisymmetric in both bilinears, while the one associated to $\rep{15}'_\chi$ is fully symmetric. The $\rep{15}_\chi$ and $\rep{15}'_\chi$ representations further differ at the level of their representation indices, with the former exhibiting mixed symmetry and the latter being fully symmetric.
    \item We next consider the chiral irreducible representations whose tensors involve six light-quark indices. In these cases, the $\rep{27}_\chi$ and $\rep{1}_\chi$ representations feature one symmetric and two antisymmetric bilinears, whereas the $\rep{28}_\chi$ representation is fully symmetric. This pattern is once again consistent with the symmetry properties of the corresponding representation indices, with $\rep{27}_\chi$ exhibiting mixed symmetry and $\rep{28}_\chi$ being fully symmetric.
\end{itemize}

With the relevant chiral tensors and their symmetry patterns identified, we now proceed to construct the corresponding operator structures for the various fixed chirality assignments. For a given representation, the operator structure is obtained by contracting the color indices of the bilinears with the appropriate color structure, as dictated by the symmetry properties of the corresponding chiral tensor. 

As an illustrative example, we consider the $\rep{15}'_R \otimes \rep{6}_L$ representation. Since in this case all three bilinears are symmetric in flavor space, the corresponding operator must be contracted with the fully symmetric color tensor $T^{SSS}$, and can therefore be written as
\begin{equation}
	\rep{15}'_R \otimes \rep{6}_L
    \sim [\cX_{\rep{15}'_R}\lvert^{abcd}]_{ijk\ell}[\cX_{\rep6_L}\lvert^{\dot a\dot b}]_{mn}
	\underbrace{[\cQ_R^{\alpha \beta}]^{ij}[\cQ_R^{\gamma\rho}]^{k\ell}[\cQ_L^{\sigma\tau}]^{mn}\,T^{SSS}_{\{\alpha\beta\}\{\gamma\rho\}\{\sigma\tau\}}}_{\widetilde\cO_{RRL}^{ijk\ell mn}}\,,
\end{equation}
where, using Eq.~\eqref{eq:sixq_ident_bilinears}, the product of three bilinears and the color tensor is identified with the six-quark operator defined in Eq.~\eqref{eq:chiral_ops_defs} and where we employ the convention that undotted (dotted) indices refer to right-handed (left-handed) chiral components.\footnote{For operator structures involving color tensors of the type $T^{AAS}$, an appropriate reordering of indices is required in order to match the canonical form of the operators defined in Eq.~\eqref{eq:chiral_ops_defs}. This is reflected in the fact that some of the corresponding chiral tensors in Table~\ref{tab:chiral_basis_op_count} appear with a modified index ordering.}

\begin{table}[t]
\centering
\scalebox{0.88}{
\begin{tabular}{c@{\hspace{0.7cm}}>{\raggedright\arraybackslash}p{2.5cm}@{\hspace{3.5cm}}ccccccc}
\toprule
\multirow{1}{*}{\textbf{Irrep}}
&\multirow{1}{*}{\textbf{~~~Operator structure}}
&$\bm{u^2d^4}$
&$\bm{u^2d^3 s}$
&$\bm{u^2d^2s^2}$
&$\bm{u^2ds^3}$
&$\bm{u^2s^4}$
&\multirow{1}{*}{\textbf{Total}}
\\
\midrule
\addlinespace[0.2cm]
$\repbar6_R\otimes\rep6_L$
&$[\cX_{\repbar 6_R}\lvert_{ab}]_{ijk\ell}\,[\cX_{\rep6_L}\lvert^{\dot a\dot b}]_{mn}\,\cO_{RRL}^{ijk\ell mn}$&1&3&6&3&1&14
\\[0.3cm]
$\rep{15}_R\otimes\repbar3_L$&$[\cX_{\rep{15}_R}\lvert\ud{ab}{c}]_{mnk\ell}\,[\cX_{\repbar3_L}\lvert_{\dot a}]_{ij}\,\cO_{LRR}^{ijk\ell mn}$&1&4&6&4&1&16
\\[0.3cm]
$\rep3_R\otimes\repbar3_L$&$[\cX_{\rep3_R}\lvert^a]_{mnk\ell}\,[\cX_{\repbar3_L}\lvert_{\dot a}]_{ij}\,\cO_{LRR}^{ijk\ell mn}$&0&1&3&1&0&5
\\[0.3cm]
$\rep{15}_R'\otimes\rep6_L$&$[\cX_{\rep{15}'_R}\lvert^{abcd}]_{ijk\ell}\,[\cX_{\rep6_L}\lvert^{\dot a\dot b}]_{mn}\,\widetilde\cO_{RRL}^{ijk\ell mn}$&3&5&6&5&3&22
\\[0.3cm]
\noalign{\vskip 0.0cm}
\cdashline{1-8}[.4pt/2pt]
\noalign{\vskip 0.2cm}
$\rep1_R\otimes\rep1_L$&$[\cX_{\rep1_R}]_{mnijk\ell}\,\cO_{RRR}^{ijk\ell mn}$&0&0&1&0&0&1
\\[0.3cm]
$\rep{27}_R\otimes\rep1_L$&$[\cX_{\rep{27}_R}\lvert\ud{ab}{cd}]_{mnijk\ell}\,\cO_{RRR}^{ijk\ell mn}$&1&2&3&2&1&9
\\[0.3cm]
$\rep{28}_R\otimes\rep1_L$&$[\cX_{\rep{28}_R}\lvert^{abcdef}]_{ijk\ell mn}\,\widetilde\cO_{RRR}^{ijk\ell mn}$&1&1&1&1&1&5
\\[0.2cm]
\bottomrule
\end{tabular}
}
\caption{Overview of the non-redundant operator structures classified according to their chiral irreducible representations. For each representation, we display the corresponding operator structure expressed in terms of the chiral tensors (see Table~\ref{tab:chiral_reps_tensors_overview}), together with the number of independent operators for different light-quark flavor compositions. The complete operator basis, obtained by resolving the individual components of the chiral irreducible representations and selecting the non-redundant structures for each light-quark flavor composition, is presented in Appendix~\ref{app:complete_ch_op}. Undotted (dotted) indices correspond to right-handed (left-handed) chiral components. The operators related by the exchange $L \leftrightarrow R$ are obtained directly from the structures shown here and are not listed separately.}
\label{tab:chiral_basis_op_count}
\end{table}

Applying this procedure to all relevant representations, one obtains the complete non-redundant set of operator structures, which we collect in Table~\ref{tab:chiral_basis_op_count}. A few remarks are in order:
\begin{itemize}
    \item The operators are organized according to definite chirality assignments. In particular, one can distinguish operators of the same chirality ($RRR$ and $LLL$), from those of mixed chirality ($RRL$ and $RLL$). For operators of the same chirality, only the chiral irreducible representations $\rep{27}_R$, $\rep{28}_R$ and $\rep1_R$ are allowed. In the case of mixed chirality, the admissible color tensors (see Eq.~\eqref{eq:ch_color_tensors}) restrict the possible chiral representations to $\repbar6_R\otimes\rep6_L$, $\rep{15}_R\otimes\repbar3_L$, $\rep3_R\otimes\repbar3_L$ and $\rep{15}_R'\otimes\rep6_L$. The operator structures corresponding to the complementary chirality assignments are obtained directly by the exchange $L\leftrightarrow R$ and are therefore not shown explicitly.
    \item The operator structures displayed in Table~\ref{tab:chiral_basis_op_count} form a minimal set, in the sense that other possible operator representations can be constructed but either vanish identically or are redundant. This follows from relations among the color tensors, originating from Schouten identities, combined with Fierz and group-theoretical relations, as discussed in Appendix~\ref{app:constr_chiral_irreps}.
    \item In addition, Table~\ref{tab:chiral_basis_op_count} provides the counting of independent operators for each light-quark flavor composition. Both the individual multiplicities and the total number of operators, amounting to $72+72_{L\leftrightarrow R}$, are found to be in agreement with the results obtained using \texttt{Sym2Int}~\cite{Fonseca:2017lem,Fonseca:2019yya}, {\color{black}and also with Ref.~\cite{Li:2020tsi},} thereby providing a non-trivial consistency check of the completeness of the operator basis.
\end{itemize}

Finally, the full operator basis, obtained by resolving the individual components of the chiral irreducible representations, is presented in Appendix~\ref{app:complete_ch_op} and it provides the starting point for the B$\chi$PT matching discussed in Section~\ref{sec:matching_BxPT}.

\subsection{SMEFT Projection onto Chiral Basis}
\label{sec:projection_SMEFT_ch_bas}
The chiral irreducible representations introduced above provide a complete classification of the operator structures under $\SU(3)_R \times \SU(3)_L$. We next examine their origin from a top-down perspective in UV completions.

From a top-down perspective, these operators arise from integrating out heavy degrees of freedom in UV completions, leading to higher-dimensional interactions, in particular dimension-nine six-quark operators within the SMEFT framework~\cite{Heeck:2026dmh,Baldes:2011mh,Arnold:2012sd,Beneito:2025ond,Chen:2022gjd}. Establishing a systematic mapping between these operators and the chiral irreducible representations identified above is therefore essential for consistently propagating the effects of UV physics through SMEFT to the chiral effective theory, thereby enabling a direct connection between the underlying UV dynamics and hadronic observables.

At the SMEFT level, the effective Lagrangian can be expressed as
\begin{equation}
	\cL_{\sscript{SMEFT}}^{{\scalebox{0.56}{$|\Delta B|=2$}}}\supset \sum_{i=1}^5 \cC_i\cO_i\,,
\end{equation}
where the Wilson coefficients have mass dimension $[\cC_i]=-5$, and encode the effects of UV physics. The five dimension-nine operators relevant for our analysis are given by~\cite{Liao:2020jmn}\footnote{We note that the projection of these SMEFT operators onto the chiral basis is not one-to-one. In particular, some dimension-nine LEFT operators (in particular those of the LLL type) arise from higher-dimensional operators in the SMEFT after electroweak symmetry breaking, once insertions of the Higgs vacuum expectation value are taken into account.}
\begin{equation}\label{eq:dim9_SMEFT_ops_reordered} 
\begin{alignedat}{4}
&[\cO_{d^4u^2}^{(1)}]^{prstvw}
&&=\varepsilon_{\alpha\gamma\sigma}\varepsilon_{\beta\rho\tau}
(u^{\alpha p}_{R} d^{\beta r}_{R})
(u^{\gamma s}_{R} d^{\rho t}_{R})
(d^{\sigma v}_{R} d^{\tau w}_{R})\,,
\\[3pt]
&[\cO_{d^4u^2}^{(2)}]^{prstvw}
&&=\varepsilon_{\alpha\rho\sigma}\varepsilon_{\beta\gamma\tau}
(u^{\alpha p}_{R} d^{\beta r}_{R})
(u^{\gamma s}_{R} d^{\rho t}_{R})
(d^{\sigma v}_{R} d^{\tau w}_{R})\,,
\\[3pt]
&[\cO_{d^3Q^2u}]^{prstvw}
&&=\varepsilon_{\alpha\gamma\sigma}\varepsilon_{\beta\rho\tau}
\varepsilon_{ij}
(u_{R}^{\alpha p}d_{R}^{\beta r})
(d_{R}^{\gamma s}d_{R}^{\rho t})
(Q_{L}^{\sigma i v}Q_{L}^{\tau j w})\,,
\\[3pt]
&[\cO_{d^2Q^4}^{(1)}]^{prstvw}
&&=\varepsilon_{\alpha\gamma\sigma}\varepsilon_{\beta\rho\tau}
\varepsilon_{ik}\varepsilon_{j\ell}
(d^{\alpha p}_{R} d^{\beta r}_{R})
(Q_{L}^{\gamma i s}Q^{\rho j t}_{L})
(Q_{L}^{\sigma k v}Q_{L}^{\tau\ell w})\,,
\\
&[\cO_{d^2Q^4}^{(2)}]^{prstvw}
&&=\varepsilon_{\alpha\gamma\tau}\varepsilon_{\beta\rho\sigma}
\varepsilon_{ik}\varepsilon_{j\ell}
(d^{\alpha p}_{R} d^{\beta r}_{R})
(Q_{L}^{\gamma i s}Q^{\rho j t}_{L})
(Q_{L}^{\sigma k v}Q_{L}^{\tau\ell w})\,.
\end{alignedat}
\end{equation}
In these expressions, $\alpha,\beta,\gamma,\rho,\sigma,\tau$ denote color indices, $i,j,k,\ell$ are electroweak $\SU(2)_{\rm L}$ indices, and $p,r,s,t,v,w$ label flavor. As in Section~\ref{sec:chiral_ops_reps_constr}, we adopt a two-component notation, where $u_R$, $d_R$, and $Q_L$ are treated as Weyl fermions. The operator counting as a function of the number of active flavors, as well as the flavor relations among different operator components can be found in Ref.~\cite{Liao:2020jmn}.

Given the set of relevant SMEFT operators, we now outline the procedure used to construct the mapping onto the chiral basis. The mapping proceeds in two steps. In the first step, the SMEFT operators are rewritten by explicitly resolving the $\SU(2)_{\rm L}$ indices, thereby expressing them in a form where the electroweak structure is no longer manifest, corresponding to the matching onto the LEFT operators. The resulting LEFT structures are collected in Table~\ref{tab:SMEFT_ch_match_LEFT_strs}. In the second step, we restrict to the light-quark sector and match the resulting LEFT structures onto the chiral irreducible representations constructed in Section~\ref{sec:chiral_ops_reps_constr}. This establishes the complete mapping from the SMEFT operators to their chiral and hadronic realizations.

Inspecting the SMEFT operators in Eq.~\eqref{eq:dim9_SMEFT_ops_reordered}, one observes that they belong to three distinct chirality configurations, namely $RRR$, $RRL$ and $RLL$, which directly correspond to the classes analyzed in the chiral construction. The $RRR$ structures retain their form in the matching, while $RRL$ and $RLL$ operators require the explicit decomposition of the $\SU(2)_{\rm L}$ indices described above in order to match onto the LEFT basis. In what follows, we treat the SMEFT operators collectively within each chirality class and make use of the color-tensor identities summarized in Appendix~\ref{app:color_tensors} to express the resulting structures in terms of the $T^{SSS}$ and $T^{AAS}$ tensors.

As a final remark, it is worth emphasizing that the matching presented here does not assume any particular choice of flavor basis. The matching relations corresponding to a specific flavor alignment can be obtained by performing the appropriate CKM rotations. For instance, in the up-quark mass basis one has $d_{Lw}\to (Vd_L)_w$, which induces a corresponding rotation of the flavor indices in the SMEFT Wilson coefficients.

\begin{table}[t]
\centering
\renewcommand{\arraystretch}{1.2}
\scalebox{0.93}{
\begin{tabular}{c@{\hspace{1.0cm}}c@{\hspace{1.0cm}}c@{\hspace{1.0cm}}c}
\toprule
\multirow{1}{*}{\textbf{SMEFT operator}}
&\multirow{1}{*}{$\bm\chi$ \textbf{type}}
&\multirow{1}{*}{\textbf{LEFT structure}}
&\multirow{1}{*}{\textbf{Definition}}
\\
\midrule
\addlinespace[0.2cm]
$[\cO_{d^4u^2}^{(1,2)}]^{prstvw}$
&$RRR$
&$[L_{d^4u^2}]_{prstvw}^{\alpha\beta\gamma\rho\sigma\tau}$
&$(u^\alpha_{Rp}d^\beta_{Rr})(u^\gamma_{Rs}d^\rho_{Rt})(d^\sigma_{Rv}d^\tau_{Rw})$
\\[0.2cm]
\noalign{\vskip 0.0cm}
\cdashline{1-4}[.4pt/2pt]
\noalign{\vskip 0.2cm}
\vspace{+0.2cm}
\multirow{1}{*}{\vspace{-0.0cm}$[\cO_{d^3Q^2u}]^{prstvw}$}
&\multirow{1}{*}{\vspace{-0.0cm}$RRL$}
&$[L_{d^3Q^2u}]^{\alpha\beta\gamma\rho\sigma\tau}_{prstvw}$
&$(u^\alpha_{Rp} d^\beta_{Rr})(d^\gamma_{Rs}d^\rho_{Rt})(u^\sigma_{Lv} d^\tau_{Lw})$
\\[0.0cm]
\noalign{\vskip 0.0cm}
\cdashline{1-4}[.4pt/2pt]
\noalign{\vskip 0.2cm}
\vspace{+0.2cm}
\multirow{2}{*}{\vspace{-0.3cm}$[\cO_{d^2Q^4}^{(1,2)}]^{prstvw}$}
&\multirow{2}{*}{\vspace{-0.3cm}$RLL$}
&$[L^{(1)}_{d^2Q^4}]^{\alpha\beta\gamma\rho\sigma\tau}_{prstvw}$
&$(d^\alpha_{Rp} d^\beta_{Rr})(u^\gamma_{Ls} u^\rho_{Lt})(d^\sigma_{Lv}d^\tau_{Lw})$
\\[0.2cm]
&
&$[L^{(2)}_{d^2Q^4}]^{\alpha\beta\gamma\rho\sigma\tau}_{prstvw}$
&$(d^\alpha_{Rp} d^\beta_{Rr})(u^\gamma_{Ls} d^\rho_{Lt})(d^\sigma_{Lv}u^\tau_{Lw})$
\\[0.2cm]
\bottomrule
\end{tabular}
}
\caption{Overview of the LEFT structures entering the mapping from SMEFT onto the chiral basis. The first column lists the relevant SMEFT operators, while the second column indicates their chirality class. The third and fourth columns display the corresponding LEFT operator structures obtained after explicitly expanding over $\SU(2)_{\rm L}$ indices, together with their definitions.} 
\label{tab:SMEFT_ch_match_LEFT_strs}
\end{table}

\paragraph{$\bm{RRR}$ type.} We begin with the $RRR$ chirality class, which involves two SMEFT operators
\begin{equation}
	\cL_{\sscript{SMEFT}}^{{\scalebox{0.56}{$|\Delta B|=2$}}}\supset
	[\cC_{d^4u^2}^{(1)}]_{prstvw}[\cO_{d^4u^2}^{(1)}]^{prstvw}+[\cC_{d^4u^2}^{(2)}]_{prstvw}[\cO_{d^4u^2}^{(2)}]^{prstvw}\,.
\end{equation}
Using the color-tensor identities listed in Appendix~\ref{app:color_tensors}, this Lagrangian can be rewritten as
\begin{equation}\label{eq:RRR_type_Lag_trafo}
	\begin{alignedat}{2}
		\cL_{\sscript{LEFT}}^{{\scalebox{0.56}{$|\Delta B|=2$}}}&\supset
		\frac{1}{4}[\cC_{d^4u^2}^{(1)}+\cC_{d^4u^2}^{(2)}]_{prstvw}T^{SSS}_{\{\alpha\beta\}\{\gamma\rho\}\{\sigma\tau\}}[L_{d^4u^2}]_{prstvw}^{\alpha\beta\gamma\rho\sigma\tau}
		\\[3pt]&
		+\frac{1}{4}[\cC_{d^4u^2}^{(1)}-\cC_{d^4u^2}^{(2)}]_{prstvw}T^{AAS}_{[\alpha\beta][\gamma\rho]\{\sigma\tau\}}[L_{d^4u^2}]_{prstvw}^{\alpha\beta\gamma\rho\sigma\tau}
		\\[3pt]&
		+\frac{1}{4}\lzv [\cC_{d^4u^2}^{(1)}+\cC_{d^4u^2}^{(2)}]_{prstvw}+[\cC_{d^4u^2}^{(1)}-\cC_{d^4u^2}^{(2)}]_{stprvw} \dzv T^{AAS}_{[\alpha\beta][\sigma\tau]\{\gamma\rho\}}[L_{d^4u^2}]_{prstvw}^{\alpha\beta\gamma\rho\sigma\tau}\,.
	\end{alignedat}
\end{equation}
Restricting to the light-quark sector, we fix the right-handed up-quark index to $1$, while the down-quark indices are identified with $1,2$ ($2,3$) in the LEFT (chiral) basis. With this identification, the operators appearing in the first two lines can be mapped onto the chiral basis as
\begin{equation}\label{eq:RRR_iden_1}
	\begin{alignedat}{2}
	T^{SSS}_{\{\alpha\beta\}\{\gamma\rho\}\{\sigma\tau\}}[L_{d^4u^2}]_{1r1tvw}^{\alpha\beta\gamma\rho\sigma\tau}&\to \widetilde\cO_{RRR}^{1(r+1)1(t+1)(v+1)(w+1)}\,,
	\\[2pt]
	T^{AAS}_{[\alpha\beta][\gamma\rho]\{\sigma\tau\}}[L_{d^4u^2}]_{1r1tvw}^{\alpha\beta\gamma\rho\sigma\tau}&\to\cO_{RRR}^{1(r+1)1(t+1)(v+1)(w+1)}\,,
	\end{alignedat}
\end{equation}
while the contribution in the third line of Eq.~\eqref{eq:RRR_type_Lag_trafo} can be written as
\begin{equation}\label{eq:RRR_iden_2}
	\begin{alignedat}{2}
		T^{AAS}_{[\alpha\beta][\sigma\tau]\{\gamma\rho\}}[L_{d^4u^2}]_{prstvw}^{\alpha\beta\gamma\rho\sigma\tau}
		&=T^{AAS}_{[\alpha\beta][\gamma\rho]\{\sigma\tau\}}(u_{R}^\alpha d_{Rr}^\beta)(d_{Rv}^\gamma d_{Rw}^\rho)(u_{R}^\sigma d_{Rt}^\tau)
		\\[3pt]&\to \cO_{RRR}^{1(r+1)(v+1)(w+1)1(t+1)}\,.
	\end{alignedat}
\end{equation}
Implementing these identifications, the Lagrangian in Eq.~\eqref{eq:RRR_type_Lag_trafo} can be recast as
\begin{equation}
	\begin{alignedat}{2}
		\cL_{\chi}^{{\scalebox{0.56}{$|\Delta B|=2$}}}&\supset
		\frac{1}{4}[\cC_{d^4u^2}^{(1)}+\cC_{d^4u^2}^{(2)}]_{1r1tvw}\,\widetilde\cO_{RRR}^{1(r+1)1(t+1)(v+1)(w+1)}
		\\[2pt]&
		+\frac{1}{4}[\cC_{d^4u^2}^{(1)}-\cC_{d^4u^2}^{(2)}]_{1r1tvw}\,\cO_{RRR}^{1(r+1)1(t+1)(v+1)(w+1)}
		\\[2pt]&
		+\frac{1}{4}\lzv [\cC_{d^4u^2}^{(1)}+\cC_{d^4u^2}^{(2)}]_{1r1tvw}+[\cC_{d^4u^2}^{(1)}-\cC_{d^4u^2}^{(2)}]_{1t1rvw} \dzv\cO_{RRR}^{1(r+1)(v+1)(w+1)1(t+1)}\,.
	\end{alignedat}
\end{equation}
The matching between the SMEFT Wilson coefficients and the coefficients of the chiral irreducible representations is summarized in Table~\ref{tab:SMEFT_proj_Od4u2}.

\paragraph{$\bm{RRL}$ type.} For the $RRL$ class, the starting point is the SMEFT Lagrangian involving a single operator
\begin{equation}
	\cL_{\sscript{SMEFT}}^{{\scalebox{0.56}{$|\Delta B|=2$}}}\supset [\cC_{d^3Q^2u}]_{prstvw}[\cO_{d^3Q^2u}]^{prstvw}\,.
\end{equation}
Decomposing the $\SU(2)_{\rm L}$ indices and applying the color-tensor identities we have
\begin{equation}
	\begin{alignedat}{2}
		\cL_{\sscript{LEFT}}^{{\scalebox{0.56}{$|\Delta B|=2$}}}&\supset
		\frac{1}{4}\lzv [\cC^{\phantom{a}}_{d^3Q^2u}]_{prstvw}-[\cC^{\phantom{a}}_{d^3Q^2u}]_{prstwv} \dzv T^{SSS}_{\{\alpha\beta\}\{\gamma\rho\}\{\sigma\tau\}}[L_{d^3Q^2u}]^{\alpha\beta\gamma\rho\sigma\tau}_{prstvw}		
		\\[3pt]&
		+\frac{1}{4}\lzv [\cC^{\phantom{a}}_{d^3Q^2u}]_{prstvw}-[\cC^{\phantom{a}}_{d^3Q^2u}]_{prstwv} \dzv T^{AAS}_{[\alpha\beta][\gamma\rho]\{\sigma\tau\}}[L_{d^3Q^2u}]^{\alpha\beta\gamma\rho\sigma\tau}_{prstvw}
		\\[3pt]&
		+\frac{1}{4}\lzv [\cC^{\phantom{a}}_{d^3Q^2u}]_{prstvw}+[\cC^{\phantom{a}}_{d^3Q^2u}]_{prstwv} \dzv T^{ASA}_{[\alpha\beta]\{\gamma\rho\}[\sigma\tau]}[L_{d^3Q^2u}]^{\alpha\beta\gamma\rho\sigma\tau}_{prstvw}
		\\[3pt]&
		+\frac{1}{4}\lzv [\cC^{\phantom{a}}_{d^3Q^2u}]_{prstvw}+[\cC^{\phantom{a}}_{d^3Q^2u}]_{prstwv}\dzv T^{SAA}_{\{\alpha\beta\}[\gamma\rho][\sigma\tau]}[L_{d^3Q^2u}]^{\alpha\beta\gamma\rho\sigma\tau}_{prstvw}\,.
	\end{alignedat}
\end{equation}
As in the $RRR$ case, upon restricting to the light-quark indices and performing the identification analogous to Eqs.~\eqref{eq:RRR_iden_1} and \eqref{eq:RRR_iden_2}, we obtain
\begin{equation}
	\begin{alignedat}{2}
		\cL_{\chi}^{{\scalebox{0.56}{$|\Delta B|=2$}}}&\supset
		\frac{1}{4}\lzv [\cC^{\phantom{a}}_{d^3Q^2u}]_{1rst1w}-[\cC^{\phantom{a}}_{d^3Q^2u}]_{1rstw1} \dzv\,\widetilde\cO_{RRL}^{1(r+1)(s+1)(t+1)1(w+1)}
		\\[3pt]
		&+\frac{1}{4}\lzv [\cC^{\phantom{a}}_{d^3Q^2u}]_{1rst1w}-[\cC^{\phantom{a}}_{d^3Q^2u}]_{1rstw1} \dzv\,\cO_{RRL}^{1(r+1)(s+1)(t+1)1(w+1)}
		\\[3pt]
		&+\frac{1}{4}\lzv [\cC^{\phantom{a}}_{d^3Q^2u}]_{1rst1w}+[\cC^{\phantom{a}}_{d^3Q^2u}]_{1rstw1} \dzv\,\cO_{LRR}^{1(w+1)1(r+1)(s+1)(t+1)}
		\\[3pt]
		&+\frac{1}{4}\lzv [\cC^{\phantom{a}}_{d^3Q^2u}]_{1rst1w}+[\cC^{\phantom{a}}_{d^3Q^2u}]_{1rstw1} \dzv\,\cO_{LRR}^{1(w+1)(s+1)(t+1)1(r+1)}
		\,.
	\end{alignedat}
\end{equation}
The mapping between the SMEFT Wilson coefficients and the coefficients of the chiral operators is summarized in Table~\ref{tab:SMEFT_proj_Od3Q2u}.

\paragraph{$\bm{RLL}$ type.} Lastly, for the $RLL$ class, the SMEFT Lagrangian involves two operators
\begin{equation}
	\cL_{\sscript{SMEFT}}^{{\scalebox{0.56}{$|\Delta B|=2$}}}\supset
	[\cC_{d^2Q^4}^{(1)}]_{prstvw}[\cO_{d^2Q^4}^{(1)}]^{prstvw}+[\cC_{d^2Q^4}^{(2)}]_{prstvw}[\cO_{d^2Q^4}^{(2)}]^{prstvw}
	\,.
\end{equation}
Upon resolving the $\SU(2)_{\rm L}$ and the color indices, these can be rewritten as
\begin{equation}\label{eq:RLL_lagr_LEFT}
{\scalebox{0.95}{$
\begin{alignedat}{2}
			\cL_{\sscript{LEFT}}^{{\scalebox{0.56}{$|\Delta B|=2$}}}&\supset
		\frac{1}{4}[\cC_{d^2Q^4}^{(1+2)}]_{prstvw}\,T^{SSS}_{\{\alpha\beta\}\{\gamma\rho\}\{\sigma\tau\}}[L^{(1)}_{d^2Q^4}]^{\alpha\beta\gamma\rho\sigma\tau}_{prstvw}
		-
		\frac{1}{4}[\cC_{d^2Q^4}^{(1+2)}]_{prstvw}\,T^{SSS}_{\{\alpha\beta\}\{\gamma\rho\}\{\sigma\tau\}}[L^{(2)}_{d^2Q^4}]^{\alpha\beta\gamma\rho\sigma\tau}_{prstvw}
		\\[3pt]
		&
		+\frac{1}{4}[\cC_{d^2Q^4}^{(1+2)}]_{prstvw}\,T^{AAS}_{[\alpha\beta][\gamma\rho]\{\sigma\tau\}}[L^{(1)}_{d^2Q^4}]^{\alpha\beta\gamma\rho\sigma\tau}_{prstvw}
		-\frac{1}{4}[\cC_{d^2Q^4}^{(1+2)}]_{prstvw}\,T^{AAS}_{[\alpha\beta][\gamma\rho]\{\sigma\tau\}}[L^{(2)}_{d^2Q^4}]^{\alpha\beta\gamma\rho\sigma\tau}_{prstvw}
		\\[3pt]
		&+\frac{1}{4}[\cC_{d^2Q^4}^{(1-2)}]_{prstvw}\,T^{ASA}_{[\alpha\beta]\{\gamma\rho\}[\sigma\tau]}[L^{(1)}_{d^2Q^4}]^{\alpha\beta\gamma\rho\sigma\tau}_{prstvw}
		-\frac{1}{4}[\cC_{d^2Q^4}^{(1-2)}]_{prstvw}\,T^{ASA}_{[\alpha\beta]\{\gamma\rho\}[\sigma\tau]}[L^{(2)}_{d^2Q^4}]^{\alpha\beta\gamma\rho\sigma\tau}_{prstvw}
		\\[3pt]
		&+\frac{1}{4}[\cC_{d^2Q^4}^{(1-2)}]_{prstvw}\,T^{SAA}_{\{\alpha\beta\}[\gamma\rho][\sigma\tau]}[L^{(1)}_{d^2Q^4}]^{\alpha\beta\gamma\rho\sigma\tau}_{prstvw}
		-\frac{1}{4}[\cC_{d^2Q^4}^{(1-2)}]_{prstvw}\,T^{SAA}_{\{\alpha\beta\}[\gamma\rho][\sigma\tau]}[L^{(2)}_{d^2Q^4}]^{\alpha\beta\gamma\rho\sigma\tau}_{prstvw}
		\,,
\end{alignedat}
$}}
\end{equation}
where, for brevity, we introduce
\begin{equation}
	\begin{alignedat}{2}
		[\cC_{d^2Q^4}^{(1+2)}]_{prstvw}&=[\cC_{d^2Q^4}^{(1)}]_{prstvw}+[\cC_{d^2Q^4}^{(1)}]_{prvwst}+[\cC_{d^2Q^4}^{(2)}]_{prstvw}+[\cC_{d^2Q^4}^{(2)}]_{prvwst}\,,
		\\[3pt]
		[\cC_{d^2Q^4}^{(1-2)}]_{prstvw}&=[\cC_{d^2Q^4}^{(1)}]_{prstvw}+[\cC_{d^2Q^4}^{(1)}]_{prvwst}-[\cC_{d^2Q^4}^{(2)}]_{prstvw}-[\cC_{d^2Q^4}^{(2)}]_{prvwst}\,.
	\end{alignedat}
\end{equation}
Applying the light-quark restriction and identifying the chiral structures, the Lagrangian in Eq.~\eqref{eq:RLL_lagr_LEFT} can be rewritten as
\begin{equation}
	\begin{alignedat}{2}
		\cL_{\chi}^{{\scalebox{0.56}{$|\Delta B|=2$}}}&\supset
		\frac{1}{4}[\cC_{d^2Q^4}^{(1+2)}]_{pr11vw}\,\widetilde\cO_{LLR}^{11(v+1)(w+1)(p+1)(r+1)}
		-\frac{1}{4}[\cC_{d^2Q^4}^{(1+2)}]_{pr1tv1}\,\widetilde\cO_{LLR}^{1(t+1)(v+1)1(p+1)(r+1)}
		\\[3pt]&
		+\frac{1}{4}[\cC_{d^2Q^4}^{(1+2)}]_{pr11vw}\,\cO_{RLL}^{(p+1)(r+1)11(v+1)(w+1)}
		-\frac{1}{4}[\cC_{d^2Q^4}^{(1+2)}]_{pr1tv1}\,\cO_{RLL}^{(p+1)(r+1)1(t+1)(v+1)1}
		\\[3pt]&
		+\frac{1}{4}[\cC_{d^2Q^4}^{(1-2)}]_{pr11vw}\,\cO_{RLL}^{(p+1)(r+1)(v+1)(w+1)11}
		-\frac{1}{4}[\cC_{d^2Q^4}^{(1-2)}]_{pr1tv1}\,\cO_{RLL}^{(p+1)(r+1)(v+1)11(t+1)}
		\\[3pt]&
		+\frac{1}{4}[\cC_{d^2Q^4}^{(1-2)}]_{pr11vw}\,\cO_{LLR}^{11(v+1)(w+1)(p+1)(r+1)}
		-\frac{1}{4}[\cC_{d^2Q^4}^{(1-2)}]_{pr1tv1}\,\cO_{LLR}^{1(t+1)(v+1)1(p+1)(r+1)}\,,
	\end{alignedat}
\end{equation}
where the mapping between the SMEFT Wilson coefficients and the coefficients of the chiral operators is summarized in Tables~\ref{tab:SMEFT_proj_Od2Q4_part1} and \ref{tab:SMEFT_proj_Od2Q4_part2}.

\section{Matching onto Baryon Chiral Perturbation Theory}
\label{sec:matching_BxPT}

In this section, we first review the chiral fields and building blocks relevant for the construction, followed by the derivation of the corresponding operators and matching procedure.

\subsection{Review of Chiral Fields and Lagrangian}
\label{sec:BxPT_fram_and_defs}
In Section~\ref{sec:chiral_sym_overview}, we introduced the chiral symmetry structure of QCD and discussed its spontaneous breaking pattern. In particular, we identified the organization of the light-quark fields and the emergence of the diagonal $\SU(3)_V$ subgroup as the relevant symmetry of the vacuum. We now extend this discussion to energies below the QCD confinement scale, where the appropriate degrees of freedom are hadronic fields, namely the light pseudoscalar mesons and baryons.

In this regime, the dynamics of the light pseudoscalar mesons can be described using the CCWZ formalism~\cite{Coleman:1969sm,Callan:1969sn}, which provides a systematic construction of the effective theory associated with the spontaneous breaking $\SU(3)_R \times \SU(3)_L \to \SU(3)_V$. The Goldstone bosons are parametrized in terms of the matrix-valued fields
\begin{equation}
    \Sigma(x)\equiv e^{2i \pi^a(x) T^a/f_\pi}\,,
    \qquad \xi(x)\equiv e^{i \pi^a(x) T^a/f_\pi}\,,
\end{equation}
where $f_\pi$ is the pion decay constant $f_\pi=130.41(20)\,\mev$~\cite{ParticleDataGroup:2024cfk}, $T^a$ are the generators of $\SU(3)$, and
\begin{equation}
\pi(x)\equiv\pi^a(x) T^a =
\begin{bmatrix}
\frac{1}{\sqrt{2}}\pi^0+\frac{1}{\sqrt{6}}\eta & \pi^+ & K^+ \\
\pi^- & -\frac{1}{\sqrt{2}}\pi^0+\frac{1}{\sqrt{6}}\eta & K^0 \\
K^- & \bar K^0 & -\frac{2}{\sqrt{6}}\eta
\end{bmatrix}\,.
\end{equation}
Under chiral transformations, these fields transform as
\begin{equation}
    \Sigma(x)\rightarrow L\Sigma(x)R^\dag\,,
    \qquad
    \xi(x)\rightarrow L\xi(x)U^\dag(x)=U(x)\xi(x)R^\dag\,,
\end{equation}
where $U(x)\in \SU(3)_V$ is a compensating field that depends non-linearly on $\xi(x)$, $L$, and $R$. Under parity, the fields transform as
\begin{equation}\label{eq:BxPT_parity_trafo_meson}
    \pi^a(t,\vec{x})\rightarrow -\pi^a(t,-\vec{x})
    \quad\implies\quad
    \Sigma(x)\rightarrow \Sigma^\dag(t,-\vec{x})\,,
    \qquad
    \xi(x)\rightarrow \xi^\dag(t,-\vec{x})\,.
\end{equation}

In addition to the mesonic sector, the low-energy theory also contains baryonic degrees of freedom. These are incorporated by introducing the octet baryon field
\begin{equation}
    B(x)\equiv B^a(x) T^a=
    \begin{bmatrix}
        \frac{1}{\sqrt{2}}\Sigma^0+\frac{1}{\sqrt{6}}\Lambda & \Sigma^+ & p \\
        \Sigma^- & -\frac{1}{\sqrt{2}}\Sigma^0+\frac{1}{\sqrt{6}}\Lambda & n \\
        \Xi^- & \Xi^0 & -\frac{2}{\sqrt{6}}\Lambda
    \end{bmatrix}\,,
\end{equation}
which transforms under chiral symmetry as
\begin{equation}
    B(x)\rightarrow e^{i\alpha_B}U(x)B(x)U^\dag(x)\,,
\end{equation}
where $U(x)$ is the same compensating transformation as above, and $\alpha_B$ denotes the baryon number phase. It is also convenient to define the chiral projections $B_{L,R}=P_{L,R}B$, which transform under parity as
\begin{equation}
    B_{R,L}(t,\vec{x})\rightarrow \gamma^0 B_{L,R}(t,-\vec{x})\,.
\end{equation}

In order to construct the operators relevant for the chiral matching, it is necessary to identify suitable combinations of meson and baryon fields that transform in a controlled way under the chiral symmetry group, allowing for a systematic group-theoretical classification of the resulting operator structures. To this end, we consider composite objects formed by dressing the baryon field with the $\xi(x)$ field, namely $\xi B\xi$, $\xi^\dag B\xi^\dag$, $\xi B\xi^\dag$ and $\xi^\dag B \xi$. By construction, these combinations exhibit well-defined transformation properties under $\SU(3)_R \times \SU(3)_L$, which can be readily derived from the transformation rules of $\xi(x)$ and $B(x)$:
\begin{equation}\label{eq:xi_B_xi_traf_rules}
{\scalebox{0.98}{$    \xi B \xi \to L(\xi B \xi)R^\dag\,,
    \quad
    \xi^\dag B \xi^\dag \to R(\xi^\dag B \xi^\dag)L^\dag\,,
    \quad
    \xi B \xi^\dag \to L(\xi B \xi^\dag)L^\dag\,,
    \quad
    \xi^\dag B \xi \to R(\xi^\dag B \xi)R^\dag\,.  $}}  
\end{equation}
In addition, the transformation rules in Eq.~\eqref{eq:xi_B_xi_traf_rules} allow for a direct identification of the chiral representation under which each of the composite fields transforms:
\begin{equation}\label{eq:BxPT_objects_charge_p1}
    \xi B \xi \sim (\repbar 3_R,\rep3_L)\,,
    \quad \xi^\dag B \xi^\dag \sim (\rep 3_R,\repbar3_L)\,,
    \quad \xi B \xi^\dag \sim (\rep 1_R,\rep8_L)\,,
    \quad \xi^\dag B \xi \sim (\rep 8_R,\rep1_L)\,.
\end{equation}
Finally, we introduce the axial-vector quantity $\xi_\mu \equiv i(\xi^\dagger\partial_\mu\xi-\xi\partial_\mu\xi^\dagger)$, which transforms in the adjoint representation under $\SU(3)_V$. In terms of these ingredients, the leading-order baryon chiral perturbation theory Lagrangian can be written as~\cite{Claudson:1981gh,Bernard:1995dp}\footnote{Interaction terms involving vector currents of the form $\bar B \gamma^\mu B$ vanish at leading order~\cite{Claudson:1981gh}.}
\begin{equation}\label{eq:BxPT_L0}
    \begin{alignedat}{2}
        \cL_{\sscript{B\chi PT}}^{0}&=
        \frac{f_\pi^2}{8}\,\Tr\!\Big[\partial_\mu\Sigma\,\partial^\mu\Sigma^\dagger\Big]
        +\Tr\!\Big[ \bar B(i\slashed{\partial}-M_B)B \Big]
        -\frac{D}{2}\Tr\!\Big[ \bar B\gamma^\mu\gamma^5\{\xi_\mu,B\} \Big]
        \\[3pt]&
        -\frac{F}{2}\Tr\!\Big[ \bar B\gamma^\mu\gamma^5[\xi_\mu,B] \Big]\,,
    \end{alignedat}
\end{equation}
Expanding Eq.~\eqref{eq:BxPT_L0} to first order in the pseudoscalar fields yields the familiar derivative meson-baryon interactions. Restricting to interaction terms involving two baryons and one meson the relevant Lagrangian can be written in a compact form as
\begin{equation}\label{eq:BxPT_L0_compact_expr}
    \cL_{\sscript{B\chi PT}}^{0}\supset \frac{1}{f_\pi}\sum_{f_1,f_2}\sum_{h} \cC_{f_1 f_2 h}^\sscript{LO} \bar f_1 \gamma^\mu \gamma^5 f_2\, \partial_\mu h\,,
\end{equation}
where the sum runs over baryon fields $f_{1,2}$ and pseudoscalar mesons $h$, and $\cC_{f_1 f_2 h}^\sscript{LO}$ denotes the leading-order coefficient. Upon expansion, the compact expression above reproduces the familiar leading-order Lagrangian in the form
\begin{equation}\label{eq:BxPT_L0_expanded}
\begin{alignedat}{2}
\cL_{\sscript{B\chi PT}}^{0}\supset
&-\frac{D+3F}{\sqrt6\,f_\pi}\,(\bar{\Lambda}\gamma^\mu\gamma^5 n)\,\partial_\mu \bar K^0
-\frac{D+3F}{\sqrt6\,f_\pi}\,(\bar{\Lambda}\gamma^\mu\gamma^5 p)\,\partial_\mu K^-
+\frac{D+F}{f_\pi}\,(\overline p\gamma^\mu\gamma^5 n)\,\partial_\mu\pi^+
\\[0.2cm]
&+\frac{3F-D}{2\sqrt6\,f_\pi}\left(\overline p \gamma^\mu\gamma^5 p+\overline n \gamma^\mu\gamma^5 n\right)\partial_\mu\eta
+\frac{D+F}{2\sqrt2\,f_\pi}\left(\overline p\gamma^\mu\gamma^5 p-\overline n\gamma^\mu\gamma^5 n\right)\partial_\mu\pi^0+\hermc\,,
\end{alignedat}
\end{equation}
where we retain only the interaction terms involving the $p$, $n$ and $\Lambda$ baryons, which constitute the relevant degrees of freedom for the phenomenological analysis in Section~\ref{sec:pheno_applications}. For the mesonic sector, the interactions involving $\eta$ and $\pi^0$ are Hermitian, such that the corresponding Hermitian conjugate terms yield an overall factor of $2$, which accounts for the normalization of these contributions. The axial couplings are extracted as $F \simeq 0.46$ and $D \simeq 0.80$~\cite{Bali:2022qja,Aoki:2008ku,ParticleDataGroup:2024cfk}.

\subsection{Construction of B$\chi$PT Operators} \label{sec:construction_BChiPT_ops}
With the B$\chi$PT framework established, we now extend this formalism to the $|\Delta B|=2$ sector and construct the corresponding hadronic realizations of the chiral irreducible representations identified in Section~\ref{sec:chiral_ops_reps_constr}. The guiding principle is to preserve the chiral transformation properties of the underlying quark-level operators, ensuring a consistent matching between the two descriptions.

\begin{table}[t]
\centering
\renewcommand{\arraystretch}{1.2}
\scalebox{0.75}{
\begin{tabular}{c@{\hspace{0.7cm}}>{\raggedright\arraybackslash}p{1.0cm}@{\hspace{2.2cm}}c}
\toprule
\multirow{1}{*}{\textbf{Irrep}}
&\multirow{1}{*}{\textbf{Operator}}
&\multirow{1}{*}{\textbf{Definition}}
\\
\midrule
\addlinespace[0.2cm]
\multirow{2}{*}{\vspace{-0.3cm}$\repbar6_R\otimes\rep6_L$}
&$[\mathcal N^{(1)}_{\repbar6_R\otimes\rep6_L}]\ud{\dot a\dot b}{ab}$
&$\frac{1}{4}\lzs [\xi \bar B^c\xi]\ud{\dot a}{a}[\xi B\xi]\ud{\dot b}{b}+[\xi \bar B^c\xi]\ud{\dot a}{b}[\xi B\xi]\ud{\dot b}{a} \dzs$
\\[0.2cm]
&$[\mathcal N^{(2)}_{\repbar6_R\otimes\rep6_L}]\ud{\dot a\dot b}{ab}$
&$\frac{1}{4}\lzs [\xi \bar B^cB\xi]\ud{\dot a}{a}[\Sigma]\ud{\dot b}{b}+[\xi \bar B^cB\xi]\ud{\dot a}{b}[\Sigma]\ud{\dot b}{a} \dzs$
\\[0.2cm]
\noalign{\vskip 0.0cm}
\cdashline{1-3}[.4pt/2pt]
\noalign{\vskip 0.2cm}
\vspace{+0.2cm}
\multirow{2}{*}{\vspace{-0.3cm}$\rep{15}_R\otimes\repbar3_L$}
&$[\mathcal N^{(1)}_{\rep{15}_R\otimes\repbar3_L}]\ud{ab}{c,\dot a}$
&$\frac{1}{4}\Big[ [\xi^\dag \bar B^c\xi]\ud{a}{c}[\xi^\dag B\xi^\dag]\ud{b}{\dot a}+[\xi^\dag \bar B^c\xi]\ud{b}{c}[\xi^\dag B\xi^\dag]\ud{a}{\dot a} \Big]
-\frac{1}{16}\Big[\delta\ud{a}{c}[\xi^\dag \bar B^cB\xi^\dag]\ud{b}{\dot a}+\delta\ud{b}{c}[\xi^\dag \bar B^cB\xi^\dag]\ud{a}{\dot a}\Big]
$
\\
&$[\mathcal N^{(2)}_{\rep{15}_R\otimes\repbar3_L}]\ud{ab}{c,\dot a}$
&$\frac{1}{4}\Big[ [\xi^\dag \bar B^cB\xi]\ud{a}{c}[\Sigma^\dag]\ud{b}{\dot a}+[\xi^\dag \bar B^cB\xi]\ud{b}{c}[\Sigma^\dag]\ud{a}{\dot a} \Big]
-\frac{1}{16}\Big[ \delta\ud{a}{c}[\xi^\dag \bar B^cB\xi^\dag]\ud{b}{\dot a}+\delta\ud{b}{c}[\xi^\dag \bar B^cB\xi^\dag]\ud{a}{\dot a} \Big]$
\\
\noalign{\vskip 0.2cm}
\cdashline{1-3}[.4pt/2pt]
\noalign{\vskip 0.2cm}
\multirow{3}{*}{\vspace{-0.4cm}$\rep3_R\otimes\repbar3_L$}
&$[\mathcal N^{(1)}_{\rep3_R\otimes\repbar3_L}]\ud{a}{\dot a}$
&$\frac{1}{2}[\xi^\dag \bar B^cB\xi^\dag]\ud{a}{\dot a}$
\\[0.2cm]
&$[\mathcal N^{(2)}_{\rep3_R\otimes\repbar3_L}]\ud{a}{\dot a}$
&$\frac{1}{8}\varepsilon^{abc}\varepsilon_{\dot a\dot b\dot c}[\xi \bar B^c\xi]\ud{\dot b}{b}[\xi B\xi]\ud{\dot c}{c}$
\\[0.2cm]
&$[\mathcal N^{(3)}_{\rep3_R\otimes\repbar3_L}]\ud{a}{\dot a}$
&$\frac{1}{8}\varepsilon^{abc}\varepsilon_{\dot a\dot b\dot c}[\xi \bar B^cB\xi]\ud{\dot b}{b}[\Sigma]\ud{\dot c}{c}$
\\[0.2cm]
\noalign{\vskip 0.0cm}
\cdashline{1-3}[.4pt/2pt]
\noalign{\vskip 0.2cm}
$\rep{15}_R'\otimes \rep6_L$
&$[\mathcal N_{\rep{15}_R'\otimes \rep6_L}]^{abcd,\dot a \dot b}$
&$\frac{1}{48}\varepsilon^{\dot a \dot d \dot e}\varepsilon^{\dot b \dot f \dot g}\sum_{\sigma\in S_4}[\xi^\dag \bar B^c\xi^\dag]\ud{\sigma(a)}{\dot d}[\Sigma^\dag]\ud{\sigma(b)}{\dot e}[\xi^\dag B\xi^\dag]\ud{\sigma(c)}{\dot f}[\Sigma^\dag]\ud{\sigma(d)}{\dot g}$
\\[0.2cm]
\noalign{\vskip 0.0cm}
\cdashline{1-3}[.4pt/2pt]
\noalign{\vskip 0.2cm}
$\rep1_R\otimes\rep1_L$
&$\mathcal N_{\rep1_R\otimes\rep1_L}$&$\frac{1}{2}\Tr[\bar B^cB]$
\\[0.2cm]
\noalign{\vskip 0.0cm}
\cdashline{1-3}[.4pt/2pt]
\noalign{\vskip 0.2cm}
\multirow{2}{*}{$\rep{27}_R\otimes\rep1_L$}
&\multirow{2}{*}{$[\mathcal N_{\rep{27}_R\otimes\rep1_L}]\ud{ab}{cd}$}
&$\frac{1}{4}\Big[[\xi^\dag \bar B^c \xi]\ud{a}{c}[\xi^\dag B\xi]\ud{b}{d}+[\xi^\dag \bar B^c \xi]\ud{a}{d}[\xi^\dag B\xi]\ud{b}{c}\Big]
+\frac{1}{80}\Big[\delta\ud{a}{c}\delta\ud{b}{d}+\delta\ud{b}{c}\delta\ud{a}{d}\Big]\Tr[\bar B^c B]$
\\[3pt]
&&$-\frac{1}{20}\Big[\delta\ud{a}{c}[\xi^\dag \bar B^c B\xi]\ud{b}{d}+\delta\ud{b}{c}[\xi^\dag \bar B^c B\xi]\ud{a}{d}+\delta\ud{a}{d}[\xi^\dag \bar B^c B\xi]\ud{b}{c}+\delta\ud{b}{d}[\xi^\dag \bar B^c B\xi]\ud{a}{c} \Big]$
\\[0.2cm]
\noalign{\vskip 0.0cm}
\cdashline{1-3}[.4pt/2pt]
\noalign{\vskip 0.2cm}
\vspace{+0.2cm}
\multirow{2}{*}{$\rep{28}_R\otimes\rep1_L$}
&\multirow{2}{*}{$[\mathcal N_{\rep{28}_R\otimes\rep1_L}]^{abcdef}$}
&$\frac{1}{(4\pi f_\pi)^2}\frac{1}{1440}\varepsilon^{\dot a\dot b\dot c}\varepsilon^{\dot d\dot e\dot f}
\sum_{\sigma\in S_6}[\xi^\dag \bar B^c\xi]\ud{\sigma(a)}{\dot a}[\Sigma^\dag]\ud{\sigma(b)}{\dot b}[(\partial_\mu\Sigma)^\dag]\ud{\sigma(c)}{\dot c}$
\\[-0.2cm]
&&$\times [\xi^\dag B\xi]\ud{\sigma(d)}{\dot d}[\Sigma^\dag]\ud{\sigma(e)}{\dot e}[(\partial_\mu\Sigma)^\dag]\ud{\sigma(f)}{\dot f}$
\\[0.2cm]
\bottomrule
\end{tabular}
}
\caption{Overview of the B$\chi$PT realizations of the non-redundant chiral irreducible representations identified in Section~\ref{sec:chiral_ops_reps_constr}. The first column lists the chiral representations, the second column labels the corresponding B$\chi$PT operator structures, and the third column provides their explicit expressions in terms of building blocks carrying baryon number $+1$ and $+2$. The parity-conjugate structures can be obtained by the exchange $L \leftrightarrow R$, accompanied by $\xi \leftrightarrow \xi^\dag$ and $\Sigma \leftrightarrow \Sigma^\dag$. Undotted indices correspond to right-handed components, while dotted indices denote left-handed ones.}
\label{tab:BxPT_fields}
\end{table}

To this end, we again rely on group-theoretical considerations, together with the building blocks introduced above, which transform in definite representations of $\SU(3)_R \times \SU(3)_L$. In addition to the objects defined in Eq.~\eqref{eq:BxPT_objects_charge_p1}, whose baryon number is $+1$, one can construct composite fields carrying baryon number $+2$, which will serve as the additional basic ingredients for the operator construction. These are given by
\begin{equation}\label{eq:BxPT_objects_charge_p2}
    \xi \bar B^c B \xi \sim (\repbar 3_R,\rep3_L)\,,
    \quad \xi \bar B^c B \xi^\dag \sim (\rep 1_R,\rep8_L)\,,
    \quad \xi^\dag \bar B^c B \xi \sim (\rep 8_R,\rep1_L)\,,
    \quad \xi^\dag \bar B^c B \xi^\dag \sim (\rep 3_R,\repbar3_L)\,.
\end{equation}
With these objects at hand, together with the list of admissible chiral irreducible representations (see Table~\ref{tab:chiral_basis_op_count}), one can construct explicit realizations of the corresponding $|\Delta B|=2$ operators in B$\chi$PT. As a general strategy, we first identify all structures that can be formed without derivatives or quark-mass-spurion insertions.\footnote{We emphasize that this choice is primarily guided by simplicity and it is not meant to imply a strict hierarchy, especially in view of the known limitations in the convergence of the B$\chi$PT expansion in the kinematical regime of interest~\cite{Jenkins:1990jv,Jenkins:1991es,Krause:1990xc}.} In certain cases, however, such leading constructions vanish identically, and non-vanishing realizations can only be obtained by including derivatives or quark-mass spurions. The resulting non-redundant operator structures are summarized in Table~\ref{tab:BxPT_fields}. In what follows, we briefly outline the technical steps involved in constructing the corresponding operator structures:
\begin{itemize}
    \item For the $\repbar6_R\otimes\rep6_L$, the construction relies on the decompositions $\repbar3_R\otimes\repbar3_R\supset\repbar6_R$ and $\rep3_L\otimes\rep3_L\supset\rep6_L$, obtained by symmetrizing over the corresponding chiral indices. Examining the available building blocks, one finds two independent B$\chi$PT realizations, corresponding to inequivalent ways of forming symmetric combinations consistent with the required transformation properties.
    \item For the $\rep{15}_R\otimes\repbar3_L$, the construction proceeds by forming the $\rep{15}_R$ component through the decomposition $\rep8_R\otimes\rep3_R\supset \rep{15}_R$, while the $\repbar3_L$ component is already provided by the available building blocks. Examining the possible combinations with these transformation properties, one finds two independent realizations. 
    \item For the $\rep3_R\otimes \repbar3_L$,  one immediately observes from the transformation properties of the $|\Delta B|=2$ building blocks in Eq.~\eqref{eq:BxPT_objects_charge_p2} that the $\xi^\dag\bar B^c B\xi^\dag$ transforms in the desired representation, providing one direct realization. In addition, further realizations can be constructed using the decompositions $\repbar3_R\otimes\repbar3_R\supset\rep3_R$ and $\rep3_L\otimes\rep3_L\supset\repbar3_L$, where the projection onto the triplet (antitriplet) is obtained by contracting the corresponding indices with the Levi-Civita tensor, yielding two additional independent realizations.
    \item For the $\rep{15}_R'\otimes \rep6_L$, there exists a single independent realization. In this case, in addition to the $|\Delta B|=2$ building blocks, the construction requires two insertions of $\Sigma^\dag$ in order to achieve the correct transformation properties. The $\rep{15}_R'$ component is obtained from the decomposition $\rep3_R^{\otimes 4}\supset\rep{15}_R'$, and is selected by fully symmetrizing over the fundamental indices. For the left-handed chirality, the $\rep6_L$ is extracted from the decomposition of the form $\repbar3_L^{\otimes4}\supset\rep3_L\otimes\rep3_L\supset\rep6_L$, yielding in total the required transformation properties.
    \item For the $\rep1_R\otimes\rep1_L$, the B$\chi$PT realization is given by the singlet contraction $\Tr[\bar B^c B]$, which is invariant under $\SU(3)_R \times \SU(3)_L$ using the cyclicity of the trace.
    \item For the $\rep{27}_R\otimes\rep1_L$, the construction can be achieved using two insertions of the $\xi^\dag B\xi$,  object each of which transforms as $\rep8_R$. The desired representation is then obtained through the decomposition $\rep8_R\otimes\rep8_R\supset\rep{27}_R$, while the singlet structure in the left-handed sector is trivially preserved. The projection onto $\rep{27}_R$ is realized by symmetrizing over the adjoint indices and removing the singlet and octet components, ensuring a traceless tensor structure. Alternatively, one may consider a construction based on two insertions of $\xi B\xi$ accompanied by two insertions of $\Sigma^\dag$. However, this leads to an equivalent structure, as the identity $\Sigma^\dag \xi B\xi=\xi^\dag B\xi$ allows one to rewrite the resulting operator in terms of the form above.
    \item For the $\rep{28}_R\otimes\rep1_L$, a naive construction starts from two insertions of $\xi^\dag B\xi$ supplemented by four insertions of $\Sigma^\dag$. In this case, the desired chiral structure would follow from the decomposition $\rep3_R^{\otimes6}\supset\rep{28}_R$ and $\rep3_L^{\otimes6}\supset\rep1_L$, obtained by fully symmetrizing over the fundamental indices in the right-handed sector, while contracting the left-handed indices using the Levi-Civita tensor. However, this construction vanishes identically. A non-vanishing realization can instead be obtained by replacing two of the $\Sigma^\dag$ insertions by $(\partial_\mu\Sigma)^\dag$, or, alternatively, with spurion insertions such as $M^\dag$. Since the latter are parametrically suppressed, we restrict to the derivative construction. This leads to a non-vanishing structure reproducing the required transformation properties. Given the presence of derivatives, this operator is normalized with two inverse powers of $4\pi f_\pi$, the natural cutoff scale of B$\chi$PT.
\end{itemize}

\subsection{Matching Procedure}
After constructing the chiral irreducible representations and their corresponding B$\chi$PT realizations, we can now assemble the effective Lagrangian at the hadronic level. Starting from the classification of operators in terms of chiral irreducible representations, the effective Lagrangian can be written as
\begin{equation}\label{eq:chi_lagr_general}
    \cL_{\chi}^{{\scalebox{0.56}{$|\Delta B|=2$}}}=\sum_\cR \sum_I \cC_{\cR}^I\cO_\cR^I\,,
    \qquad
    \cO_\cR^I=[\cX_\cR\lvert^I]_{ijk\ell mn}\cO_{\chi_1\chi_2\chi_3}^{ijk\ell mn}\,,
\end{equation}
where the sum runs over all non-redundant chiral irreducible representations $\cR$ and, for each representation, over the corresponding independent components labeled by $I$. The coefficients $\cC_{\cR}^I$ denote the Wilson coefficients associated with the given chiral structure, while $\cO_\cR^I$ represent the corresponding operators (see Appendix~\ref{app:complete_ch_op_basis}).

In order for the effective Lagrangian in Eq.~\eqref{eq:chi_lagr_general} to be invariant under $\SU(3)_R \times \SU(3)_L$, the Wilson coefficients $\cC_{\cR}^I$ are promoted to spurion tensors transforming in the conjugate representation $\cR^*$.\footnote{The spurion method is standard $\chi$PT: for instance, the quark mass matrix $M$, which explicitly breaks chiral symmetry, is promoted to a spurion transforming as $M\to LMR^\dag$. Similarly, electromagnetic effects can be incorporated by promoting the quark charge matrix $Q$ to a spurion transforming as $Q \to L Q L^\dag = R Q R^\dag$, allowing for the construction of invariant operators proportional to $e^2$, which encode electromagnetic symmetry breaking, for instance in the pion mass splitting~\cite{Pich:1995bw,Das:1967it,Urech:1994hd,Neufeld:1995mu}.} In this way, each term in Eq.~\eqref{eq:chi_lagr_general} is constructed as an invariant contraction between $\cC_{\cR}^I$ and $\cO_\cR^I$, ensuring that the full Lagrangian transforms as a singlet under the chiral symmetry group. 

Using the chiral irreducible representations and their corresponding realizations in B$\chi$PT, we can now go ahead and formulate the matching between the chiral operator basis and the hadronic theory. The key point is that a given chiral operator $\cO_\cR^I$ may admit one or more independent realizations in B$\chi$PT, such that the mapping can be written as
\begin{equation}\label{eq:ch_BxPT_operator_mapping}
    \cO_{\cR}^I\longrightarrow \sum_k \alpha_{\cR}^{(k)}\,[\cN_\cR^{(k)}]^I\,,
    \qquad
    \cN_\cR^{(k)}\equiv \cN_\cR^{(k)}(B,\xi)\,,
\end{equation}
where $[\cN_\cR^{(k)}]^I$ denotes an independent B$\chi$PT realization of the same chiral structure, and the index $k$ labels the distinct hadronic realizations when more than one is present (see Table~\ref{tab:BxPT_fields}). The coefficients $\alpha_{\cR}^{(k)}$ are low-energy constants (LECs) that encode the non-perturbative QCD dynamics associated with each realization and are, in general, determined using non-perturbative methods such as lattice QCD~\cite{Aoki:2013yxa,Aoki:2017puj,Yoo:2021gql}. At present, however, only a limited subset of these quantities is known. In particular, four independent LECs have been computed on the lattice, providing quantitative input for a corresponding subset of operator structures, with numerical values at $\mu=2\,\gev$ given by~\cite{Rinaldi:2018osy,Rinaldi:2019thf}\footnote{{\color{black}{Since the known LECs are all found to be of the same order of magnitude, \(\alpha_\cR^{(k)}\sim 10^{-4}\,\gev^6\), we assume that the remaining LECs are expected to be of comparable size.}}}
\begin{equation}\label{eq:LEC_in_BxPT}
\begin{alignedat}{8}
    &\alpha_{\repbar6_R\otimes\rep6_L}^{(1)}&&=25(6)\times 10^{\eminus5}\,\gev^6\,,\quad\qquad &&\alpha_{\rep{15}_R\otimes\repbar{3}_L}^{(1)}&&=-48(9)\times 10^{\eminus5}\,\gev^6\,,
    \\[3pt]
    &\alpha_{\rep{15}'_R\otimes\rep{6}_L}&&=2.12(96)\times 10^{\eminus5}\,\gev^6\,,\quad\qquad &&\alpha_{\rep{27}_R\otimes\rep1_L}&&=23(7)\times 10^{\eminus5}\,\gev^6\,.
    \end{alignedat}
\end{equation}
The operator mapping given by Eq.~\eqref{eq:ch_BxPT_operator_mapping} then directly yields the effective Lagrangian at the hadronic level
\begin{equation}\label{eq:DelB2_compact_Lagr_map}
    \cL_{\sscript{B\chi PT}}^{{\scalebox{0.56}{$|\Delta B|=2$}}}=\sum_\cR\sum_I \cC_\cR^I \lzs \sum_k \alpha_\cR^{(k)}\,[\cN_\cR^{(k)}]^I \dzs\,.
\end{equation}
After expansion, the Lagrangian can be expressed in a form analogous to the leading-order B$\chi$PT Lagrangian in Eq.~\eqref{eq:BxPT_L0_compact_expr}
\begin{equation}\label{eq:DeltaB_BxPT_lagr_comp_exp_f12h}
    \begin{alignedat}{2}
    \cL_{\sscript{B\chi PT}}^{{\scalebox{0.56}{$|\Delta B|=2$}}}&\supset 
    \sum_{f_1\leq f_2}\cC_{f_1f_2}(\bar f_1^cf_2)
    +\dfrac{1}{f_\pi}\sum_{f_1\leq f_2}\sum_h\cC_{f_1f_2}^h(\bar f_1^cf_2)h
    +\dfrac{1}{f_\pi^2}\sum_{f_1\leq f_2}\sum_{h_1\leq h_2}\cC_{f_1f_2}^{h_1h_2}(\bar f_1^cf_2)h_1h_2
    \\&
    +\dfrac{1}{(4\pi f_\pi)^2}\dfrac{1}{f_\pi^2}\sum_{f_1\leq f_2}\sum_{h_1\leq h_2}\widetilde\cC_{f_1f_2}^{h_1h_2}(\bar f_1^c f_2)\partial_\mu h_1\partial^\mu h_2
    +\hermc
    \,,
    \end{alignedat}
\end{equation}
where $f_{1,2}$ denote Dirac spinor fields describing baryons, while $h_{1,2}$ label scalar meson fields. We explicitly restrict the sums to ordered pairs, $f_1\leq f_2$ and $h_1\leq h_2$, such that each operator appears only once in the Lagrangian. This prescription avoids double counting of identical structures related by the interchange of fermionic or mesonic fields, which are equivalent due to the symmetry of the bilinear $\bar f_1^c f_2$ and the commutativity of the scalar fields. 

An additional important feature of the decomposition in Eq.~\eqref{eq:DeltaB_BxPT_lagr_comp_exp_f12h} concerns its behavior under parity. The leading-order B$\chi$PT Lagrangian in Eq.~\eqref{eq:BxPT_L0_compact_expr}, being inherited from QCD, preserves parity and therefore restricts the baryon-number conserving interactions to operators involving an even number of meson fields. In contrast, the $|\Delta B|=2$ operators considered here arise from generic new-physics dynamics and are not required to respect this symmetry. As a result, the corresponding effective Lagrangian admits both parity-even and parity-odd structures, allowing for interactions with both even and odd numbers of meson fields.

Although the general construction above allows for both parity-even and parity-odd structures, it is useful to examine the consequences of imposing parity as an additional {\color{black}requirement of the theory to respect}. At the level of the underlying fermion bilinears, the parity transformations read
\begin{equation}
    (\psi_L\chi_L) \overset{P}{\longrightarrow} -(\psi_R\chi_R)\,,
    \qquad 
    (\psi_R\chi_R) \overset{P}{\longrightarrow} -(\psi_L\chi_L)\,,
\end{equation}
such that operators transforming in a given chiral representation $\cR$ are mapped onto those of the parity-conjugated representation $\cR_P$, obtained by exchanging $L \leftrightarrow R$. Requiring invariance of the effective Lagrangian under this transformation fixes the parity transformation properties of the spurions $\cC_\cR^I$ and the corresponding B$\chi$PT realizations:
\begin{equation}
    \cC_{\cR}^I \overset{P}{\longrightarrow} -\,\cC_{\cR_P}^I\,,
    \qquad
    [\cN_{\cR}^{(k)}]^I \overset{P}{\longrightarrow} -\,[\cN_{\cR_{P}}^{(k)}]^I\,,
\end{equation}
and consequently this enforces the relations $\alpha_{\cR}^{(k)}=\alpha_{\cR_{P}}^{(k)}${\color{black}, which is a necessary requirement of QCD alone (encoded in the LECs)}.

In full generality, the Lagrangian in Eq.~\eqref{eq:DeltaB_BxPT_lagr_comp_exp_f12h} admits 14 independent coefficients of the type $\cC_{f_1f_2}$, 88 of the type $\cC_{f_1f_2}^h$, and 342 of the type $\cC_{f_1f_2}^{h_1h_2}$ and $\widetilde\cC_{f_1f_2}^{h_1h_2}$. For all interaction terms, we adopt a normalization in which the corresponding coefficients carry mass dimension one, which proves convenient for the phenomenological analysis. With this choice, all coefficients can be expressed uniformly in terms of the underlying six-quark Wilson coefficients, of mass dimension $-5$, and low-energy constants. The full set of coefficients can be extracted using the supplementary \texttt{Mathematica} notebook provided with this work. For completeness, we also report in Appendix~\ref{app:Cf1f2_coeffs} the full set of coefficients $\cC_{f_1f_2}$, together with a selection of $\cC_{f_1f_2}^{h_1h_2}$ and $\widetilde\cC_{f_1f_2}^{h_1h_2}$ relevant for the phenomenological analysis.

\section{Phenomenological Applications}
\label{sec:pheno_applications}

Having introduced the EFT framework relevant for $|\Delta B|=2$ transitions at the various stages of the low-energy expansion, namely SMEFT, LEFT, and B$\chi$PT, we now apply it to the two main classes of $|\Delta B|=2$ processes probed experimentally. These are baryon oscillations, with most experimental efforts focusing on $n-\bar n$ oscillations, and dinucleon decay, in which two nucleons bound inside a nucleus decay into a two-meson final state.

In this section, we study both classes of BSM processes within the formalism developed in this work. Whereas baryon oscillations have been extensively discussed in the literature (see Refs.~\cite{Phillips:2014fgb,Mohapatra:2009wp} for theoretical and experimental reviews), we place particular emphasis on $|\Delta B|=2$ dinucleon decay. As will become clear, our framework shows that dinucleon decay is not only complementary to baryon oscillations, but can also probe a wider region of the parameter space of the UV theory responsible for $|\Delta B|=2$.

\subsection{Baryon Oscillations}
\label{sec:baryon_oscillations_pheno}
Baryon oscillations have historically provided the primary probe of the $|\Delta B|=2$ sector. In these processes, an electrically neutral baryon, most commonly the neutron $n$, transforms into its antiparticle.\footnote{Analogous analyses have also been carried out for $\Lambda$ oscillations in Refs.~\cite{Kang:2009xt,Bittar:2024nrn,Beneito:2025ond}.} Note that such transitions can only occur for neutral baryons, as dictated by electric charge conservation. Experimentally, they can be probed in two main settings: free-baryon beam experiments and bound systems, where the baryon is bound inside a nucleus. The former set direct bounds on the oscillation parameter, typically denoted by $\delta m_{\cB}$, where $\cB$ is a generic neutral baryon, such as $n$ or $\Lambda$. This parameter arises from the $|\Delta B|=2$ contribution in the following Lagrangian
\begin{equation}\label{eq:bbbar}
    \mathcal{L}\supset -m_\cB\,\bar \cB \cB -\frac12 \, \delta m_\cB\Big( \cB^\intercal C \cB + \bar{\cB} C \bar{\cB}^\intercal \Big)\,,
\end{equation}
where $m_\cB$ is the canonical baryon mass. The study of baryon--antibaryon oscillations is naturally formulated within a quantum-mechanical two-state system. Following Refs.~\cite{Mohapatra:2009wp,Beneito:2025ond}, the time evolution of a baryon state $\mathcal{B}$ and its antiparticle $\bar{\mathcal{B}}$ is governed by the Schr\"odinger equation
\begin{equation}
    i \frac{\partial}{\partial t}
    \begin{bmatrix}
        \mathcal{B} \\
        \bar{\mathcal{B}}
    \end{bmatrix}
    =
    H_{\sscript{eff}}
    \begin{bmatrix}
        \mathcal{B} \\
        \bar{\mathcal{B}}
    \end{bmatrix} \,,
\end{equation}
where the effective Hamiltonian governing the oscillation can be written as~\cite{Mohapatra:1980de,Cowsik:1980np,Kuo:1980ew}
\begin{equation}
    H_{\sscript{eff}} =
    \begin{bmatrix}
        M_1 & \delta m_\cB \\
        \delta m_\cB & M_2
    \end{bmatrix} \, .
\end{equation}
Here $M_i=m_\cB-i\lambda/2+\alpha_i$, where $\lambda^{-1}=\tau_{\cB}$ is the mean lifetime of the free baryon, and $\alpha_i$ parametrizes the baryon’s interaction with the environment, such as the external magnetic
field. Upon diagonalizing $H_{\sscript{eff}}$, a non-vanishing $\delta m_\cB$ is found to induce oscillations between the baryon and antibaryon states, with the corresponding transition probability given by
\begin{equation}
    P(\mathcal{B} \to \bar{\mathcal{B}})
    =
    \frac{\delta m_\cB^2}{\delta m_\cB^2 + (\Delta M/2)^2}
    \sin^2\!\Bigl(\sqrt{\delta m_\cB^2 + (\Delta M/2)^2}\; t\Bigr)
    e^{\eminus\lambda t} \, ,
\end{equation}
where $\Delta M = M_1 - M_2$. The key quantity controlling the oscillation is therefore $\delta m_\cB$, whose relation to the EFT framework developed in this work is discussed below. We restrict our attention to neutron and $\Lambda$ oscillations, for which the corresponding experimental bounds were introduced above. Nevertheless, the same general framework can in principle be extended to other neutral baryon systems, such as $\Xi$ and $\Sigma^0$. In particular, as demonstrated in Ref.~\cite{Beneito:2025ond}, starting from the nonleptonic weak Lagrangian, both the tree-level and meson-loop (pionic and kaonic) contributions can be systematically computed, thereby enabling indirect constraints to be inferred from the experimental bounds on $n-\bar n$ oscillations.

Previous analyses of $n-\bar n$ oscillations were formulated in the $\SU(2)$ chiral limit~\cite{Rao:1982gt,Buchoff:2015qwa,Rinaldi:2018osy,Rinaldi:2019thf,Bijnens:2017xrz}, while here we extend the treatment to $\SU(3)$, providing a unified framework that also incorporates $\Lambda-\bar\Lambda$ oscillations. In particular, using the notation of Eq.~\eqref{eq:DeltaB_BxPT_lagr_comp_exp_f12h}, the quark-level chiral irreducible representations that contribute to these baryon oscillations are identified through the matching relations listed in Appendix~\ref{app:Cf1f2_coeffs}. In the notation adopted throughout this work, the oscillation parameter $\delta m_\cB$ is related to the coefficient $\cC_{\cB\cB}$ appearing in Eq.~\eqref{eq:DeltaB_BxPT_lagr_comp_exp_f12h} through
\begin{equation}
    \delta m_\cB \equiv 2\,\cC_{\cB\cB}\,,
\end{equation}
where $\cB$ denotes the baryon undergoing the oscillation. The factor of $2$ arises from the normalization of the leading $|\Delta B|=2$ term in the effective Lagrangian in Eq.~\eqref{eq:DeltaB_BxPT_lagr_comp_exp_f12h}. The contributions to the different baryon oscillation channels can be summarized as follows:
\begin{itemize}
    \item $\bm{n-\bar n}$: The contributions to this channel arise from the irreps $\rep{15}_R\otimes\repbar{3}_L$, $\rep{15}_R'\otimes\rep{6}_L$, $\repbar{6}_R\otimes\rep{6}_L$, and $\rep{27}_R\otimes\rep{1}_L$, together with their $L\leftrightarrow R$ counterparts. For the $\rep{15}_R\otimes\repbar{3}_L$ and $\repbar{6}_R\otimes\rep{6}_L$ classes, only the first respective B$\chi$PT realization shown in Table~\ref{tab:BxPT_fields} contributes to this process, namely that labeled by $(1)$.
    \item $\bm{\Lambda-\bar\Lambda}$: In this case, all irreps contribute, with the exception of $\rep{28}_R\otimes\rep{1}_L$ and its $L\leftrightarrow R$ counterpart, including all of the corresponding B$\chi$PT realizations displayed in Table~\ref{tab:BxPT_fields}. As discussed in Section~\ref{sec:construction_BChiPT_ops}, B$\chi$PT operators involving irrep $\rep{28}_R\otimes\rep{1}_L$ can only be constructed through insertions of $M^\dagger$ or two mesonic derivative fields $\partial_\mu\Sigma^\dagger$. {\color{black}{Since the realizations involving $M^\dagger$ are quark-mass suppressed compared to those with two mesonic derivatives, we neglect the former while retaining the latter in the analysis.}}
\end{itemize}

From the experimental side, current free-particle searches reach sensitivities of $\delta m_n \leq 7.8 \times 10^{\eminus33}\,\gev$ for $n-\bar n$ oscillations~\cite{Baldo-Ceolin:1994hzw}, and $\delta m_\Lambda \leq 2.1 \times 10^{\eminus18}\,\gev$ for $\Lambda-\bar\Lambda$ oscillations~\cite{BESIII:2024gcd}. However, the strongest present constraint on a $|\Delta B|=2$ oscillation parameter comes from searches for neutrons bound in ${}^{16}$O nuclei, yielding $\delta m_n \leq 1.4 \times 10^{\eminus33}\,\mathrm{GeV}$~\cite{Super-Kamiokande:2020bov}. These experimental sensitivities are expected to improve significantly in the coming years. In particular, next-generation large-volume detectors such as Hyper-Kamiokande~\cite{Hyper-Kamiokande:2018ofw} are expected to improve the intranuclear sensitivity, while the dedicated HIBEAM/NNBAR~\cite{Addazi:2020nlz} program at the European Spallation Source is designed to greatly extend the reach for free $n-\bar n$ oscillations.

\subsection{Dinucleon Decays}
While baryon oscillations provide a direct probe of $|\Delta B|=2$ interactions at the single-particle level, dinucleon decay processes offer a complementary and highly sensitive probe of the same underlying physics. The Wilson coefficients relevant for oscillations also enter dinucleon decay amplitudes, which are sensitive to a broader set of contributions and often yield the strongest experimental constraints. In these transitions, two nucleons within a nucleus annihilate into mesonic final states, giving rise to distinctive nuclear decay signatures. The sensitivity of such processes is enhanced by the high nucleon density in nuclei, which increases the probability of short-distance two-body interactions.

In this section, we first establish a general framework for the computation of the relevant tree-level diagrams in B$\chi$PT, which we then use to extract constraints on the effective couplings from existing experimental limits.

\subsubsection{Computation of Scattering Amplitudes}

\begin{figure}[t]
    \centering
    \includegraphics[width=0.95\linewidth]{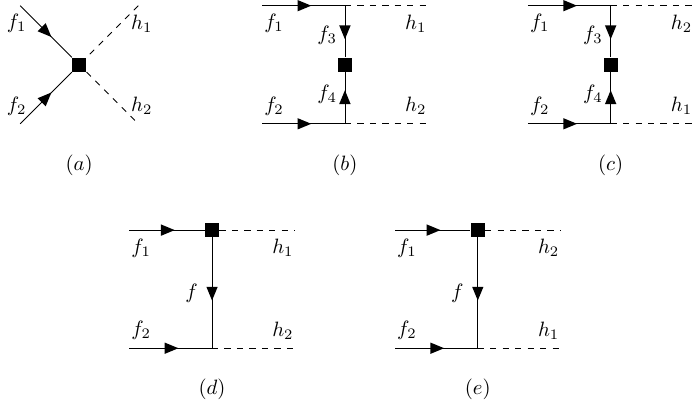}
    \caption{{\color{black}{Tree-level topologies contributing to dinucleon decay.}} The black square denotes the insertion of a $|\Delta B|=2$ operator. The fermionic fields $f_i$ correspond to baryons, while $h_i$ denote mesonic scalar degrees of freedom in B$\chi$PT.}
    \label{fig:tree_lev_feyn_diags_dinuc}
\end{figure}
The relevant tree-level topologies contributing to dinucleon decay are shown in Figure~\ref{fig:tree_lev_feyn_diags_dinuc}.\footnote{One could also envisage an $s$-channel contribution in which the $|\Delta B|=2$ interaction is mediated by a single-meson exchange, involving a vertex proportional to $\cC_{f_1 f_2}^h$ and a subsequent meson propagator. The exchanged meson would then have to couple to the external states through a three-meson interaction, which is forbidden by parity invariance (see Eq.~\eqref{eq:BxPT_parity_trafo_meson}).} These include contact interactions as well as $t$- and $u$-channel diagrams. In what follows, we examine each contribution separately, specifying the relevant interaction terms and deriving the associated matrix elements.

\paragraph{Contact interaction.} We begin with the contact interaction, corresponding to diagram $(a)$ in Figure~\ref{fig:tree_lev_feyn_diags_dinuc}. The relevant contribution arises from the following $|\Delta B|=2$ operators
\begin{equation}
    \cL_{\sscript{B\chi PT}}^{{\scalebox{0.56}{$|\Delta B|=2$}}}\supset \dfrac{1}{f_\pi^2}\sum_{f_1\leq f_2}\sum_{h_1\leq h_2}\cC_{f_1f_2}^{h_1h_2}(\bar f_1^cf_2)h_1h_2
    +\dfrac{1}{16\pi^2f_\pi^4}\sum_{f_1\leq f_2}\sum_{h_1\leq h_2}\widetilde\cC_{f_1f_2}^{h_1h_2}(\bar f_1^c f_2)\partial_\mu h_1\partial^\mu h_2
    \,.
\end{equation}
Using $\bar\psi^c=\psi^\intercal C$, the corresponding matrix element reads
\begin{equation}\label{eq:dinuc_matr_contact}
    i\cM_a(f_1f_2\to h_1h_2)=\frac{i S_{f_1f_2}^{h_1h_2}}{f_\pi^2}\lzs \cC_{f_1f_2}^{h_1h_2}-\widetilde \cC_{f_1f_2}^{h_1h_2}\frac{s-m_{h_1}^2-m_{h_2}^2}{32\pi^2f_\pi^2} \dzs\big[ u_1^\intercal(p_1) C u_2(p_2) \big]\,,
\end{equation}
where $u_1(p_1)$ and $u_2(p_2)$ denote the Dirac spinors associated with the incoming fermions $f_1$ and $f_2$, with momenta $p_1$ and $p_2$, respectively, and $s$ denotes the Mandelstam variable $s=(p_1+p_2)^2=(k_1+k_2)^2$, with $k_1$ and $k_2$ being the outgoing momenta of the mesonic fields $h_1$ and $h_2$, respectively. The factor $S_{f_1f_2}^{h_1h_2}\equiv (1+\delta_{f_1f_2})(1+\delta_{h_1 h_2})$ accounts for identical-particle combinatorics in the absence of factorial normalization in the Lagrangian.

\paragraph{$\bm t$- and $\bm u$-channel.} For the $t$- and $u$-channel contributions, corresponding to diagrams $(b)$ and $(c)$ in Figure~\ref{fig:tree_lev_feyn_diags_dinuc}, the relevant interactions are given by
\begin{equation}\label{eq:rel_int_for_tu_chan_add}
    \cL_{\sscript{B\chi PT}}\supset
    \frac{1}{f_\pi}\sum_{f_1\leq f_2}\sum_h \cC_{f_1 f_2 h}^\sscript{LO} (\bar f_1 \gamma^\mu \gamma^5 f_2)\, \partial_\mu h
    +\sum_{f_1\leq f_2}\cC_{f_1 f_2}(\bar f_1^c f_2)\,,
\end{equation}
where the first term corresponds to the leading-order baryon-number conserving B$\chi$PT interaction, while the second term encodes the local $|\Delta B|=2$ operator. In order to consistently construct the fermion exchange diagrams and match the fermion flow along the internal propagators, it is convenient to also express the baryon-number conserving interaction in its charge-conjugated form:
\begin{equation}\label{eq: BchiPT at LO with charge conjugation}
    \cL_{\sscript{B\chi PT}}^0\supset
    \frac{1}{f_\pi}\sum_{f_1\leq f_2}\sum_h \cC_{f_1 f_2 h}^\sscript{LO} (\bar f_2^c \gamma^\mu \gamma^5 f_1^c)\, \partial_\mu h\,,
\end{equation}
where we make use of the identity $\bar\psi_1\Gamma\psi_2=\bar\psi_2^c(C\,\Gamma^\intercal\, C^{\eminus1}) \psi_1^c$, together with $C\gamma^5\gamma^\mu C^{\eminus1}=(\gamma^5\gamma^\mu)^\intercal$. {\color{black}{Using these interaction terms, the matrix elements take the form}}
\begin{equation}\label{eq:t_and_u_channels_dinuc}
{\scalebox{0.92}{$
    \begin{alignedat}{2}
        i\cM_b(f_1f_2\to h_1h_2)&=\frac{i}{f_\pi^2}\sum_{f_3,f_4}\frac{\cC_{f_3f_1 h_1}^\sscript{LO}\cC_{f_3f_4}\cC_{f_4f_2h_2}^\sscript{LO}}{(t-m_{f_3}^2)(t-m_{f_4}^2)}
        S_{f_3f_4}\big[ u_1^\intercal(p_1)C\slashed k_1(\slashed p_t+m_{f_3})(\slashed p_t+m_{f_4})\slashed k_2 u_2(p_2) \big]\,,
        \\[3pt]
        i\cM_c(f_1f_2\to h_1h_2)&=\frac{i}{f_\pi^2}\sum_{f_3,f_4}\frac{\cC_{f_3f_1h_2}^\sscript{LO}\cC_{f_3f_4}\cC_{f_4f_2h_1}^\sscript{LO}}{(u-m_{f_3}^2)(u-m_{f_4}^2)}
        S_{f_3f_4}\big[ u_1^\intercal(p_1)C\slashed k_2(\slashed p_u+m_{f_3})(\slashed p_u+m_{f_4})\slashed k_1 u_2(p_2) \big]\,,
    \end{alignedat}
$}}
\end{equation}
where the Mandelstam invariants are defined as $t=p_t^2=(p_1-k_1)^2=(p_2-k_2)^2$ and $u=p_u^2=(p_1-k_2)^2=(p_2-k_1)^2$. In these expressions, the sums over $f_3$ and $f_4$ run over the internal fermionic states appearing in the exchange diagrams, whereas $S_{f_3f_4}\equiv1+\delta_{f_3f_4}$ accounts for identical-particle combinatorics. 

In addition, this structure shows that dinucleon decay amplitudes access a broader set of coefficient combinations than baryon oscillations alone. In this context, a technical point worth mentioning concerns the propagator structure associated with the charge-conjugated fermion line. Using the identity $CS_F^\intercal(p)C^{\eminus1}=S_F(-p)$, one may rewrite the transposed propagator in terms of a standard one with opposite momentum~\cite{Denner:1992vza,Paraskevas:2018mks}. This, however, does not imply a physical reversal of momentum flow: the sign change is compensated at the level of the Fourier integration, and the propagator enters the amplitude in its usual form (see Appendix~\ref{app:QFT_comp_matrix_elem}). Finally, the $t$- and $u$-channel amplitudes are related by crossing symmetry, more precisely, the $u$-channel expression follows from the $t$-channel one by exchanging the two outgoing mesons, $k_1\leftrightarrow k_2$, together with the replacement $t\leftrightarrow u$. 

{\color{black}{Lastly, for the $t$- and $u$-channel contributions corresponding to diagrams $(d)$ and $(e)$ in Figure~\ref{fig:tree_lev_feyn_diags_dinuc}, the relevant interaction terms read
\begin{equation}
    \cL_{\sscript{B\chi PT}}\supset
    \frac{1}{f_\pi}\sum_{f_1\leq f_2}\sum_h \cC_{f_1 f_2 h}^\sscript{LO} (\bar f_1 \gamma^\mu \gamma^5 f_2)\, \partial_\mu h
    +\dfrac{1}{f_\pi}\sum_{f_1\leq f_2}\sum_h\cC_{f_1f_2}^h(\bar f_1^cf_2)h\,,
\end{equation}
and the matrix elements are given by
\begin{equation}
\begin{alignedat}{2}
i\cM_d(f_1f_2\to h_1h_2)&=\frac{i}{f_\pi^2}\sum_f\frac{\cC_{ff_2h_2}^\sscript{LO}\cC^{h_1}_{f_1f}}{t-m_f^2}\big[ u_1^\intercal(p_1)C (\slashed p_t-m_f)\slashed k_2\gamma^5 u_2(p_2) \big]
\,,
\\[3pt]
i\cM_e(f_1f_2\to h_1h_2)&=\frac{i}{f_\pi^2}\sum_f\frac{\cC_{ff_2h_1}^\sscript{LO}\cC^{h_2}_{f_1f}}{u-m_f^2}\big[ u_1^\intercal(p_1)C (\slashed p_u-m_f)\slashed k_1\gamma^5 u_2(p_2) \big]
\,.
\end{alignedat}
\end{equation}
With the individual contributions at hand, the spin-averaged squared amplitude can be decomposed as
}}
\begin{equation}\label{eq:amp_squared_fully_general_FG}
{\scalebox{0.97}{
$
\begin{alignedat}{2}
        |\cM|^2&=
        \frac{1}{4f_\pi^4}\sum_{f_3, f_4}\sum_{f_3', f_4'}\big( \cC_{f_3f_1h_1}^\sscript{LO}\cC_{f_3f_4} \cC_{f_4f_2h_2}^\sscript{LO} \big) \big( \cC_{f_3'f_1h_1}^\sscript{LO}\cC_{f_3'f_4'} \cC_{f_4'f_2h_2}^\sscript{LO} \big)S_{f_3f_4}S_{f_3'f_4'}\,[\mathcal F_t]_{f_1f_2h_1h_2}^{f_3f_4f_3'f_4'}
        \\[4pt]
        &+\frac{1}{4f_\pi^4}\sum_{f_3, f_4}\sum_{f_3', f_4'}\big( \cC_{f_3f_1h_2}^\sscript{LO}\cC_{f_3f_4} \cC_{f_4f_2h_1}^\sscript{LO} \big) \big( \cC_{f_3'f_1h_2}^\sscript{LO}\cC_{f_3'f_4'} \cC_{f_4'f_2h_1}^\sscript{LO} \big)S_{f_3f_4}S_{f_3'f_4'}\,[\mathcal F_u]_{f_1f_2h_1h_2}^{f_3f_4f_3'f_4'}
        \\[4pt]
        &+\frac{1}{2f_\pi^4}\sum_{f_3, f_4}\sum_{f_3', f_4'}\big( \cC_{f_3f_1h_1}^\sscript{LO}\cC_{f_3f_4} \cC_{f_4f_2h_2}^\sscript{LO} \big)\big( \cC_{f_3'f_1h_2}^\sscript{LO}\cC_{f_3'f_4'} \cC_{f_4'f_2h_1}^\sscript{LO} \big)S_{f_3f_4}S_{f_3'f_4'}\,[\mathcal F_{tu}]_{f_1f_2h_1h_2}^{f_3f_4f_3'f_4'}
        \\[4pt]
        &+\frac{S_{f_1f_2}^{h_1h_2}}{2f_\pi^4}\lzs \cC_{f_1f_2}^{h_1h_2}-\widetilde \cC_{f_1f_2}^{h_1h_2}\frac{s-m_{h_1}^2-m_{h_2}^2}{32\pi^2f_\pi^2} \dzs\sum_{f_3, f_4}\big( \cC_{f_3f_1h_1}^\sscript{LO}\cC_{f_3f_4} \cC_{f_4f_2h_2}^\sscript{LO} \big)S_{f_3f_4}\,[\mathcal G_t]_{f_1f_2h_1h_2}^{f_3f_4}
        \\[4pt]
        &+\frac{S_{f_1f_2}^{h_1h_2}}{2f_\pi^4}\lzs \cC_{f_1f_2}^{h_1h_2}-\widetilde \cC_{f_1f_2}^{h_1h_2}\frac{s-m_{h_1}^2-m_{h_2}^2}{32\pi^2f_\pi^2} \dzs\sum_{f_3, f_4}\big( \cC_{f_3f_1h_2}^\sscript{LO}\cC_{f_3f_4} \cC_{f_4f_2h_1}^\sscript{LO} \big)S_{f_3f_4}\,[\mathcal G_u]_{f_1f_2h_1h_2}^{f_3f_4}
        \\[4pt]
        &+\frac{(S_{f_1f_2}^{h_1h_2})^2}{2f_\pi^4}\lzs \cC_{f_1f_2}^{h_1h_2}-\widetilde \cC_{f_1f_2}^{h_1h_2}\frac{s-m_{h_1}^2-m_{h_2}^2}{32\pi^2f_\pi^2} \dzs^2 \big[s-(m_{f_1}+m_{f_2})^2\big]
        \\[4pt]
        &{\color{black}{
        +\frac{1}{4f_\pi^4}\sum_{f,f'}\cC^\sscript{LO}_{ff_2h_2}\cC_{f_1f}^{h_1}\cC^\sscript{LO}_{f'f_2h_2}\cC_{f_1f'}^{h_1}[\mathcal H_t]^{ff'}_{f_1f_2h_1h_2}
        +\frac{1}{4f_\pi^4}\sum_{f,f'}\cC^\sscript{LO}_{ff_2h_1}\cC_{f_1f}^{h_2}\cC^\sscript{LO}_{f'f_2h_1}\cC_{f_1f'}^{h_2}[\mathcal H_u]^{ff'}_{f_1f_2h_1h_2}}}
        \\[4pt]
        &{\color{black}{+\frac{1}{2f_\pi^4}\sum_{f,f'}\cC^\sscript{LO}_{ff_2h_2}\cC_{f_1f}^{h_1}\cC^\sscript{LO}_{f'f_2h_1}\cC_{f_1f'}^{h_2}[\mathcal H_{tu}]^{ff'}_{f_1f_2h_1h_2}
        }}
        \,,
    \end{alignedat}
$
}}
\end{equation}
where, for simplicity, the Wilson coefficients are taken to be real, and $\mathcal F_j$, $\mathcal G_j$ and $\mathcal H_j$ denote kinematic functions encoding the spin-summed Dirac structures and their dependence on the external momenta. Their analytical structure is provided in Appendix~\ref{app:F_and_G_func_matr_elem}. In practice, these functions enter the evaluation of the decay width introduced below, where they can be computed for specific transitions and kinematical configurations. Lastly, from this decomposition, one can directly identify the dependence on the relevant $|\Delta B|=2$ coefficients. In particular, the contributions associated with the exchange diagrams probe the coefficients $\cC_{f_3f_4}$, which include the combinations entering baryon oscillations, {\color{black}{as well as $\cC_{f_1f_2}^{h_1}$}}, while the contact interaction is governed by $\cC_{f_1f_2}^{h_1h_2}$ and $\widetilde\cC_{f_1f_2}^{h_1h_2}$, which contribute exclusively to dinucleon decay processes {\color{black}at tree-level}. The decomposition in Eq.~\eqref{eq:amp_squared_fully_general_FG} thus makes clear that dinucleon decay processes are sensitive both to the coefficients entering baryon oscillations and to additional contact interactions.

We conclude this part by noting that the scattering amplitudes considered here may, in principle, receive sizable contributions from higher-order terms in the B$\chi$PT expansion in Eq.~\eqref{eq:BxPT_L0_expanded}, in particular from operators involving additional derivatives. This is due to the fact that the characteristic energy scale of the process, set by the nucleon masses $m_{p,n}$, lies close to the chiral cutoff scale $\Lambda_\chi\sim 4\pi f_\pi\sim\cO(1\,\gev)$. In standard applications, such effects can be systematically addressed within the heavy B$\chi$PT formulation~\cite{Jenkins:1990jv,Fuchs:2003qc,Gegelia:1999gf,Becher:1999he}, which is well suited for processes involving soft external momenta. In the present case, however, this approach is not directly applicable, as the outgoing mesons can carry energies of order $m_{p,n}$, i.e. comparable to the EFT cutoff. In addition, extending heavy B$\chi$PT to baryon-number-violating processes introduces conceptual subtleties, making its application to the processes considered here non-trivial.\footnote{The heavy B$\chi$PT formulation has been applied in Ref.~\cite{Bijnens:2017xrz} to compute NLO corrections to $n-\bar n$ transitions, where separate degrees of freedom are introduced to describe $n$ and $\bar n$ excitations about the mass shell. From a top-down perspective, however, such constructions require some care, as they do not straightforwardly map onto the relativistic B$\chi$PT framework with a single Dirac field. This reflects a more general subtlety in extending heavy B$\chi$PT to baryon-number-violating processes.}

{\color{black}{Consequently, in view of these convergence issues, and taking into account the additional uncertainties associated with the relevant low-energy constants and nuclear matrix elements, the numerical results presented in the following section should be interpreted as order-of-magnitude estimates, rather than precision predictions. Improving this situation would require dedicated non-perturbative input for the relevant hadronic matrix elements, together with a more controlled treatment of the corresponding nuclear matrix elements in the dinucleon case.}}

\subsubsection{Numerical Evaluation of Decay Rates}
\label{sec:pheno_num_constrs_exps}
As a first step, we begin by specifying the kinematical configuration relevant for the dinucleon decay processes. The initial state consists of two baryons bound inside a nucleus, and a natural and well-motivated approximation is to treat them as non-relativistic particles with small but non-vanishing spatial momenta of order the Fermi momentum $k_F\sim\cO(200\,\mev)$. In this setup, their four-momenta can be written in the center-of-mass frame as
\begin{equation}
    p_1\simeq( E_{f_1},k_F,0,0) \,,
    \qquad 
    p_2\simeq( E_{f_2},- k_F,0,0)\,,
    \qquad 
    E_{f_{1,2}}=\sqrt{m_{f_{1,2}}^2+k_F^2}\,,
\end{equation}
where the spatial components reflect the fact that the total three-momentum of the pair is approximately zero in the nuclear rest frame. Momentum conservation then implies that the outgoing mesons are produced approximately back-to-back. Their energies are fixed by the available invariant mass of the initial state, while their three-momenta have equal magnitude and opposite direction, up to corrections induced by the Fermi motion. Specifically, we have
\begin{equation}
    E_{h_1}=\frac{s+m_{h_1}^2-m_{h_2}^2}{2\sqrt s}\,,
    \qquad
    E_{h_2}=\frac{s+m_{h_2}^2-m_{h_1}^2}{2\sqrt s}\,,
    \qquad
    |\vec k_1|=|\vec k_2|=\frac{\lambda^{1/2}(s,m_{h_1}^2,m_{h_2}^2)}{2\sqrt s}\,,
\end{equation}
where $\lambda(a,b,c)\equiv a^2+b^2+c^2-2ab-2ac-2bc$ and $s=(E_{f_1}+E_{f_2})^2$. The remaining Mandelstam invariants can then be expressed as
\begin{equation}
    \begin{alignedat}{2}
        t&=m_{f_1}^2+m_{h_1}^2-2E_{f_1}E_{h_1}+ \frac{k_F}{\sqrt s}\,\lambda^{1/2}(s,m_{h_1}^2,m_{h_2}^2)\cos\theta \,,
        \\[3pt]
        u&=m_{f_1}^2+m_{h_2}^2-2E_{f_1}E_{h_2}-\frac{k_F}{\sqrt s}\,\lambda^{1/2}(s,m_{h_1}^2,m_{h_2}^2)\cos\theta\,,
    \end{alignedat}
\end{equation}
where $\theta$ denotes the scattering angle between $\vec p_1$ and $\vec k_1$.

To connect the squared amplitude derived above with experimentally constrained observables, we introduce the intranuclear decay rate. The exact rate is determined by the many-body nuclear matrix element of the $|\Delta B|=2$ operator between the initial and final nuclear states. In the short-range approximation, where the $|\Delta B|=2$ interaction is treated as a contact operator at nuclear scales, the transition factorizes into a product of a local two-nucleon annihilation probability and the density of nucleon pairs inside the nucleus as follows
\begin{equation}\label{eq:Gamma_A_pair}
\Gamma_A(N_1N_2 \to h_1 h_2) \simeq \int d^3 r \; n_{N_1N_2}(\rep r)\, v_{\sscript{rel}}\,\sigma(N_1N_2 \to h_1 h_2)\,,
\end{equation}
where $\Gamma_A$ denotes the intranuclear decay rate of a nucleus with mass number $A$, $n_{N_1N_2}(\rep r)$ is the local density of nucleon pairs, $v_{\sscript{rel}}$ is the relative velocity of the annihilating nucleons, and $\sigma(N_1N_2 \to h_1h_2)$ is the cross section for the corresponding elementary two-nucleon process. The latter quantity can be expressed as follows
\begin{equation}
v_{\sscript{rel}}\,\sigma(N_1N_2\to h_1 h_2)
= \frac{1}{64\pi}\frac{\lambda^{1/2}(s,m_{h_1}^2,m_{h_2}^2)}{ s\,E_{f_1} E_{f_2}}\int d(\cos\theta)\,|\mathcal M(N_1N_2\to h_1 h_2)|^2\,,
\end{equation}
where $|\mathcal M(N_1N_2\to h_1 h_2)|^2$ is the squared amplitude for the elementary decay process. Approximating the spatial integral over the local pair density by an effective constant proportional to the average nucleon density $\rho_N$, the intranuclear decay rate becomes
\begin{equation}\label{eq:Gamma_A_pair_final_exp}
\Gamma_A(N_1N_2 \to h_1 h_2) \simeq \frac{\kappa_A^{N}}{64\pi s}\frac{\rho_N}{E_{f_1} E_{f_2}}\lambda^{1/2}(s,m_{h_1}^2,m_{h_2}^2)\int d(\cos\theta)\,|\mathcal M(N_1N_2\to h_1 h_2)|^2\,,
\end{equation}
where $\kappa_A^{N}$ is a dimensionless coefficient, typically of $\cO(1)$, encoding the nuclear-structure dependence of the initial two-nucleon configuration, including pair counting as well as spin and isospin factors. In the phenomenological analysis below, we adopt the standard approximation used in the dinucleon-decay literature, taking $\kappa_A^{N}\simeq9$ and $\rho_N\simeq 2\times10^{\eminus3}\,\gev^3$~\cite{Goity:1994dq,Aitken:2017wie,Beneito:2025ond}.\footnote{The numerical values of $\kappa_A^N$ and $\rho_N$ should be regarded as order-of-magnitude inputs. They effectively parametrize nuclear-structure effects such as pair counting, spin--isospin factors, and the replacement of the local two-nucleon distribution by an average density. This approximation is therefore one of the dominant sources of theoretical uncertainty in the intranuclear decay-rate estimate.}

\begin{table}[t]
    \centering
\centering

\renewcommand{\arraystretch}{1.2}

\scalebox{0.685}{

\begin{tabular}{l@{\hspace{1.0cm}}c@{\hspace{1.4cm}}lc@{\hspace{1.4cm}}lc@{\hspace{1.4cm}}lc@{\hspace{1.4cm}}lc}

\toprule

\multirow{1}{*}{~~~\textbf{Decay}}

&\multirow{1}{*}{$\bm{\Gamma_A\,[\mathrm{GeV}]}$}

&\multicolumn{2}{l}{$\bm{\gamma_1\,[\gev^{\eminus1}]}$~~}

&\multicolumn{2}{l}{$\bm{\gamma_2\,[\gev^{\eminus1}]}$~~}

&\multicolumn{2}{l}{$\bm{\gamma_3\,[\gev^{\eminus1}]}$~~}

&\multicolumn{2}{l}{{\color{black}{$\bm{\gamma_4\,[\gev^{\eminus1}]}$}}~~}

\\

\midrule

\addlinespace[0.2cm]


$pp\to \pi^+\pi^+$

&$3.0\times 10^{\eminus64}$

&$[\gamma_1]^{\pi^+\pi^+}_{pp}$&$2.88$

&$[\gamma_2]_{pp}^{\pi^+\pi^+}$&$1.27$

&$[\gamma_3]_{pp}^{\pi^+\pi^+}$&$1.32$

&$[\gamma_4]_{pp}^{\pi^+\pi^+}$&$0.02$

\\[0.2cm]

$pp\to {\scalebox{0.75}{$K$}}^+{\scalebox{0.75}{$K$}}^+$

&$1.2\times 10^{\eminus64}$

&$[\gamma_1]^{{\scalebox{0.65}{$K$}}^+ {\scalebox{0.65}{$K$}}^+}_{pp}$&$1.48$

&$[\gamma_2]_{pp}^{{\scalebox{0.65}{$K$}}^+ {\scalebox{0.65}{$K$}}^+}$&$0.29$

&$[\gamma_3]_{pp}^{{\scalebox{0.65}{$K$}}^+  {\scalebox{0.65}{$K$}}^+}$&$1.32$

&$[\gamma_4]_{pp}^{{\scalebox{0.65}{$K$}}^+  {\scalebox{0.65}{$K$}}^+}$&$0.17$

\\[0.2cm]

$pn\to \pi^+\pi^0$

&$1.2\times 10^{\eminus64}$

&$[\gamma_1]^{\pi^+\pi^0}_{pn}$&$0.10$

&$[\gamma_2]^{\pi^+\pi^0}_{pn}$&$0.06$

&$[\gamma_3]^{\pi^+\pi^0}_{pn}$&$0.08$

&$[\gamma_4]^{\pi^+\pi^0}_{pn}$&$0.02$

\\[0.2cm]

$nn\to \pi^0\pi^0$

&$0.5\times 10^{\eminus64}$

&$[\gamma_1]^{\pi^0\pi^0}_{nn}$&$0.04$

&$[\gamma_2]^{\pi^0\pi^0}_{nn}$&$0.16$

&$[\gamma_3]^{\pi^0\pi^0}_{nn}$&$1.31$

&$[\gamma_4]^{\pi^0\pi^0}_{nn}$&$0.002$

\\[0.2cm]

\noalign{\vskip 0.0cm}

\cdashline{1-10}[.4pt/2pt]

\noalign{\vskip 0.2cm}

$nn\to  {\scalebox{0.75}{$K$}}^+{\scalebox{0.75}{$K$}}^{\eminus}$

&$1.2\times 10^{\eminus64}$

&$[\gamma_1]^{{\scalebox{0.65}{$K$}}^+ {\scalebox{0.65}{$K$}}^{\eminus}}_{nn}$&$0.01$

&$[\gamma_2]^{{\scalebox{0.65}{$K$}}^+ {\scalebox{0.65}{$K$}}^{\eminus}}_{nn}$&$0.03$

&$[\gamma_3]^{{\scalebox{0.65}{$K$}}^+ {\scalebox{0.65}{$K$}}^{\eminus}}_{nn}$&$0.29$

&$[\gamma_4]^{{\scalebox{0.65}{$K$}}^+ {\scalebox{0.65}{$K$}}^{\eminus}}_{nn}$&$0.06$

\\[0.2cm]

$nn\to \pi^+\pi^{\eminus}$

&$3.0\times 10^{\eminus64}$

&$[\gamma_1]^{\pi^+\pi^{\eminus}}_{nn}$&--

&$[\gamma_2]^{\pi^+\pi^{\eminus}}_{nn}$&--

&$[\gamma_3]^{\pi^+\pi^{\eminus}}_{nn}$&0.33

&$[\gamma_4]^{\pi^+\pi^{\eminus}}_{nn}$&$0.07$

\\[0.2cm]

\bottomrule

\end{tabular}

}
    \caption{Overview of the dinucleon decay channels considered in this work. The first column lists the transition, while the second column reports the corresponding decay width $\Gamma_A$, inferred from the current experimental lower bounds on the nuclear lifetime for the given process~\cite{Heeck:2019kgr}. The remaining columns display the numerical values of the coefficients $\gamma_i$, evaluated in the kinematical configuration specified above. The channels below the dashed line exhibit qualitatively distinct behavior: the $nn \to K^+ K^{\eminus}$ channel is sensitive to Wilson coefficients associated with $\Sigma$-baryon transitions, which are not accessible through baryon--antibaryon oscillation observables. By contrast, $nn\to \pi^+\pi^{\eminus}$ channel depends only on direct contact contributions. See the text for further discussion.}
    \label{tab:Dinucleon_channels_numerics}
\end{table}

With these inputs and the kinematical configuration specified, the intranuclear decay rate can be evaluated for all dinucleon decay channels. At present, however, only five such processes have been experimentally constrained~\cite{Heeck:2019kgr,Super-Kamiokande:2015jbb,Super-Kamiokande:2014hie}, and are listed in Table~\ref{tab:Dinucleon_channels_numerics}. Using the leading-order B$\chi$PT Lagrangian defined in Eq.~\eqref{eq:BxPT_L0_expanded}, the corresponding decay widths for the first four channels can be expressed as follows 
\begin{equation}\label{eq:first_4_decay_rates}
{\scalebox{0.9}{
$
\begin{alignedat}{8}
        &\Gamma_A(pp\to \pi^+\pi^+)&&\simeq \cC_{nn}^2\,[\gamma_1]^{\pi^+\pi^+}_{pp}+\Delta_{pp}^{\pi^+\pi^+}\cC_{nn}\,[\gamma_2]^{\pi^+\pi^+}_{pp}+(\Delta_{pp}^{\pi^+\pi^+})^2\,[\gamma_3]^{\pi^+\pi^+}_{pp}
        +(\cC_{np}^{\pi^+})^2\,[\gamma_4]^{\pi^+\pi^+}_{pp}
        \,,
        \\[4pt]
        &\Gamma_A(pp\to {\scalebox{0.78}{$K$}}^+{\scalebox{0.78}{$K$}}^+)&&\simeq \cC_{\Lambda\Lambda}^2\,[\gamma_1]^{{\scalebox{0.65}{$K$}}^+ {\scalebox{0.65}{$K$}}^+}_{pp}+\Delta_{pp}^{{\scalebox{0.65}{$K$}}^+ {\scalebox{0.65}{$K$}}^+}\cC_{\Lambda\Lambda}\,[\gamma_2]^{{\scalebox{0.65}{$K$}}^+ {\scalebox{0.65}{$K$}}^+}_{pp}+(\Delta_{pp}^{{\scalebox{0.65}{$K$}}^+ {\scalebox{0.65}{$K$}}^+})^2[\gamma_3]^{{\scalebox{0.65}{$K$}}^+ {\scalebox{0.65}{$K$}}^+}_{pp}
        +(\cC_{\Lambda p}^{{\scalebox{0.65}{$K$}}^+})^2\,[\gamma_4]^{{\scalebox{0.65}{$K$}}^+ {\scalebox{0.65}{$K$}}^+}_{pp}
        \,,
        \\[4pt]
        &\Gamma_A(pn\to \pi^+\pi^0)&&\simeq \cC_{nn}^2\,[\gamma_1]^{\pi^+\pi^0}_{pn}-\Delta_{pn}^{\pi^+\pi^0}\cC_{nn}\,[\gamma_2]^{\pi^+\pi^0}_{pn}+(\Delta_{pn}^{\pi^+\pi^0})^2\,[\gamma_3]^{\pi^+\pi^0}_{pn}
        +\frac{(\bar\cC_{pn}^\pi)^2}{8}\,[\gamma_4]^{\pi^+\pi^0}_{pn}
        \,,
        \\[4pt]
        &\Gamma_A(nn\to \pi^0\pi^0)&&\simeq \cC_{nn}^2\,[\gamma_1]^{\pi^0\pi^0}_{nn}+\Delta_{nn}^{\pi^0\pi^0}\cC_{nn}\,[\gamma_2]^{\pi^0\pi^0}_{nn}+(\Delta_{nn}^{\pi^0\pi^0})^2\,[\gamma_3]^{\pi^0\pi^0}_{nn}
        +(\cC_{nn}^{\pi^0})^2\,[\gamma_4]^{\pi^0\pi^0}_{nn}
        \,,
    \end{alignedat}
$
}}
\end{equation}
where $(\bar\cC_{pn}^\pi)^2=(\cC_{pn}^{\pi^+})^2+8(\cC_{nn}^{\pi^0})^2$, and where we define\footnote{In the definition of $\Delta_{f_1f_2}^{h_1h_2}$, the chiral Lagrangian coefficients depend on $\bar h_i$ rather than $h_i$, reflecting the convention that all hadrons are taken to be incoming.}
\begin{equation}\label{eq:delta_C_tilde_C_kin_enh}
    \Delta_{f_1f_2}^{h_1h_2}=\cC_{f_1f_2}^{\bar h_1\bar h_2}-\widetilde \cC_{f_1f_2}^{\bar h_1\bar h_2}\frac{s-m_{h_1}^2-m_{h_2}^2}{32\pi^2f_\pi^2}\,.
\end{equation}
The coefficients $[\gamma_i]_{f_1f_2}^{h_1h_2}$ listed in Table~\ref{tab:Dinucleon_channels_numerics} encode the kinematical dependence of the squared amplitude, including the integration over the scattering angle and the associated phase-space factors, as well as the overall prefactors specific to each transition.\footnote{We note that the estimate for $\Gamma_A(pp\to {\scalebox{0.78}{$K$}}^+{\scalebox{0.78}{$K$}}^+)$ in Ref.~\cite{Beneito:2025ond} is numerically consistent with our result upon neglecting the $\Delta_{pp}^{\pi^+\pi^+}$ contributions.}

As can be seen from Eq.~\eqref{eq:first_4_decay_rates}, both the exchange contributions and their interference terms in the squared amplitudes are governed by the coefficients $\cC_{nn}$ and $\cC_{\Lambda\Lambda}$, which coincide with the combinations entering baryon oscillations, as discussed in Section~\ref{sec:baryon_oscillations_pheno}. In addition, the contact interaction coefficients $\cC_{f_1f_2}^{h_1h_2}$ and $\widetilde\cC_{f_1f_2}^{h_1h_2}$ are unique to dinucleon decay processes.

In contrast to the four channels in Eq.~\eqref{eq:first_4_decay_rates}, the decay rate for the $nn\to K^+K^{\eminus}$ channel exhibits a distinct qualitative structure, opening up sensitivity to the $\cC_{\Sigma^+\Sigma^{\eminus}}$ coefficient, as can be understood from the leading-order B$\chi$PT Lagrangian. While the interactions in Eq.~\eqref{eq:BxPT_L0_expanded} are expressed in terms of $p$, $n$, and $\Lambda$, capturing the dominant contributions to baryon oscillations, the full Lagrangian also contains additional couplings, particularly in the $\Sigma$ sector. For the $nn \to K^+K^{\eminus}$ transition, the relevant leading-order interaction reads
\begin{equation}
\begin{alignedat}{2}
\cL_{\sscript{B\chi PT}}^{0}\supset\frac{D-F}{f_\pi}\,(\bar\Sigma^{\eminus}\gamma^\mu\gamma^5 n)\,\partial_\mu K^{\eminus}+\hermc\,.
\end{alignedat}
\end{equation}
Using this interaction, the decay rate for the $nn \to K^+K^{\eminus}$ process can be written as
\begin{equation}\label{eq:nn_to_KK_decay_rate}
{\scalebox{0.87}{
$
\begin{alignedat}{2}
\Gamma_A(nn\to {\scalebox{0.78}{$K$}}^+{\scalebox{0.78}{$K$}}^{\eminus})\simeq\frac{\cC^2_{\Sigma^+\Sigma^{\eminus}}}{4}[\gamma_1]^{{\scalebox{0.65}{$K$}}^+ {\scalebox{0.65}{$K$}}^{\eminus}}_{nn}+\Delta_{nn}^{{\scalebox{0.65}{$K$}}^+ {\scalebox{0.65}{$K$}}^{\eminus}}\cC_{\Sigma^+\Sigma^{\eminus}}\,[\gamma_2]^{{\scalebox{0.65}{$K$}}^+ {\scalebox{0.65}{$K$}}^{\eminus}}_{nn}+(\Delta_{nn}^{{\scalebox{0.65}{$K$}}^+ {\scalebox{0.65}{$K$}}^{\eminus}})^2[\gamma_3]^{{\scalebox{0.65}{$K$}}^+ {\scalebox{0.65}{$K$}}^{\eminus}}_{nn}
+(\cC_{\Sigma n}^{{\scalebox{0.65}{$K$}}})^2[\gamma_4]^{{\scalebox{0.65}{$K$}}^+ {\scalebox{0.65}{$K$}}^{\eminus}}_{nn}\,.
\end{alignedat}
$
}}
\end{equation}
It is worth emphasizing that, unlike $\cC_{nn}$ and $\cC_{\Lambda\Lambda}$, the $\cC_{\Sigma^+\Sigma^{\eminus}}$ coefficient appearing in Eq.~\eqref{eq:nn_to_KK_decay_rate} is not associated with baryon–antibaryon oscillations, due to the mismatch in the quantum numbers of the $\Sigma^\pm$ states. Instead, it enters as a transition coefficient, contributing through intermediate $\Sigma$ exchange in the amplitude and is therefore accessible only in dinucleon decay processes. While such a contribution is fully allowed at the level of the internal propagator, the mass splitting $m_{\Sigma^+}\neq m_{\Sigma^{\eminus}}$ prevents the corresponding transition from occurring on shell. 

Finally, the $nn\to \pi^+\pi^{\eminus}$ channel is characterized by a particularly simple structure. In this case, only the direct contact contribution is present, as reflected in Table~\ref{tab:Dinucleon_channels_numerics}, where the coefficients $[\gamma_1]^{\pi^+\pi^{\eminus}}_{nn}$ and $[\gamma_2]^{\pi^+\pi^{\eminus}}_{nn}$ do not appear. As a result, the intranuclear decay rate can be written as
\begin{equation}
    \Gamma_A(nn\to \pi^+\pi^{\eminus})\simeq (\Delta_{nn}^{\pi^+\pi^{\eminus}})^2 [\gamma_3]^{\pi^+\pi^{\eminus}}_{nn}+(\cC_{pn}^{\pi^0})^2\,[\gamma_4]^{\pi^+\pi^{\eminus}}_{nn}
    \,.
\end{equation}

In addition to the channels discussed above, a variety of other processes can probe Wilson coefficients that are not accessible directly through baryon–antibaryon oscillations. In particular, transitions such as $pp\to \pi^+K^+$ and $pn\to \pi^+K^0$ are sensitive to $\cC_{n\Lambda}$, providing well-motivated targets for future experimental searches and access to complementary directions in parameter space. More generally, the framework developed here also enables the systematic computation of dinucleon decay processes with multi-meson final states. These results provide a broader theoretical basis for future searches, which could probe regions of the $|\Delta B|=2$ parameter space not accessible through oscillation experiments.

\subsection{Final Results}
\label{sec:pheno_final_res}

\begin{table}[t]
    \centering
\renewcommand{\arraystretch}{1.2}
\scalebox{0.815}{
\begin{tabular}{l@{\hspace{1.3cm}}c@{\hspace{1.3cm}}l}
\toprule
\multirow{1}{*}{~~~\textbf{Decay}}
&\multirow{1}{*}{\textbf{Combination}}
&\multirow{1}{*}{~~~~~\textbf{Constraint}}
\\
\midrule
\addlinespace[0.2cm]
$pp\to \pi^+\pi^+$
&$\cC_{nn}^2+0.44\,\Delta_{pp}^{\pi^+\pi^+}\cC_{nn}+0.46\,(\Delta_{pp}^{\pi^+\pi^+})^2+0.01\,(\cC_{np}^{\pi^+})^2$
&$\lesssim 1.04\times10^{\eminus 64}\,\gev^2$
\\[0.2cm]
$pp\to {\scalebox{0.75}{$K$}}^+{\scalebox{0.75}{$K$}}^+$
&$\cC_{\Lambda\Lambda}^2+0.20\,\Delta_{pp}^{{\scalebox{0.65}{$K$}}^+ {\scalebox{0.65}{$K$}}^+}\cC_{\Lambda\Lambda}+0.89\,(\Delta_{pp}^{{\scalebox{0.65}{$K$}}^+ {\scalebox{0.65}{$K$}}^+})^2
+0.11\,(\cC_{\Lambda p}^{{\scalebox{0.65}{$K$}}^+})^2$
&$\lesssim 8.10\times 10^{\eminus 65}\,\gev^2$
\\[0.2cm]
$pn\to \pi^+\pi^0$
&$\cC_{nn}^2-0.60\,\Delta_{pn}^{\pi^+\pi^0}\cC_{nn}+0.80\,(\Delta_{pn}^{\pi^+\pi^0})^2+0.03\,(\bar\cC_{pn}^\pi)^2$
&$\lesssim 1.20\times10^{\eminus63}\,\gev^2$
\\[0.2cm]
$nn\to \pi^0\pi^0$
&$\cC_{nn}^2+4.00\,\Delta_{nn}^{\pi^0\pi^0}\cC_{nn}+32.8\,(\Delta_{nn}^{\pi^0\pi^0})^2+0.05\,(\cC_{nn}^{\pi^0})^2$
&$\lesssim 1.25\times 10^{\eminus63}\,\gev^2$
\\[0.2cm]
\noalign{\vskip 0.0cm}
\cdashline{1-3}[.4pt/2pt]
\noalign{\vskip 0.2cm}
$nn\to {\scalebox{0.75}{$K$}}^+{\scalebox{0.75}{$K$}}^{\eminus}$
&$\cC^2_{\Sigma^+\Sigma^{\eminus}}+12.0\,\Delta_{nn}^{{\scalebox{0.65}{$K$}}^+ {\scalebox{0.65}{$K$}}^{\eminus}}\cC_{\Sigma^+\Sigma^{\eminus}}+116\,(\Delta_{nn}^{{\scalebox{0.65}{$K$}}^+ {\scalebox{0.65}{$K$}}^{\eminus}})^2+6.00\,(\cC_{\Sigma n}^{{\scalebox{0.65}{$K$}}})^2$
&$\lesssim 4.80\times10^{\eminus 62}\,\gev^2$
\\[0.2cm]
$nn\to \pi^+\pi^{\eminus}$
&$(\Delta_{nn}^{\pi^+\pi^{\eminus}})^2+0.21\,(\cC_{pn}^{\pi^0})$
&$\lesssim 9.09\times10^{\eminus 64}\,\gev^2$
\\[0.2cm]
\bottomrule
\end{tabular}
}
    \caption{Overview of constraints from dinucleon decay channels. The first column lists the transition, the second shows the corresponding combination of Wilson coefficients, and the third reports the resulting bounds.}
    \label{tab:dinuc_constr_num}
\end{table}

The resulting constraints from dinucleon decay on the relevant combinations of the B$\chi$PT Wilson coefficients are summarized in Table~\ref{tab:dinuc_constr_num}. Where appropriate, the combinations are normalized such that the contribution proportional to $\cC_{\cB\cB}$ does not carry additional numerical prefactors, in contrast to the interference terms and the contact contribution. These results illustrate that dinucleon decay, through the exchange diagrams, provides indirect sensitivity to the Wilson coefficients entering $n-\bar n$ and $\Lambda-\bar\Lambda$ oscillations. As discussed in Section~\ref{sec:baryon_oscillations_pheno}, the bounds on the corresponding coefficients extracted from baryon–antibaryon oscillations are given by 
\begin{equation}\label{eq: bounds coming from oscillation}
    \cC_{nn}^\sscript{osc}\lesssim 0.7\times10^{\eminus33}\,\gev\,,
    \qquad
    \cC_{\Lambda\Lambda}^\sscript{osc}\lesssim1.1\times10^{\eminus18}\,\gev\,.
\end{equation}
Focusing on the two channels that provide the strongest overall constraints, namely $pp\to \pi^+\pi^+$ and $pp\to K^+K^+$, and retaining only the contributions proportional to $\cC_{nn}^2$ and $\cC_{\Lambda\Lambda}^2$, respectively, one can extract the bounds
\begin{equation}
    \cC_{nn}^\sscript{din}\lesssim 1.02\times10^{\eminus32}\,\gev\,,
    \qquad
    \cC_{\Lambda\Lambda}^\sscript{din}\lesssim 9.0\times10^{\eminus 33}\,\gev\,.
\end{equation}
In general, however, the contributions proportional to $\Delta_{f_1f_2}^{h_1h_2}$ cannot be neglected, as both $\Delta_{f_1f_2}^{h_1h_2}$ and $\cC_{f_1f_2}$ depend on the same underlying Wilson coefficients in the chiral basis. For instance, from Eqs.~\eqref{eq:C1} and~\eqref{eq:Cffhh_eq15}, one finds that $\cC_{nn}=-\Delta_{pp}^{\pi^+\pi^+}$ upon neglecting contributions from the $\rep{15}'_R\otimes\rep{6}_L$ and $\rep{28}_R\otimes\rep1_L$ representations. In this case, the resulting bound is not significantly modified. By contrast, an analogous relation for \(\cC_{\Lambda\Lambda}\) would lead to an improvement of the corresponding bound at the level of $\sim 30\%$.

Comparing the bounds obtained here with those in Eq.~\eqref{eq: bounds coming from oscillation}, we find that $n-\bar n$ oscillations yield a constraint on $\cC_{nn}$ that is stronger by about one order of magnitude, whereas dinucleon decay provides a significantly stronger bound on $\cC_{\Lambda\Lambda}$, improving upon the oscillation limit by roughly twelve orders of magnitude, in agreement with Ref.~\cite{Beneito:2025ond}.

The constraints derived above can be further interpreted within the EFT framework developed in this work. In particular, the B$\chi$PT coefficients constrained here are related, through the matching procedure, to the underlying short-distance dynamics and can therefore be traced back to the scale of the UV theory. As outlined in Appendix~\ref{app:Cf1f2_coeffs}, the B$\chi$PT coefficients, which carry mass dimension one, can be expressed as the product of a low-energy constant and the coefficient multiplying the corresponding chiral operator. For instance, in the case of $\cC_{nn}$, the matching onto the chiral basis can be written as
\begin{equation}
\begin{alignedat}{2}
\cC_{nn}&
= \frac{\alpha_{\rep{15}_R\otimes\repbar3_L}^{(1)}}{2\Lambda_1^5}
+\frac{11\,\alpha_{\rep{15}'_R\otimes\rep6_L}}{24\,\Lambda_2^5}
+\frac{\alpha_{\repbar6_R\otimes\rep6_L}^{(1)}}{2\,\Lambda_3^5}
+\frac{\alpha_{\rep{27}_R\otimes\rep1_L}}{2\,\Lambda_4^5}\,.
\end{alignedat}
\end{equation}
For simplicity, and to provide an illustrative estimate, we assume that all coefficients belonging to the same chiral irreducible representation are associated with a common effective scale $\Lambda_i$, independently of the specific component. Using the values of the LECs quoted in Eq.~\eqref{eq:LEC_in_BxPT}, together with the bounds derived above, we obtain that the corresponding scales lie in the range $\Lambda_i\sim\cO(10^3\,\tev)$, for $\cO(1)$ couplings, with mild variation across different individual contributions. This result provides an estimate of the sensitivity to the underlying dimension-nine operators, and hence to the characteristic scale of the corresponding UV completions. 

{\color{black}{Lastly, we stress that this interpretation benefits from the explicit matching derived in this work, which relates the constrained B$\chi$PT coefficients to the underlying short-distance operator coefficients in a systematic way. At the same time, the numerical extraction of the corresponding scale should be understood within the limitations discussed above. In particular, the relevant hadronic matrix elements are evaluated in a kinematic regime close to the range of applicability of B$\chi$PT, and the resulting bounds therefore inherit the associated non-perturbative uncertainties. Consequently, the quoted values of $\Lambda_i$ should be regarded as indicative sensitivity estimates. Nevertheless, since the decay rates scale as inverse powers of the underlying dimension-nine scale, unknown corrections to the matrix elements are expected to have a reduced impact on the inferred scale, entering only through the corresponding fifth root.}}

\section{Conclusions} 
\label{sec:conclusions}
Processes involving $|\Delta B|=2$ constitute a particularly powerful probe of baryon-number violation. They offer direct sensitivity to physics beyond the Standard Model and its low-energy manifestations, which in turn can be complementary and, in several respects, qualitatively distinct with respect to $|\Delta B|=1$ effects. In this work, we developed a systematic effective field theory framework that provides a coherent description of these processes across energy scales.

Starting from the underlying quark-level operators, we classified them into irreducible representations of the $\SU(3)_R \times \SU(3)_L$ chiral symmetry group and constructed a complete, non-redundant operator basis governing $|\Delta B|=2$ transitions at low energies. Following this, we pursued two complementary directions. First, we established the connection to the UV by projecting the relevant dimension-nine SMEFT operators onto the chiral basis. Second, we matched the chiral operators onto baryon chiral perturbation theory, obtaining their representation in terms of hadronic degrees of freedom. We also provided a simple \texttt{Mathematica} notebook that allows for the direct extraction of all relevant hadronic couplings in terms of our basis coefficients.

{\color{black}{Having developed the low-energy Lagrangian for $|\Delta B|=2$, we proceeded to study the associated phenomenology. In this context, we considered two classes of observables: baryon–antibaryon oscillations and dinucleon decays. The former can be treated straightforwardly within our framework, allowing for a direct extraction of bounds on the relevant Wilson coefficients, in particular for the $n-\bar n$ and $\Lambda-\bar\Lambda$ transitions. For dinucleon decay, we provided a systematic computation of the relevant scattering amplitudes within the B$\chi$PT framework, going beyond more approximate treatments adopted in previous studies. Given the kinematics of these processes, however, the convergence of the chiral expansion is not fully controlled, and the resulting decay rates should therefore be interpreted as order-of-magnitude estimates rather than precision predictions. Within this interpretation, they can nevertheless be used to confront the $|\Delta B|=2$ operators with experimental constraints and to assess the corresponding sensitivity.}}

A key outcome of this analysis is that the same Wilson coefficients entering baryon-antibaryon oscillations also contribute to dinucleon decay processes. This enables the consistent extraction of indirect bounds on these coefficients, including directions in parameter space that remain poorly constrained or even unconstrained by oscillation searches, such as those associated with $\Sigma$ transitions or, more generally, transitions between different baryons (e.g. $\cC_{n\Lambda}$).

For $\cC_{\Lambda\Lambda}$, the limits obtained from dinucleon decay are significantly stronger than those derived from $\Lambda-\bar\Lambda$ oscillations, improving upon them by roughly twelve orders of magnitude. In addition, we showed that dinucleon decay provides a probe of $\cC_{nn}$ that is comparable in sensitivity to neutron-antineutron oscillations, with the important distinction that the latter constrains only a limited region of parameter space. More generally, a key outcome of this work is that dinucleon decay observables can place competitive constraints across a broad portion of the parameter space relevant for $|\Delta B|=2$ interactions. In this context, we have identified several promising channels for future experimental investigation, in particular dinucleon decays with at least two light mesons (pions or kaons) in the final state. 

Several directions for future work naturally emerge from our analysis. On the theoretical side, the framework developed here provides a systematic pipeline that can also be extended beyond the EFT picture to specific UV completions, allowing for a detailed study of their low-energy signatures under different assumptions. In particular, incorporating flavor structures within concrete models would offer a promising avenue to further constrain $|\Delta B|=2$ interactions, similarly to what has been done in both baryon-number-violating and baryon-number-conserving contexts~\cite{Beneito:2025fzf,Moreno-Sanchez:2025bzz,Palavric:2024gvu,Greljo:2023bdy,Greljo:2023adz,Kosnik:2025srw,Heeck:2026dmh,Isidori:2025rci,Bresciani:2026acy}. 

In addition, a key ingredient for improving the quantitative precision of the matching, especially between the chiral and baryon chiral perturbation theory descriptions, is the determination of the relevant low-energy constants, which would benefit from dedicated efforts within the lattice-QCD community. {\color{black}{More broadly, the chiral basis constructed here provides a systematic starting point for lattice-QCD calculations of the relevant matrix elements for dinucleon decay, paving the way toward more quantitatively controlled predictions for these observables. On the experimental side, improved sensitivity to dinucleon-decay processes could play a key role in the search for baryon-number-violating signals.}} From a flavor perspective, these processes offer a broader discovery potential than the more extensively studied neutron–antineutron oscillations.

{\color{black}{Another complementary direction would be to explore dispersive methods for the hadronic amplitudes entering dinucleon decay~\cite{Remmen:2020uze,Azatov:2021ygj,Colangelo:2025sah}. Such approaches could be particularly useful in the kinematic regime where the outgoing mesons carry sizable energies and the convergence of the chiral expansion becomes limited. By exploiting analyticity and unitarity, together with experimental information on meson–meson and meson–baryon scattering where available, dispersive techniques may help constrain the energy dependence of the amplitudes and account for final-state-interaction effects beyond the leading-order B$\chi$PT treatment. In this sense, they could provide an additional handle on the long-distance hadronic dynamics and complement future lattice-QCD determinations of the relevant matrix elements.}}

In conclusion, the observation of baryon-number violation would unequivocally signal a new era in fundamental physics. Historically, considerable effort has been devoted, both theoretically and experimentally, to the search for $|\Delta B|=1$ signals. However, there is no fundamental principle that forbids the leading baryon-number-violating effects from arising at $|\Delta B|=2$. The framework developed in this work provides a systematic and precise theoretical description of $|\Delta B|=2$ processes, which could be probed in future experiments.

\section*{Acknowledgements}
We thank Luca Di Luzio, Mart\'in Gonz\'alez-Alonso, V\'ictor Mart\'in Lozano, Antonio Pich and Suraj Prakash for useful discussions and for carefully reading the manuscript. We are also grateful to Nicola Barbieri for testing the \texttt{Mathematica} notebook. The work of ABB is funded by the grant CIACIF/2021/061 of the ``Generalitat Valenciana'' and also by the Spanish ``Agencia Estatal de Investigaci\'on'' through MICIN/AEI/10.13039/501100011033. The work of AP is supported by MCIU/AEI/10.13039/501100011033 (grants CEX2023-001292-S and PID2023-146220NB-I00). The work of AS received funding from the INFN Iniziativa Specifica APINE.

\clearpage
\appendix
\section{Color Tensors}
\label{app:color_tensors}
We consider the most general set of color tensors that combine six quark fields into color-singlet operators. Such tensors can be constructed as linear combinations of products of two Levi-Civita symbols of the form $\varepsilon_{\alpha\beta\gamma}\varepsilon_{\rho\sigma\tau}$. This defines a vector space spanned by ten (redundant) basis elements corresponding to all possible index assignments. Each of these elements can be represented as a vector in a $3^6$-dimensional space. By organizing them into a $10\times 3^6$ matrix, one can determine the number of linearly independent tensors by computing its rank. We find that the dimension of this vector space is five, implying that there exist at most five independent color tensors. A convenient choice of basis is given by
\begin{equation}\label{eq:app_def_col_tensors}
    \begin{alignedat}{4}
        &T^{SSS}_{\{\alpha\beta\}\{\gamma\rho\}\{\sigma\tau\}}&&=\varepsilon_{\alpha\gamma\sigma}\varepsilon_{\beta\rho\tau}+\varepsilon_{\beta\gamma\sigma}\varepsilon_{\alpha\rho\tau}+\varepsilon_{\alpha\rho\sigma}\varepsilon_{\beta\gamma\tau}+\varepsilon_{\alpha\gamma\tau}\varepsilon_{\beta\rho\sigma}\,,
        \\
        &T^{AAS}_{[\alpha\beta][\gamma\rho]\{\sigma\tau\}}&&=\varepsilon_{\alpha\beta\sigma}\varepsilon_{\gamma\rho\tau}+\varepsilon_{\alpha\beta\tau}\varepsilon_{\gamma\rho\sigma}\,,
        \\
        &T^{ASA}_{[\alpha\beta]\{\gamma\rho\}[\sigma\tau]}&&=\varepsilon_{\alpha\beta\gamma}\varepsilon_{\sigma\tau\rho}+{\varepsilon_{\alpha\beta\rho}\varepsilon_{\sigma\rho\tau}}\,, 
        \\
        &T^{SAA}_{\{\alpha\beta\}[\gamma\rho][\sigma\tau]}&&=\varepsilon_{\sigma\tau\alpha}\varepsilon_{\gamma\rho\beta}+\varepsilon_{\sigma\tau\beta}\varepsilon_{\gamma\rho\alpha}\,,
        \\
        &T^{AAA}_{[\alpha\beta][\gamma\rho][\sigma\tau]}&&=\varepsilon_{\alpha\beta\sigma}\varepsilon_{\gamma\rho\tau}-\varepsilon_{\alpha\beta\tau}\varepsilon_{\gamma\rho\sigma}\,.
\end{alignedat}
\end{equation}
This basis, already derived in Ref.~\cite{Buchoff:2015qwa}, is particularly well suited for identifying chiral irreducible representations in the construction of the chiral Lagrangian, as the corresponding color tensors exhibit definite (anti)symmetry properties within each color pair. In addition, it is orthogonal with respect to the standard scalar product in $\mathbb{R}^{3^6}$. The tensors satisfy the following identities
\begin{equation}
    \begin{alignedat}{5}
    &T^{SSS}_{\{\alpha\beta\}\{\gamma\rho\}\{\sigma\tau\}}
    &&=T^{SSS}_{\{\gamma\rho\}\{\alpha\beta\}\{\sigma\tau\}}
    &&=T^{SSS}_{\{\sigma\tau\}\{\gamma\rho\}\{\alpha\beta\}}\,,
    \\[4pt]
    &T^{AAS}_{[\alpha\beta][\gamma\rho]\{\sigma\tau\}}
    &&=T^{AAS}_{[\gamma\rho][\alpha\beta]\{\sigma\tau\}}
    &&=T^{ASA}_{[\alpha\beta]\{\sigma\tau\}[\gamma\rho]}=T^{SAA}_{\{\sigma\tau\}[\alpha\beta][\gamma\rho]}\,,
    \\[4pt]
    &T^{AAA}_{[\alpha\beta][\gamma\rho][\sigma\tau]}
    &&=-T^{AAA}_{[\gamma\rho][\alpha\beta][\sigma\tau]}
    &&=-T^{AAA}_{[\alpha\beta]  [\sigma\tau][\gamma\rho]}\,.
\end{alignedat}
\end{equation}
They also satisfy the following Schouten-type identities~\cite{Buchoff:2015qwa}
\begin{equation}\label{eq:TXXX_color_exchange_symmetries}
    \begin{alignedat}{4}
    &T^{SSS}_{\{\gamma\beta\}\{\alpha\rho\}\{\sigma\tau\}}&&=-\frac{1}{2}T^{SSS}_{\{\alpha\beta\}\{\gamma\rho\}\{\sigma\tau\}}-\frac{3}{2}T^{AAS}_{[\alpha\beta][\gamma\rho]\{\sigma\tau\}}\,,
    \\
    &T^{AAS}_{[\gamma\beta][\alpha\rho]\{\sigma\tau\}}&&=-\frac{1}{2}T^{SSS}_{\{\alpha\beta\}\{\gamma\rho\}\{\sigma\tau\}}+\frac{1}{2}T^{AAS}_{[\alpha\beta][\gamma\rho]\{\sigma\tau\}}\,,
    \\
    &T^{AAS}_{[\alpha\beta][\sigma\rho]\{\gamma\tau\}}&&=-\frac{1}{2}T^{AAS}_{[\alpha\beta][\gamma\rho]\{\sigma\tau\}}-\frac{1}{2}T^{AAS}_{[\alpha\beta][\sigma\tau]\{\gamma\rho\}}+T^{AAA}_{[\alpha\beta][\gamma\rho][\sigma\tau]}\,,
    \\
    &T^{AAA}_{[\gamma\beta][\alpha\rho][\sigma\tau]}&&=+\frac{1}{2}T^{AAS}_{[\alpha\beta][\sigma\tau]\{\gamma\rho\}}-\frac{1}{2}T^{AAS}_{[\gamma\rho][\sigma\tau]\{\alpha\beta\}}\,, 
\end{alignedat}
\end{equation}
among many others. Finally, the above relations can be inverted to express the basis tensors in terms of the color tensors. This representation proves particularly useful for projecting SMEFT operators onto the chiral basis: 
\begin{equation}
    \varepsilon_{\alpha\gamma\sigma}\varepsilon_{\beta\rho\tau}=\frac{1}{4}T^{SSS}_{\{\alpha\beta\}\{\gamma\rho\}\{\sigma\tau\}}+\frac{1}{4}T^{AAS}_{[\alpha\beta][\gamma\rho]\{\sigma\tau\}}+\frac{1}{4}T^{ASA}_{[\alpha\beta]\{\gamma\rho\}[\sigma\tau]}+\frac{1}{4}T^{SAA}_{\{\alpha\beta\}[\gamma\rho][\sigma\tau]}\,.
\end{equation}
Analogous relations follow from permutations of the indices, for instance
\begin{equation}
\begin{alignedat}{4}
    &\varepsilon_{\alpha\gamma\tau}\varepsilon_{\beta\rho\sigma}&&=\frac{1}{4}T^{SSS}_{\{\alpha\beta\}\{\gamma\rho\}\{\sigma\tau\}}+\frac{1}{4}T^{AAS}_{[\alpha\beta][\gamma\rho]\{\sigma\tau\}}-\frac{1}{4}T^{ASA}_{[\alpha\beta]\{\gamma\rho\}[\sigma\tau]}-\frac{1}{4}T^{SAA}_{\{\alpha\beta\}[\gamma\rho][\sigma\tau]}\,,
    \\[3pt]
    &\varepsilon_{\alpha\rho\sigma}\varepsilon_{\beta\gamma\tau}&&=\frac{1}{4}T^{SSS}_{\{\alpha\beta\}\{\gamma\rho\}\{\sigma\tau\}}-\frac{1}{4}T^{AAS}_{[\alpha\beta][\gamma\rho]\{\sigma\tau\}}+\frac{1}{4}T^{ASA}_{[\alpha\beta]\{\gamma\rho\}[\sigma\tau]}-\frac{1}{4}T^{SAA}_{\{\alpha\beta\}[\gamma\rho][\sigma\tau]}\,.
\end{alignedat}
\end{equation}

\section{Group-Theoretical Aspects of Chiral Representations}
\label{app:constr_chiral_irreps}
In this Appendix, we collect the group-theoretical ingredients underlying the construction of the chiral irreducible representations discussed in the main text. As a starting point, we consider the decomposition of tensor products of fundamental representations relevant for four- and six-quark structures. For the four-quark case, the relevant decomposition is given by
\begin{equation}
    \rep3_\chi^{\otimes4}=\rep{15}'_\chi\oplus\rep{15}_\chi^{\oplus 3}\oplus\repbar6_\chi^{\oplus2}\oplus\rep3_\chi^{\oplus 3}\,, \label{eq:4qirreps}
\end{equation}
while for the six-quark case, the corresponding decomposition reads
\begin{equation}
    \rep3_\chi^{\otimes6}=\rep{28}_\chi\oplus \rep{35}_\chi^{\oplus5}\oplus\rep{27}_\chi^{\oplus9}\oplus\rep{10}_\chi^{\oplus10}\oplus\repbar{10}_\chi^{\oplus5}\oplus\rep8_\chi^{\oplus16}\oplus\rep1_\chi^{\oplus5}\,, \label{eq:6qirreps}
\end{equation}
where $\rep3_\chi^{\otimes n}$ denotes the $n$-fold tensor product of the fundamental representation, while multiplicities are denoted by $\cR_\chi^{\oplus n}$, indicating the direct sum of $n$ instances of the same irreducible representation $\cR_\chi$. Although the above decompositions contain a large number of irreducible representations, not all of them are realized in the operator structures once identities in Lorentz and color space are taken into account (see Appendix~\ref{app:color_tensors}). For this reason, it is more convenient to work directly at the level of the quark bilinears $\cQ_\chi$ defined in Eq.~\eqref{eq:sixq_ident_bilinears}.

As indicated in Eq.~\eqref{eq:def2Qs}, the quark bilinear $\cQ_\chi$ can be decomposed into components that are symmetric and antisymmetric under the exchange of color and flavor indices, corresponding to the sextet and triplet representations, respectively. With this in mind, the relevant decompositions for two insertions of the quark bilinears are given by
\begin{equation}
    \repbar{3}_\chi\otimes \repbar{3}_\chi = \rep{3}_\chi\oplus \repbar{6}_\chi\,,
    \qquad
    \repbar{3}_\chi\otimes \rep{6}_\chi = \rep{3}_\chi\oplus \rep{15}_\chi\,,
    \qquad
    \rep{6}_\chi\otimes \rep{6}_\chi = \rep{15}'_\chi\oplus \rep{15}_\chi\oplus \repbar{6}_\chi\,,
\end{equation}
where $\rep{15}'_\chi$ and $\rep{15}_\chi$ correspond to the fully symmetric and mixed-symmetry representations, respectively. Analogously, for three insertions the relevant decompositions read
\begin{equation}\label{eq:redundant_transformations}
    \begin{alignedat}{2}
        \repbar{3}_\chi\otimes \repbar{3}_\chi\otimes \repbar{3}_\chi &= \repbar{10}_\chi\oplus \rep{8}^{\oplus2}_\chi\oplus \rep{1}_\chi\,,
        \\[2pt]
        \repbar{3}_\chi\otimes \repbar{3}_\chi\otimes \rep{6}_\chi &= \rep{27}_\chi\oplus \rep{10}_\chi\oplus \rep{8}^{\oplus2}_\chi\oplus \rep{1}_\chi\,,
        \\[2pt]
        \rep{6}_\chi\otimes \rep{6}_\chi\otimes \rep{6}_\chi &= \rep{35}_\chi\oplus \repbar{35}_\chi\oplus \rep{28}_\chi\oplus \rep{27}^{\oplus3}_\chi\oplus \rep{10}_\chi\oplus \repbar{10}_\chi\oplus \rep{8}^{\oplus2}_\chi\oplus \rep{1}_\chi\,.
    \end{alignedat} 
\end{equation}
Since each quark bilinear $\cQ_\chi$ transforms in the same irreducible representation in both color and flavor space, only the three combinations listed above can give rise to a color singlet. In particular, the product $\repbar{3}_\chi \otimes \rep{6}_\chi \otimes \rep{6}_\chi$ does not contain a singlet and is therefore not relevant for the construction of color-singlet operators.

In the remainder of this Appendix, we analyze in detail the construction of the individual representations and the extraction of the corresponding chiral tensors appearing in Table~\ref{tab:chiral_reps_tensors_overview}. Representations that arise in the above decompositions but are absent from Table~\ref{tab:chiral_reps_tensors_overview} can be shown to either vanish or be redundant, as demonstrated in Section~\ref{app:ch_redund}. Performing this analysis and resolving the associated redundancies leads to a construction that is naturally aligned with the color-tensor basis employed in Refs.~\cite{Buchoff:2015qwa,Rinaldi:2018osy,Rinaldi:2019thf,Chang:1980ey,Kuo:1980ew,Rao:1982gt,Rao:1983sd,Caswell:1982qs}.

\subsection{Chiral Representations for Two Quark Bilinears}

\paragraph{Construction of $\bm{\repbar6_\chi}$.} From the decomposition $\repbar3_\chi\otimes\repbar3_\chi$, the $\repbar6_\chi$ representation is obtained by symmetrizing the product of two insertions of $[\cD_{\chi}^{\alpha\beta}]_a$ as
\begin{equation}
    \repbar3_\chi\otimes\repbar3_\chi\supset \repbar6_\chi\sim [\cD_{\chi}^{\alpha\beta}]_{(a|}[\cD_{\chi}^{\gamma\rho}]_{|b)}
    =[\cX_{\repbar6_\chi}\lvert_{ab}]_{ijk\ell}\,[\cQ_\chi^{\alpha\beta}]^{ij}[\cQ_\chi^{\gamma\rho}]^{k\ell}\,, 
\end{equation}
where we define the chiral tensor as
\begin{equation}\label{eq:chir_tens_6_forB3}
[\cX_{\repbar6_\chi}\lvert_{ab}]_{ijk\ell}=\frac{1}{4}\varepsilon_{(a|ij}\varepsilon_{|b)k\ell}\,.
\end{equation}
Alternatively, from the decomposition $\rep6_\chi\otimes\rep6_\chi$, the $\repbar6_\chi$ can be extracted by contracting two $[D_\chi^{\alpha\beta}]^{ab}$ as
\begin{equation}
    \rep6_\chi\otimes\rep6_\chi\supset\repbar6_\chi\sim 
    \varepsilon_{(a|cd}\varepsilon_{|b)ef} [\cD_{\chi}^{\alpha\beta}]^{ce}[\cD_{\chi}^{\gamma\rho}]^{df}=[\cX'_{\repbar6_\chi}\lvert_{ab}]_{ijk\ell}\,[\cQ_\chi^{\alpha\beta}]^{ij}[\cQ_\chi^{\gamma\rho}]^{k\ell}\,,
\end{equation}
with the chiral tensor given as
\begin{equation}\label{eq:chir_tens_6pr_forB3}
    [\cX'_{\repbar6_\chi}\lvert_{ab}]_{ijk\ell}=\frac{1}{2}\Big(\varepsilon_{(a|ik}\varepsilon_{|b)j\ell}+\varepsilon_{(a|i\ell}\varepsilon_{|b)jk}\Big)\,.
\end{equation}
We find that this second explicit representation of the $\repbar{6}_\chi$ is linearly dependent on the one above, and is therefore redundant.



\paragraph{Construction of $\bm{\rep3_\chi}$.} From the decomposition $\rep6_\chi\otimes\repbar3_\chi$, the $\rep3_\chi$ can be extracted by direct contraction as
\begin{equation}
    \rep6_\chi\otimes\repbar3_\chi \supset\rep3_\chi\sim\frac{1}{2}[\cD_\chi^{\alpha\beta}]^{ab}[\cD_{\chi}^{\gamma\rho}]_c \delta\ud{c}{b}=[\cX_{\rep3_\chi}\lvert^a]_{ijk\ell}\,[\cQ_\chi^{\alpha\beta}]^{ij}[\cQ_\chi^{\gamma\rho}]^{k\ell}\,,
\end{equation}
where
\begin{equation}
[\cX_{\rep3_\chi}\lvert^a]_{ijk\ell}=\frac{1}{4}\delta\ud{a}{(i|}\varepsilon_{|j)k\ell}\,.
\end{equation}
Alternatively, using the decomposition $\repbar3_\chi\otimes \repbar3_\chi$, the $\rep3_\chi$ can be extracted as the antisymmetrized product of two $[D_\chi^{\alpha\beta}]_a$ insertions as
\begin{equation}
    \repbar3_\chi\otimes \repbar3_\chi\supset\rep3_\chi\sim-\frac{1}{4}\varepsilon^{abc}[\cD_\chi^{\alpha\beta}]_{b}[\cD_{\chi}^{\gamma\rho}]_c
    =[\cX'_{\rep3_\chi}\lvert^a]_{ijk\ell}\,[\cQ_\chi^{\alpha\beta}]^{ij}[\cQ_\chi^{\gamma\rho}]^{k\ell}\,,
\end{equation}
where
\begin{equation}
[\cX'_{\rep3_\chi}\lvert^a]_{ijk\ell}=\frac{1}{4}\delta\ud{a}{[i|}\varepsilon_{|j]k\ell}\,.
\end{equation}
Similarly to the extraction of the $\repbar{6}_\chi$, the alternative realization of the triplet is found to be linearly dependent on the one above and is therefore redundant.

\paragraph{Construction of $\bm{\rep{15}_\chi}$.} From the decomposition $\rep6_\chi\otimes\repbar 3_\chi$ we can construct the mixed-symmetric $\rep{15}_\chi$ irrep by taking the direct tensor product and subtracting the trace:
\begin{equation}
        \rep6_\chi\otimes\repbar 3_\chi\supset \rep{15}_\chi\sim[\cD_\chi^{\alpha\beta}]^{ab}[\cD_{\chi}^{\gamma\rho}]_c-\frac{1}{2}[\cD_\chi^{\alpha\beta}]^{(a|d}[\cD_{\chi}^{\gamma\rho}]_d \delta\ud{|b)}{c}
        =[\cX_{\rep{15}_\chi}\lvert\ud{ab}{c}]_{ijk\ell}\,[\cQ_\chi^{\alpha\beta}]^{ij}[\cQ_\chi^{\gamma\rho}]^{k\ell}\,,
\end{equation}
where the corresponding chiral tensor is defined as
\begin{equation}
    [\cX_{\rep{15}_\chi}\lvert\ud{ab}{c}]_{ijk\ell}=\frac{1}{2}\delta\ud{a}{(i|}\delta\ud{b}{|j)}\varepsilon_{ck\ell}-\frac{1}{8}\delta\ud{a}{c}\delta\ud{b}{(i|}\varepsilon_{|j)k\ell}-\frac{1}{8}\delta\ud{b}{c}\delta\ud{a}{(i|}\varepsilon_{|j)k\ell}\,.
\end{equation}
The normalization of the last two terms is chosen such that $[\cX_{\rep{15}_\chi}\lvert\ud{ab}{c}]_{ijk\ell}$ is traceless, i.e. $[\cX_{\rep{15}_\chi}\lvert\ud{ab}{b}]_{ijk\ell}=0$, as well as symmetric under the exchange $a\leftrightarrow b$, i.e. $[\cX_{\rep{15}_\chi}\lvert\ud{ab}{c}]_{ijk\ell}=[\cX_{\rep{15}_\chi}\lvert\ud{ba}{c}]_{ijk\ell}$. In total, these two constraints imply that the number of independent components is $3\times 6-3=15$.

Alternatively, one may consider constructing the $\rep{15}_\chi$ representation from the decomposition $\rep{6}_\chi \otimes \rep{6}_\chi$ through the contraction of the form
\begin{equation}
    \rep6_\chi\otimes\rep6_\chi\supset\rep{15}_\chi \sim\varepsilon_{cde} [\cD_{\chi}^{\alpha\beta}]^{(a|d}[\cD_{\chi}^{\gamma\rho}]^{|b)e}\,.
\end{equation}
However, this structure vanishes once the allowed color contractions are taken into account. In particular, the only admissible color tensor, $T^{SSS}$, is symmetric under the exchange $(\alpha,\beta)\leftrightarrow(\gamma,\rho)$, which renders the above expression identically zero.

\paragraph{Construction of $\bm{\rep{15}'_\chi}$.} From the decomposition $\rep6_\chi\otimes\rep6_\chi$, the $\rep{15}'_\chi$ is obtained by fully symmetrizing the product of two insertions of $[\cD_\chi^{\alpha\beta}]^{ab}$ as
\begin{equation}
    \rep6_\chi\otimes\rep6_\chi\supset\rep{15}'_\chi\sim[\cD_\chi^{\alpha\beta}]^{(ab}[\cD_\chi^{\gamma\rho}]^{cd)}=[\cX_{\rep{15}'_\chi}\lvert^{abcd}]_{ijk\ell}\,[\cQ_\chi^{\alpha\beta}]^{ij}[\cQ_\chi^{\gamma\rho}]^{k\ell}\,,
\end{equation}
where the fully-symmetric chiral tensor reads
\begin{equation}
    \begin{alignedat}{2}
        [\cX_{\rep{15}'_\chi}\lvert^{abcd}]_{ijk\ell}&=
        \frac{1}{6}\delta\ud{a}{(i|}\delta\ud{b}{|j)}\delta\ud{c}{(k|}\delta\ud{d}{|\ell)}
        +\frac{1}{6}\delta\ud{a}{(i|}\delta\ud{c}{|j)}\delta\ud{b}{(k|}\delta\ud{d}{|\ell)}
        +\frac{1}{6}\delta\ud{a}{(i|}\delta\ud{d}{|j)}\delta\ud{b}{(k|}\delta\ud{c}{|\ell)}
        \\[2pt]
        &\,+\frac{1}{6}\delta\ud{c}{(i|}\delta\ud{d}{|j)}\delta\ud{a}{(k|}\delta\ud{b}{|\ell)}
        +\frac{1}{6}\delta\ud{b}{(i|}\delta\ud{d}{|j)}\delta\ud{a}{(k|}\delta\ud{c}{|\ell)}
        +\frac{1}{6}\delta\ud{b}{(i|}\delta\ud{c}{|j)}\delta\ud{a}{(k|}\delta\ud{d}{|\ell)}\,.
    \end{alignedat}
\end{equation}

\subsection{Chiral Representations for Three Quark Bilinears} \label{app:cons3q}

\paragraph{Construction of $\bm{\rep{27}_\chi}$.}The $\rep{27}_\chi$ arises in the decomposition $\rep6_\chi\otimes\repbar3_\chi\otimes\repbar3_\chi\supset \rep6_\chi\otimes\repbar6_\chi$ and can be obtained by projecting out the singlet and octet components from the direct product, thereby isolating the traceless tensor corresponding to the $\rep{27}_\chi$ as
\begin{equation}
    \begin{alignedat}{2}
        \rep6_\chi\otimes\repbar6_\chi\supset\rep{27}_\chi&\sim 
        [\cD_\chi^{\alpha\beta}]^{ab}[\cD_{\chi}^{\gamma\rho}]_{(c|}[\cD_{\chi}^{\sigma\tau}]_{|d)}+\frac{1}{20}\lzm\delta\ud{a}{c}\delta\ud{b}{d}+\delta\ud{a}{d}\delta\ud{b}{c}\dzm[\cD_\chi^{\alpha\beta}]^{ef}[\cD_{\chi}^{\gamma\rho}]_{(e|}[\cD_{\chi}^{\sigma\tau}]_{|f)}
        \\&
        -\frac{1}{5}\bigg( 
        \delta\ud{a}{c}[\cD_\chi^{\alpha\beta}]^{be}[\cD_{\chi}^{\gamma\rho}]_{(d|}[\cD_{\chi}^{\sigma\tau}]_{|e)}+\delta\ud{a}{d}[\cD_\chi^{\alpha\beta}]^{be}[\cD_{\chi}^{\gamma\rho}]_{(c|}[\cD_{\chi}^{\sigma\tau}]_{|e)}
        \\
        &\hspace{+0.65cm}+\delta\ud{b}{c}[\cD_\chi^{\alpha\beta}]^{ae}[\cD_{\chi}^{\gamma\rho}]_{(d|}[\cD_{\chi}^{\sigma\tau}]_{|e)}+ 
        \delta\ud{b}{d}[\cD_\chi^{\alpha\beta}]^{ae}[\cD_{\chi}^{\gamma\rho}]_{(c|}[\cD_{\chi}^{\sigma\tau}]_{|e)}\bigg)
        \\[3pt]
        &=[\cX_{\rep{27}_\chi}\lvert\ud{ab}{cd}]_{ijk\ell mn}\,[\cQ_\chi^{\alpha\beta}]^{ij}[\cQ_\chi^{\gamma\rho}]^{k\ell}[\cQ_\chi^{\sigma\tau}]^{mn}\,,
    \end{alignedat}
\end{equation}
where the chiral tensor is given by
\begin{equation}
    \begin{alignedat}{2}
        [\cX_{\rep{27}_\chi}\lvert\ud{ab}{cd}]_{ijk\ell mn}&
        =\frac{1}{4}\delta\ud{a}{(i|}\delta\ud{b}{|j)}\varepsilon_{(c|k\ell}\varepsilon_{|d)mn}+\frac{1}{80}\lzm\delta\ud{a}{c}\delta\ud{b}{d}+\delta\ud{a}{d}\delta\ud{b}{c}\dzm\delta\ud{e}{(i|}\delta\ud{f}{|j)}\varepsilon_{(e|k\ell}\varepsilon_{|f)mn}
        \\
        &-\frac{1}{20}\bigg(
        \delta\ud{a}{c}\delta\ud{b}{(i|}\delta\ud{e}{|j)}\varepsilon_{(d|k\ell}\varepsilon_{|e)mn}
        +\delta\ud{a}{d}\delta\ud{b}{(i|}\delta\ud{e}{|j)}\varepsilon_{(c|k\ell}\varepsilon_{|e)mn}
        \\
        &\hspace{+0.71cm}+\delta\ud{b}{c}\delta\ud{a}{(i|}\delta\ud{e}{|j)}\varepsilon_{(d|k\ell}\varepsilon_{|e)mn}
        +\delta\ud{b}{d}\delta\ud{a}{(i|}\delta\ud{e}{|j)}\varepsilon_{(c|k\ell}\varepsilon_{|e)mn}
        \bigg)
        \,.
    \end{alignedat}
\end{equation}
The normalization of the last two terms is fixed such that the tensor remains traceless, i.e. $[\cX_{\rep{27}_\chi}\lvert\ud{ab}{ad}]_{ijk\ell mn}=0$. In addition, the tensor is symmetric under the exchange of indices within each pair, i.e. $(a,b)$ and $(c,d)$. Taken together, these constraints yield $6\times 6-9=27$ independent components.

Alternatively, one may extract the $\rep{27}_\chi$ starting from the decomposition $\rep6_\chi\otimes\rep6_\chi\otimes\rep6_\chi\supset \rep6_\chi\otimes\repbar6_\chi$, which leads to the same result upon implementing the replacement
\begin{equation}
    [\cD_{\chi}^{\gamma\rho}]_{(a|}[\cD_{\chi}^{\sigma\tau}]_{|b)}\quad\to \quad \varepsilon_{(a|cd}\varepsilon_{|b)ef} [\cD_{\chi}^{\gamma\rho}]^{ce}[\cD_{\chi}^{\sigma\tau}]^{df}
\end{equation}
at the level of the irreducible representation, and correspondingly
\begin{equation}
    \frac{1}{4}\varepsilon_{(a|ij}\varepsilon_{|b)k\ell}\quad\to\quad \frac{1}{2}\Big(\varepsilon_{(a|ik}\varepsilon_{|b)j\ell}+\varepsilon_{(a|i\ell}\varepsilon_{|b)jk}\Big)
\end{equation}
in the associated chiral tensor. From this, one concludes that the alternative formulation is linearly dependent on the construction given above, and is therefore redundant, as demonstrated in Appendix~\ref{app:ch_redund}.

\paragraph{Construction of $\bm{\rep{28}_\chi}$.} Using the decomposition $\rep6_\chi\otimes\rep6_\chi\otimes\rep6_\chi$, the $\rep{28}_\chi$ can be constructed by forming the fully symmetrized product of three $[\cD_\chi^{\alpha\beta}]^{ab}$ insertions as
\begin{equation}
    {\scalebox{0.96}{$\rep6_\chi\otimes\rep6_\chi\otimes\rep6_\chi\supset\rep{28}_\chi\sim[\cD_\chi^{\alpha\beta}]^{(ab}[\cD_\chi^{\gamma\rho}]^{cd}[\cD_\chi^{\sigma\tau}]^{ef)}=[\cX_{\rep{28}_\chi}\lvert^{abcdef}]_{ijk\ell mn}\,[\cQ_\chi^{\alpha\beta}]^{ij}[\cQ_\chi^{\gamma\rho}]^{k\ell}[\cQ_\chi^{\sigma\tau}]^{mn}\,,$}}
\end{equation}
where the corresponding fully-symmetric chiral tensor reads
\begin{equation}
    [\cX_{\rep{28}_\chi}\lvert^{abcdef}]_{ijk\ell mn}=
        \frac{1}{720}\sum_{\sigma\in S_6}\delta\ud{a}{\sigma(i)}\delta\ud{b}{\sigma(j)}\delta\ud{c}{\sigma(k)}\delta\ud{d}{\sigma(\ell)}\delta\ud{e}{\sigma(m)}\delta\ud{f}{\sigma(n)}\,,
\end{equation}
where the sum is taken over all the permutations of $6$ elements.
\paragraph{Construction of $\bm{\rep{1}_\chi}$.} From the decomposition $\rep6_\chi\otimes\repbar3_\chi\otimes\repbar3_\chi\supset\rep6_\chi\otimes\repbar6_\chi$, the singlet can be constructed by taking the direct contraction as
\begin{equation}\label{eq:app_singlet_first}
    \rep6_\chi\otimes\repbar3_\chi\otimes\repbar3_\chi\supset\rep1_\chi\sim\frac{1}{6}[\cD_\chi^{\alpha\beta}]^{ab}[\cD_\chi^{\gamma\rho}]_{(a|}[\cD_\chi^{\sigma\tau}]_{b)}
    =[\cX_{\rep1_\chi}]_{ijk\ell mn}\,[\cQ_\chi^{\alpha\beta}]^{ij}[\cQ_\chi^{\gamma\rho}]^{k\ell}[\cQ_\chi^{\sigma\tau}]^{mn}\,,
\end{equation}
where the chiral tensor reads
\begin{equation}
    [\cX_{\rep1_\chi}]_{ijk\ell mn}=\frac{1}{24}\varepsilon_{(i|k\ell}\varepsilon_{|j)mn}\,.
\end{equation}

In addition, there are two alternative realizations of the singlet can be considered. The first one arises from the decomposition $\repbar3_\chi\otimes\repbar3_\chi\otimes\repbar3_\chi\supset\repbar3_\chi\otimes\rep3_\chi$, where the singlet is obtained via the determinant as
\begin{equation}
    \repbar3_\chi\otimes\repbar3_\chi\otimes\repbar3_\chi\sim-\frac{1}{12}\varepsilon^{abc}[\cD_\chi^{\alpha\beta}]_{a}[\cD_\chi^{\gamma\rho}]_{b}[\cD_\chi^{\sigma\tau}]_{c}
    =[\cX^{\prime}_{\rep1_\chi}]_{ijk\ell mn}\,[\cQ_\chi^{\alpha\beta}]^{ij}[\cQ_\chi^{\gamma\rho}]^{k\ell}[\cQ_\chi^{\sigma\tau}]^{mn}\,,
\end{equation}
where 
\begin{equation}
    [\cX^{\prime}_{\rep1_\chi}]_{ijk\ell mn}=\frac{1}{24}\varepsilon_{[i|k\ell}\varepsilon_{|j]mn}\,.
\end{equation}
Finally, the decomposition $\rep6_\chi\otimes\rep6_\chi\otimes\rep6_\chi\supset\rep6_\chi\otimes\repbar6_\chi$ also allows for a singlet constructed by direct contraction as
\begin{equation}
    \rep6_\chi\otimes\rep6_\chi\otimes\rep6_\chi\sim\frac{1}{6}\varepsilon_{(a|cd}\varepsilon_{|b)ef}[\cD_\chi^{\alpha\beta}]^{ab}[\cD_\chi^{\gamma\rho}]^{ce}[\cD_\chi^{\sigma\tau}]^{df}
    =[\cX^{\prime\prime}_{\rep1_\chi}]_{ijk\ell mn}\,[\cQ_\chi^{\alpha\beta}]^{ij}[\cQ_\chi^{\gamma\rho}]^{k\ell}[\cQ_\chi^{\sigma\tau}]^{mn}\,,
\end{equation}
with the chiral tensor given by
\begin{equation}
    [\cX^{\prime\prime}_{\rep1_\chi}]_{ijk\ell mn}=\frac{1}{12}\Big(\varepsilon_{(i|km}\varepsilon_{|j)\ell n}+\varepsilon_{(i|kn}\varepsilon_{|j)\ell m}\Big)\,.
\end{equation}
The two alternative realizations of the singlet can be shown to be linearly dependent on the construction given by Eq.~\eqref{eq:app_singlet_first} and are therefore redundant.

\subsection{Redundant and Equivalent Chiral Representations}
\label{app:ch_redund}
In this subsection we provide explicit examples of the equivalence between irreps arising from the different decompositions in Eq.~\eqref{eq:redundant_transformations}. Furthermore, after imposing the allowed color contractions and applying Fierz identities in Lorentz space, some of the irreducible representations appearing in Eq.~\eqref{eq:redundant_transformations} either vanish identically or become proportional to others, rendering one of them redundant. We begin by establishing the equivalence between the two explicit representations of $\repbar6_\chi$ and $\rep3_\chi$, from which all other relevant cases follow. We then turn to the discussion of the irreducible representations that vanish.

\paragraph{Redundancy of $\bm{\repbar6_\chi}$.} We observe that the chiral tensors $\cX_{\repbar6_\chi}$ and $\cX'_{\repbar6_\chi}$ exhibit the same index symmetry structure as the color tensors $T^{AAS}$ and $T^{SSS}$, respectively. In particular, the (anti)symmetry properties of the flavor-index pairs align with those of the corresponding color indices, ensuring that their contraction does not vanish. As a result, they satisfy the following flavor-exchange relation 
\begin{equation}
    [\cX_{\repbar6_\chi}\lvert_{ab}]_{kji\ell}=-\frac{1}{4}\,[\cX'_{\repbar6_\chi}\lvert_{ab}]_{ijk\ell}+\frac{1}{2}\,[\cX_{\repbar6_\chi}\lvert_{ab}]_{ijk\ell}\,.
\end{equation}
The relation above can be lifted to the operator level by considering the corresponding contractions with the quark bilinears. Since we are interested in isolating the antisextet component, it is sufficient to focus on the part of the operator involving two quark bilinears, leaving the remaining color indices implicit. In this way, we can construct two operator structures
\begin{subequations}
\begin{align}
    &\cO_{\repbar6_\chi}^{(1)}\,\lvert_{ab}=[\cX_{\repbar6_\chi}\lvert_{ab}]_{ijk\ell}\,[\cQ^{\alpha\beta}_\chi]^{ij}[\cQ^{\gamma\rho}_\chi]^{k\ell}\,T^{AAS}_{[\alpha\beta][\gamma\rho]\{\,\cdot\,\}}\,,
    \\[3pt]
    &\cO_{\repbar6_\chi}^{(2)}\,\lvert_{ab}=[\cX'_{\repbar6_\chi}\lvert_{ab}]_{ijk\ell}\,[\cQ^{\alpha\beta}_\chi]^{ij}[\cQ^{\gamma\rho}_\chi]^{k\ell}\,T^{SSS}_{\{\alpha\beta\}\{\gamma\rho\}\{\,\cdot\,\}}\,,
    \end{align}
\end{subequations}
where the indices $a,b$ label the chiral components of the antisextet, while $\{\,\cdot\,\}$ denotes the remaining, uncontracted pair of color indices in the corresponding tensor structure. The product of quark bilinears in $\cO_{\repbar6_\chi}^{(1)}\,\lvert_{ab}$ can be rewritten using the Fierz identity as
\begin{equation}
    [\cQ^{\alpha\beta}_\chi]^{ij}[\cQ^{\gamma\rho}_\chi]^{k\ell}
    =
    -[\cQ^{\alpha\gamma}_\chi]^{ik}[\cQ^{\beta\rho}_\chi]^{j\ell}
    -[\cQ^{\alpha\rho}_\chi]^{i\ell}[\cQ^{\beta\gamma}_\chi]^{jk}\,.
\end{equation}
Combining this relation with the color-exchange relations for $T^{AAS}$ given in Eq.~\eqref{eq:TXXX_color_exchange_symmetries}, the two operator structures can be related as
\begin{equation}
    \cO_{\repbar6_\chi}^{(1)}\,\lvert_{ab}=-\frac{1}{4}\,\cO_{\repbar6_\chi}^{(2)}\,\lvert_{ab}-\frac{1}{2}\,\cO_{\repbar6_\chi}^{(1)}\,\lvert_{ab}\,,
\end{equation}
where we used the fact that, schematically, $\cX_{\repbar6_\chi} T^{SSS}=0$ and $\cX'_{\repbar6_\chi}T^{AAS}=0$ (see Eqs.~\eqref{eq:chir_tens_6_forB3} and \eqref{eq:chir_tens_6pr_forB3}). Altogether, this yields
\begin{equation}
    \cO_{\repbar6_\chi}^{(2)}\,\lvert_{ab}=-6\,\cO_{\repbar6_\chi}^{(1)}\,\lvert_{ab}\,.
\end{equation}
The above result can be applied to other chiral irreducible representations, leading to analogous equivalence relations. In particular, we find
\begin{equation}
\begin{aligned}
\rep{1}_R \otimes \rep{1}_L \;&:\quad
[\cX''_{\rep{1}_R}]_{mnijk\ell}\,\widetilde\cO_{RRR}^{ijk\ell mn}
= -6\,[\cX_{\rep{1}_R}]_{mnijk\ell}\,\cO_{RRR}^{ijk\ell mn}\,, 
\\[4pt]
\rep{27}_R \otimes \rep{1}_L \;&:\quad
[\cX'_{\rep{27}_R}\lvert\ud{ab}{cd}]_{mnijk\ell}\,\widetilde\cO_{RRR}^{ijk\ell mn}
= -6\,[\cX_{\rep{27}_R}\lvert\ud{ab}{cd}]_{mnijk\ell}\,\cO_{RRR}^{ijk\ell mn}\,, 
\\[4pt]
\repbar{6}_R \otimes \rep{6}_L \;&:\quad
[\cX'_{\repbar{6}_R}\lvert_{ab}]_{ijk\ell}\,[\cX_{\rep{6}_L}\lvert^{\dot a\dot b}]_{mn}\,\widetilde\cO_{RRL}^{ijk\ell mn}
= -6\,[\cX_{\repbar{6}_R}\lvert_{ab}]_{ijk\ell}\,[\cX_{\rep{6}_L}\lvert^{\dot a\dot b}]_{mn}\,\cO_{RRL}^{ijk\ell mn}\,.
\end{aligned}
\end{equation}

\paragraph{Redundancy of $\bm{\rep3_\chi}$.} In this case, the chiral tensors $\cX_{\rep3_\chi}$ and $\cX'_{\rep3_\chi}$ share the same index symmetry structure as $T^{AAS}$ and $T^{AAA}$, respectively, leading to the relation of the form 
\begin{equation}
    [\cX'_{\rep3_\chi}\lvert^{a}]_{kji\ell}=\frac{1}{2}\,[\cX_{\rep3_\chi}\lvert^{a}]_{ijk\ell}-\frac{1}{2}\,[\cX_{\rep3_\chi}\lvert^{a}]_{k\ell ij}\,.
\end{equation}
Restricting to the triplet component arising from two quark bilinears, the corresponding operator structures can be written as
\begin{subequations}
\begin{align}
    &\cO_{\rep3_\chi}^{(1)}\,\lvert^{a}=[\cX_{\rep3_\chi}\lvert^a]_{ijk\ell}\,[\cQ^{\alpha\beta}_\chi]^{ij}[\cQ^{\gamma\rho}_\chi]^{k\ell}\,T^{AAS}_{[\,\cdot\,][\gamma\rho]\{\alpha\beta\}}\,,
    \\[3pt]
    &\cO_{\rep3_\chi}^{(2)}\,\lvert^{a}=[\cX'_{\rep3_\chi}\lvert^a]_{ijk\ell}\,[\cQ^{\alpha\beta}_\chi]^{ij}[\cQ^{\gamma\rho}_\chi]^{k\ell}\,T^{AAA}_{[\,\cdot\,][\gamma\rho][\alpha\beta]}\,.
    \end{align}
\end{subequations}
Applying the Fierz identity, the two diquark bilinears in $\cO_{\rep3_\chi}^{(2)}\,\lvert^{a}$ can be written as
\begin{equation}
    [\cQ^{\alpha\beta}_\chi]^{ij}[\cQ^{\gamma\rho}_\chi]^{k\ell}
    =
    -[\cQ^{\alpha\gamma}_\chi]^{ik}[\cQ^{\beta\rho}_\chi]^{j\ell}
    -[\cQ^{\alpha\rho}_\chi]^{i\ell}[\cQ^{\beta\gamma}_\chi]^{jk}\,.
\end{equation}
Combining this relation with the color-exchange symmetry of $T^{AAA}$ in Eq.~\eqref{eq:TXXX_color_exchange_symmetries}, and taking into account that contractions with incompatible color and flavor symmetries vanish identically, we obtain
\begin{equation}
    \cO_{\rep3_\chi}^{(2)}\,\lvert^{a}= -\cO_{\rep3_\chi}^{(1)}\,\lvert^{a}\,.
\end{equation}
Analogously to the antisextet case, this result can be directly used to establish the corresponding equivalence relations for other chiral irreducible representations:
\begin{equation}
\begin{aligned}
\rep1_R\otimes\rep1_L \;&:\quad
[\cX'_{\rep1_R}]_{mnijk\ell}\,\widehat\cO_{RRR}^{ijk\ell mn}=-[\cX_{\rep1_R}]_{mnijk\ell}\,\cO_{RRR}^{ijk\ell mn}\,, 
\\[4pt]
\rep3_R\otimes\repbar3_L \;&:\quad
[\cX'_{\rep3_R}\lvert^a]_{mnk\ell}\,[\cX_{\repbar3_L}\lvert_{\dot a}]_{ij}\,\widehat\cO_{LRR}^{ijk\ell mn}=-[\cX_{\rep3_R}\lvert^a]_{mnk\ell}\,[\cX_{\repbar3_L}\lvert_{\dot a}]_{ij}\,\cO_{LRR}^{ijk\ell mn}\,,
\end{aligned}
\end{equation}
where we define
\begin{equation}
    \widehat\cO_{\chi_1\chi_2\chi_3}^{prstvw}
    =(q^{\alpha p} P_{\chi_1} q^{\beta r})(q^{\gamma s}  P_{\chi_2} q^{\rho t})(q^{\sigma v} P_{\chi_3} q^{\tau w})T^{AAA}_{[\alpha\beta][\gamma\rho][\sigma\tau]}\,.
\end{equation}

\paragraph{Vanishing Irreps.} We now turn to the irreducible representations that do not give rise to independent six-quark operators. The representations $\rep{35}_\chi$ and $\repbar{35}_\chi$ arise from $\rep6_\chi^{\otimes3}$, while $\rep{10}_\chi$ and $\repbar{10}_\chi$ appear in the decomposition of both $\repbar3_\chi\otimes\repbar3_\chi\otimes\rep6_\chi$ and $\rep6_\chi^{\otimes3}$. In all these cases, the corresponding chiral tensors exhibit symmetry properties under the exchange of diquark bilinears that are incompatible with the allowed color contractions once Lorentz-space Fierz identities are imposed. After rewriting all crossed contractions back into the canonical ordering of Eq.~\eqref{eq:chiral_ops_defs}, the candidate operators reduce to combinations that are symmetric in flavor and antisymmetric in color, or vice versa, and consequently vanish upon contraction with the diquark bilinears. Consequently, no operator belonging to the $\rep{35}_\chi$, $\repbar{35}_\chi$, $\rep{10}_\chi$, or $\repbar{10}_\chi$ irreps appears in the final basis of Table~\ref{tab:chiral_basis_op_count}.

For the octet, the situation is slightly different. One may first construct a tensor $X\ud{I}{J}$ with one upper and one lower adjoint index, which belongs to the reducible representation $\repbar6_\chi\otimes\rep6_\chi$, decomposing as
\begin{equation}
    \repbar6_\chi\otimes\rep6_\chi
    =\rep1_\chi\oplus\rep8_\chi\oplus\rep{27}_\chi\,.
\end{equation}
A direct inspection of the admissible contractions shows that the resulting nine-component tensor is proportional to $\delta\ud{I}{J}$, and thus lies entirely in the singlet component of $\repbar6_\chi\otimes\rep6_\chi$. Consequently, after subtracting the trace, no traceless part remains, so that the octet projection vanishes identically:
\begin{equation}
    \big[\cO_{\rep8_\chi}\big]\ud{I}{J}=0\,.
\end{equation}
Therefore, among the chiral irreducible representations displayed in Eqs.~\eqref{eq:4qirreps} and \eqref{eq:6qirreps}, only the subset listed in Table~\ref{tab:chiral_basis_op_count} remains non-vanishing and independent once gauge and Lorentz structures, as well as color and Fierz identities, are imposed.

\section{Detailed Results of Chiral Construction} \label{app:complete_ch_op}

\subsection{Complete Chiral Operator Basis}
\label{app:complete_ch_op_basis}
In Tables~\ref{tab:chiral_basis_P1}--\ref{tab:chiral_basis_P3}, we present the complete set of $|\Delta B|=2$ operators obtained by resolving the individual components of the chiral irreducible representations (see Table~\ref{tab:chiral_basis_op_count}). For each chiral representation, we explicitly label the relevant non-redundant components, and each component is associated with an independent Wilson coefficient, which we denote by $\cC_{\cR}^{abc\ldots}$, where $\cR$ labels the chiral irreducible representation and the indices $a,b,c,\ldots$ specify its components. Lastly, for each component, the corresponding operator is expressed as a linear combination of the operators defined in Eq.~\eqref{eq:chiral_ops_defs}.

\subsection{Mapping from SMEFT to Chiral Basis}
\label{app:SMEFT_chiral_map}
In Tables~\ref{tab:SMEFT_proj_Od4u2}--\ref{tab:SMEFT_proj_Od2Q4_part2}, we present the mapping of the SMEFT operators onto the chiral basis, following the prescription outlined in Section~\ref{sec:projection_SMEFT_ch_bas}. For each operator, we express the coefficients corresponding to the individual components of the chiral irreducible representations as linear combinations of the SMEFT Wilson coefficients. We also indicate the operator type, specifying the quark-field content, in order to clarify the mapping given the difference between the SMEFT/LEFT and chiral flavor indices.

\subsection{Overview of $\cC_{f_1f_2}$, $\cC_{f_1f_2}^{h_1h_2}$ and $\widetilde\cC_{f_1f_2}^{h_1h_2}$ Coefficients}
\label{app:Cf1f2_coeffs}
In Eqs.~\eqref{eq:C1}--\eqref{eq:C14}, we collect the 14 coefficients $\cC_{f_1f_2}$ introduced in Eq.~\eqref{eq:DeltaB_BxPT_lagr_comp_exp_f12h}. All coefficients are expressed as linear combinations of the low-energy constants and the Wilson coefficients of the chiral operators.
\begin{align}
\cC_{nn}&=
\frac{\alpha_{\rep{15}_R\otimes\repbar3_L}^{(1)}}{2} \cC_{\rep{15}_R\otimes\repbar3_L}^{223\dot 3}
+ \frac{\alpha_{\rep{15}'_R\otimes\rep6_L}}{12}\cC_{\rep{15}'_R\otimes\rep6_L}^{1122\dot2\dot2}  - 
\frac{\alpha_{\rep{15}'_R\otimes\rep6_L}}{8}\cC_{\rep{15}'_R\otimes\rep6_L}^{1222\dot1\dot2}  + \frac{\alpha_{\rep{15}'_R\otimes\rep6_L}}{2}\cC_{\rep{15}'_R\otimes\rep6_L}^{2222\dot1\dot1}
\nonumber 
\\&
+\frac{\alpha_{\repbar6_R\otimes\rep6_L}^{(1)}}{2}  \cC_{\repbar6_R\otimes\rep6_L}^{3 3 \dot2 \dot2}
+\frac{\alpha_{\rep{27}_R\otimes\rep1_L}}{2}  \cC_{\rep{27}_R\otimes\rep1_L}^{2 2 3 3}
\,,\label{eq:C1}
\\*
\cC_{n\Sigma^0}&=
\frac{\alpha_{\rep{15}'_R\otimes\rep{6}_L}}{6\sqrt{2}}\cC_{\rep{15}'_R \otimes \rep{6}_L}^{1122\dot{2}\dot{3}}
-\frac{\alpha_{\rep{15}'_R\otimes\rep{6}_L}}{12\sqrt{2}}\cC_{\rep{15}'_R \otimes \rep{6}_L}^{1123\dot{2}\dot{2}}
-\frac{\alpha_{\rep{15}'_R\otimes\rep{6}_L}}{4\sqrt{2}}\cC_{\rep{15}'_R \otimes \rep{6}_L}^{1222\dot{1}\dot{3}}
+\frac{\alpha_{\rep{15}'_R\otimes\rep{6}_L}}{4\sqrt{2}}\cC_{\rep{15}'_R \otimes \rep{6}_L}^{2223\dot{1}\dot{1}}
\nonumber
\\&
+\frac{\alpha_{\rep{3}_R\otimes\repbar{3}_L}^{(3)}}{8\sqrt{2}}\cC_{\rep{3}_R \otimes \repbar{3}_L}^{2\dot{3}}
+\frac{\alpha_{\rep{15}_R\otimes\repbar{3}_L}^{(2)} }{16\sqrt{2}}\cC_{\rep{15}_R \otimes \repbar{3}_L}^{121\dot{3}}
-\frac{\alpha_{\rep{15}_R\otimes\repbar{3}_L}^{(2)} }{4\sqrt{2}}\cC_{\rep{15}_R \otimes \repbar{3}_L}^{123\dot{1}}
+\frac{\alpha_{\rep{15}_R\otimes\repbar{3}_L}^{(2)} }{8\sqrt{2}}\cC_{\rep{15}_R \otimes \repbar{3}_L}^{222\dot{3}}
\nonumber
\\&
-\frac{\alpha_{\rep{15}_R\otimes\repbar{3}_L}^{(2)} }{2\sqrt{2}}\cC_{\rep{15}_R \otimes \repbar{3}_L}^{223\dot{2}}
-\frac{\alpha_{\rep{3}_R\otimes\repbar{3}_L}^{(2)}}{4\sqrt{2}}\cC_{\rep{3}_R \otimes \repbar{3}_L}^{2\dot{3}}
+\frac{3\,\alpha_{\rep{27}_R\otimes\rep{1}_L}}{10\sqrt{2}}\cC_{\rep{27}_R \otimes \rep{1}_L}^{1213}
-\frac{\sqrt{2}\,\alpha_{\rep{27}_R\otimes\rep{1}_L}}{5}\cC_{\rep{27}_R \otimes \rep{1}_L}^{2223}
\nonumber
\\&
+\frac{\alpha_{\repbar{6}_R\otimes\rep{6}_L}^{(1)}}{4\sqrt{2}}\cC_{\repbar{6}_R \otimes \rep{6}_L}^{13\dot{1}\dot{2}}
-\frac{\alpha_{\repbar{6}_R\otimes\rep{6}_L}^{(1)}}{2\sqrt{2}}\cC_{\repbar{6}_R \otimes \rep{6}_L}^{23\dot{2}\dot{2}}
+\frac{5\,\alpha_{\rep{15}_R\otimes\repbar{3}_L}^{(1)}}{16\sqrt{2}}\cC_{\rep{15}_R \otimes \repbar{3}_L}^{121\dot{3}}
+\frac{\alpha_{\rep{15}_R\otimes\repbar{3}_L}^{(1)}}{4\sqrt{2}}\cC_{\rep{15}_R \otimes \repbar{3}_L}^{123\dot{1}}
\nonumber
\\&
-\frac{3\,\alpha_{\rep{15}_R\otimes\repbar{3}_L}^{(1)}}{8\sqrt{2}}\cC_{\rep{15}_R \otimes \repbar{3}_L}^{222\dot{3}}
-\frac{\alpha_{\rep{15}_R\otimes\repbar{3}_L}^{(1)}}{2\sqrt{2}}\cC_{\rep{15}_R \otimes \repbar{3}_L}^{223\dot{2}}
-\frac{\alpha_{\repbar{6}_R\otimes\rep{6}_L}^{(2)}}{4\sqrt{2}}\cC_{\repbar{6}_R \otimes \rep{6}_L}^{23\dot{2}\dot{2}}
-\frac{\alpha_{\repbar{6}_R\otimes\rep{6}_L}^{(2)}}{2\sqrt{2}}\cC_{\repbar{6}_R \otimes \rep{6}_L}^{33\dot{2}\dot{3}}
\nonumber
\\&
-\frac{\alpha_{\rep{3}_R\otimes\repbar{3}_L}^{(1)} }{2\sqrt{2}}\cC_{\rep{3}_R \otimes \repbar{3}_L}^{2\dot{3}}
\,,
\\*
\cC_{p\Sigma^{\eminus}}&=
\frac{\alpha_{\rep{27}_R\otimes\rep{1}_L}}{5}\cC_{\rep{27}_R \otimes \rep{1}_L}^{1213}
-\frac{\alpha_{\rep{27}_R\otimes\rep{1}_L}}{10}\cC_{\rep{27}_R \otimes \rep{1}_L}^{2223}
+\frac{\alpha_{\repbar{6}_R\otimes\rep{6}_L}^{(2)}}{4}\cC_{\repbar{6}_R \otimes \rep{6}_L}^{23\dot{2}\dot{2}}
+\frac{\alpha_{\repbar{6}_R\otimes\rep{6}_L}^{(2)}}{2}\cC_{\repbar{6}_R \otimes \rep{6}_L}^{33\dot{2}\dot{3}}
\nonumber
\\&
-\frac{\alpha_{\rep{3}_R\otimes\repbar{3}_L}^{(3)}}{8}\cC_{\rep{3}_R \otimes \repbar{3}_L}^{2\dot{3}}
+\frac{3\,\alpha_{\rep{15}_R\otimes\repbar{3}_L}^{(1)}}{16}\cC_{\rep{15}_R \otimes \repbar{3}_L}^{121\dot{3}}
+\frac{\alpha_{\rep{15}_R\otimes\repbar{3}_L}^{(1)}}{4}\cC_{\rep{15}_R \otimes \repbar{3}_L}^{123\dot{1}}
-\frac{\alpha_{\rep{15}_R\otimes\repbar{3}_L}^{(1)}}{8}\cC_{\rep{15}_R \otimes \repbar{3}_L}^{222\dot{3}}
\nonumber
\\&
+\frac{\alpha_{\rep{3}_R\otimes\repbar{3}_L}^{(1)}}{2}\cC_{\rep{3}_R \otimes \repbar{3}_L}^{2\dot{3}}
+\frac{\alpha_{\rep{3}_R\otimes\repbar{3}_L}^{(2)}}{4}\cC_{\rep{3}_R \otimes \repbar{3}_L}^{2\dot{3}}
+\frac{\alpha_{\rep{15}'_R\otimes\rep{6}_L}}{12}\cC_{\rep{15}'_R \otimes \rep{6}_L}^{1122\dot{2}\dot{3}}
-\frac{\alpha_{\rep{15}'_R\otimes\rep{6}_L}}{12}\cC_{\rep{15}'_R \otimes \rep{6}_L}^{1123\dot{2}\dot{2}}
\nonumber
\\&
-\frac{\alpha_{\rep{15}'_R\otimes\rep{6}_L}}{8}\cC_{\rep{15}'_R \otimes \rep{6}_L}^{1222\dot{1}\dot{3}}
+\frac{\alpha_{\rep{15}'_R\otimes\rep{6}_L}}{24}\cC_{\rep{15}'_R \otimes \rep{6}_L}^{1223\dot{1}\dot{2}}
+\frac{\alpha_{\repbar{6}_R\otimes\rep{6}_L}^{(1)}}{4}\cC_{\repbar{6}_R \otimes \rep{6}_L}^{13\dot{1}\dot{2}}
-\frac{\alpha_{\rep{15}_R\otimes\repbar{3}_L}^{(2)}}{16}\cC_{\rep{15}_R \otimes \repbar{3}_L}^{121\dot{3}}
\nonumber
\\&
+\frac{\alpha_{\rep{15}_R\otimes\repbar{3}_L}^{(2)}}{4}\cC_{\rep{15}_R \otimes \repbar{3}_L}^{123\dot{1}}
-\frac{\alpha_{\rep{15}_R\otimes\repbar{3}_L}^{(2)}}{8}\cC_{\rep{15}_R \otimes \repbar{3}_L}^{222\dot{3}}
+\frac{\alpha_{\rep{15}_R\otimes\repbar{3}_L}^{(2)}}{2}\cC_{\rep{15}_R \otimes \repbar{3}_L}^{223\dot{2}}\,,
\end{align}

\begin{align}
\cC_{n\Xi^0}&=
\frac{\alpha_{\rep{27}_R\otimes\rep{1}_L}}{20}\cC_{\rep{27}_R \otimes \rep{1}_L}^{1111}
-\frac{3\,\alpha_{\rep{27}_R\otimes\rep{1}_L}}{20}\cC_{\rep{27}_R \otimes \rep{1}_L}^{2222}
-\frac{3\,\alpha_{\rep{27}_R\otimes\rep{1}_L}}{20}\cC_{\rep{27}_R \otimes \rep{1}_L}^{3333}
-\frac{\alpha_{\rep{3}_R\otimes\repbar{3}_L}^{(2)}}{4}\cC_{\rep{3}_R \otimes \repbar{3}_L}^{1\dot{1}}
\nonumber
\\&
+\frac{\alpha_{\repbar{6}_R\otimes\rep{6}_L}^{(1)}}{4}\cC_{\repbar{6}_R \otimes \rep{6}_L}^{23\dot{2}\dot{3}}
+\frac{\alpha_{\rep{3}_R\otimes\repbar{3}_L}^{(3)}}{4}\cC_{\rep{3}_R \otimes \repbar{3}_L}^{1\dot{1}}
+\frac{\alpha_{\rep{3}_R\otimes\repbar{3}_L}^{(3)}}{8}\cC_{\rep{3}_R \otimes \repbar{3}_L}^{2\dot{2}}
+\frac{\alpha_{\rep{3}_R\otimes\repbar{3}_L}^{(3)}}{8}\cC_{\rep{3}_R \otimes \repbar{3}_L}^{3\dot{3}}
\nonumber
\\&
+\frac{\alpha_{\rep{3}_R\otimes\repbar{3}_L}^{(1)}}{2}\cC_{\rep{3}_R \otimes \repbar{3}_L}^{2\dot{2}}
+\frac{\alpha_{\rep{3}_R\otimes\repbar{3}_L}^{(1)}}{2}\cC_{\rep{3}_R \otimes \repbar{3}_L}^{3\dot{3}}
-\frac{\alpha_{\rep{15}_R\otimes\repbar{3}_L}^{(2)}}{16}\cC_{\rep{15}_R \otimes \repbar{3}_L}^{121\dot{2}}
+\frac{\alpha_{\rep{15}_R\otimes\repbar{3}_L}^{(2)}}{4}\cC_{\rep{15}_R \otimes \repbar{3}_L}^{122\dot{1}}
\nonumber
\\&
-\frac{\alpha_{\rep{15}_R\otimes\repbar{3}_L}^{(2)}}{16}\cC_{\rep{15}_R \otimes \repbar{3}_L}^{131\dot{3}}
+\frac{3\,\alpha_{\rep{15}_R\otimes\repbar{3}_L}^{(2)}}{8}\cC_{\rep{15}_R \otimes \repbar{3}_L}^{222\dot{2}}
+\frac{3\,\alpha_{\rep{15}_R\otimes\repbar{3}_L}^{(2)}}{16}\cC_{\rep{15}_R \otimes \repbar{3}_L}^{232\dot{3}}
-\frac{\alpha_{\rep{15}_R\otimes\repbar{3}_L}^{(1)}}{16}\cC_{\rep{15}_R \otimes \repbar{3}_L}^{121\dot{2}}
\nonumber
\\&
-\frac{\alpha_{\rep{15}_R\otimes\repbar{3}_L}^{(1)}}{16}\cC_{\rep{15}_R \otimes \repbar{3}_L}^{131\dot{3}}
-\frac{\alpha_{\rep{15}_R\otimes\repbar{3}_L}^{(1)}}{8}\cC_{\rep{15}_R \otimes \repbar{3}_L}^{222\dot{2}}
+\frac{3\,\alpha_{\rep{15}_R\otimes\repbar{3}_L}^{(1)}}{16}\cC_{\rep{15}_R \otimes \repbar{3}_L}^{232\dot{3}}
-\frac{\alpha_{\rep{15}'_R\otimes\rep{6}_L}}{24}\cC_{\rep{15}'_R \otimes \rep{6}_L}^{1123\dot{2}\dot{3}}
\nonumber
\\&
+\frac{\alpha_{\rep{15}'_R\otimes\rep{6}_L}}{24}\cC_{\rep{15}'_R \otimes \rep{6}_L}^{1223\dot{1}\dot{3}}
+\frac{\alpha_{\rep{15}'_R\otimes\rep{6}_L}}{24}\cC_{\rep{15}'_R \otimes \rep{6}_L}^{1233\dot{1}\dot{2}}
-\frac{\alpha_{\rep{15}'_R\otimes\rep{6}_L}}{6}\cC_{\rep{15}'_R \otimes \rep{6}_L}^{2233\dot{1}\dot{1}}
+\frac{\alpha_{\repbar{6}_R\otimes\rep{6}_L}^{(2)}}{2}\cC_{\repbar{6}_R \otimes \rep{6}_L}^{22\dot{2}\dot{2}}
\nonumber
\\&
+\frac{\alpha_{\repbar{6}_R\otimes\rep{6}_L}^{(2)}}{4}\cC_{\repbar{6}_R \otimes \rep{6}_L}^{23\dot{2}\dot{3}}
+\frac{\alpha_{\repbar{6}_R\otimes\rep{6}_L}^{(2)}}{2}\cC_{\repbar{6}_R \otimes \rep{6}_L}^{33\dot{3}\dot{3}}
+\alpha_{\rep{1}_R\otimes\rep{1}_L}\,\cC_{\rep{1}_R \otimes \rep{1}_L}\,,
\\[0.5cm]
\cC_{n\Lambda}&=
\frac{\sqrt{3}\,\alpha_{\rep{27}_R\otimes\rep{1}_L}}{10\sqrt2}\cC_{\rep{27}_R \otimes \rep{1}_L}^{1213}
+\frac{\sqrt{3}\,\alpha_{\rep{27}_R\otimes\rep{1}_L}}{5\sqrt2}\cC_{\rep{27}_R \otimes \rep{1}_L}^{2223}
+\frac{\alpha_{\repbar{6}_R\otimes\rep{6}_L}^{(1)}}{4\sqrt{6}}\cC_{\repbar{6}_R \otimes \rep{6}_L}^{13\dot{1}\dot{2}}
+\frac{\alpha_{\repbar{6}_R\otimes\rep{6}_L}^{(1)}}{2\sqrt{6}}\cC_{\repbar{6}_R \otimes \rep{6}_L}^{23\dot{2}\dot{2}}
\nonumber
\\&
-\frac{\alpha_{\repbar{6}_R\otimes\rep{6}_L}^{(1)}}{\sqrt{6}}\cC_{\repbar{6}_R \otimes \rep{6}_L}^{33\dot{2}\dot{3}}
+\frac{5\,\alpha_{\rep{15}_R\otimes\repbar{3}_L}^{(1)}}{16\sqrt{6}}\cC_{\rep{15}_R \otimes \repbar{3}_L}^{121\dot{3}}
+\frac{\alpha_{\rep{15}_R\otimes\repbar{3}_L}^{(1)}}{4\sqrt{6}}\cC_{\rep{15}_R \otimes \repbar{3}_L}^{123\dot{1}}
+\frac{5\,\alpha_{\rep{15}_R\otimes\repbar{3}_L}^{(1)}}{8\sqrt{6}}\cC_{\rep{15}_R \otimes \repbar{3}_L}^{222\dot{3}}
\nonumber
\\&
+\frac{\alpha_{\rep{15}_R\otimes\repbar{3}_L}^{(1)}}{2\sqrt{6}}\cC_{\rep{15}_R \otimes \repbar{3}_L}^{223\dot{2}}
-\frac{\alpha_{\rep{3}_R\otimes\repbar{3}_L}^{(2)}}{4\sqrt{6}}\cC_{\rep{3}_R \otimes \repbar{3}_L}^{2\dot{3}}
+\frac{\alpha_{\rep{15}_R\otimes\repbar{3}_L}^{(2)}}{16\sqrt{6}}\cC_{\rep{15}_R \otimes \repbar{3}_L}^{121\dot{3}}
-\frac{\alpha_{\rep{15}_R\otimes\repbar{3}_L}^{(2)}}{4\sqrt{6}}\cC_{\rep{15}_R \otimes \repbar{3}_L}^{123\dot{1}}
\nonumber
\\&
+\frac{\alpha_{\rep{15}_R\otimes\repbar{3}_L}^{(2)}}{8\sqrt{6}}\cC_{\rep{15}_R \otimes \repbar{3}_L}^{222\dot{3}}
-\frac{\alpha_{\rep{15}_R\otimes\repbar{3}_L}^{(2)}}{2\sqrt{6}}\cC_{\rep{15}_R \otimes \repbar{3}_L}^{223\dot{2}}
-\frac{\alpha_{\rep{3}_R\otimes\repbar{3}_L}^{(1)}}{2\sqrt{6}}\cC_{\rep{3}_R \otimes \repbar{3}_L}^{2\dot{3}}
-\frac{\alpha_{\rep{15}'_R\otimes\rep{6}_L}}{4\sqrt{6}}\cC_{\rep{15}'_R \otimes \rep{6}_L}^{1123\dot{2}\dot{2}}
\nonumber
\\&
+\frac{\alpha_{\rep{15}'_R\otimes\rep{6}_L}}{4\sqrt{6}}\cC_{\rep{15}'_R \otimes \rep{6}_L}^{1223\dot{1}\dot{2}}
-\frac{\sqrt{3}\,\alpha_{\rep{15}'_R\otimes\rep{6}_L}}{4\sqrt2}\cC_{\rep{15}'_R \otimes \rep{6}_L}^{2223\dot{1}\dot{1}}
+\frac{\alpha_{\rep{3}_R\otimes\repbar{3}_L}^{(3)}}{8\sqrt{6}}\cC_{\rep{3}_R \otimes \repbar{3}_L}^{2\dot{3}}
-\frac{\alpha_{\repbar{6}_R\otimes\rep{6}_L}^{(2)}}{4\sqrt{6}}\cC_{\repbar{6}_R \otimes \rep{6}_L}^{23\dot{2}\dot{2}}
\nonumber
\\&
-\frac{\alpha_{\repbar{6}_R\otimes\rep{6}_L}^{(2)}}{2\sqrt{6}}\cC_{\repbar{6}_R \otimes \rep{6}_L}^{33\dot{2}\dot{3}}\,,
\\[0.5cm]
\cC_{\Xi^0\Sigma^0}&
=
\frac{5\,\alpha_{\rep{15}_R\otimes\repbar{3}_L}^{(1)}\,}{16\sqrt{2}}\cC_{\rep{15}_R \otimes \repbar{3}_L}^{131\dot{2}}
+\frac{\alpha_{\rep{15}_R\otimes\repbar{3}_L}^{(1)}}{4\sqrt{2}}\cC_{\rep{15}_R \otimes \repbar{3}_L}^{132\dot{1}}-\frac{7\,\alpha_{\rep{15}_R\otimes\repbar{3}_L}^{(1)}\,}{16\sqrt{2}}\cC_{\rep{15}_R \otimes \repbar{3}_L}^{232\dot{2}}
+\frac{3\,\alpha_{\rep{27}_R\otimes\rep{1}_L}}{10\sqrt{2}}\cC_{\rep{27}_R \otimes \rep{1}_L}^{1312}
\nonumber
\\&
-\frac{\sqrt{2}\,\alpha_{\rep{27}_R\otimes\rep{1}_L}}{5}\cC_{\rep{27}_R \otimes \rep{1}_L}^{2322}
+\frac{\alpha_{\rep{15}_R\otimes\repbar{3}_L}^{(2)}}{16\sqrt{2}}\cC_{\rep{15}_R \otimes \repbar{3}_L}^{131\dot{2}}
-\frac{\alpha_{\rep{15}_R\otimes\repbar{3}_L}^{(2)}}{4\sqrt{2}}\cC_{\rep{15}_R \otimes \repbar{3}_L}^{132\dot{1}}
-\frac{3\,\alpha_{\rep{15}_R\otimes\repbar{3}_L}^{(2)}}{16\sqrt{2}}\cC_{\rep{15}_R \otimes \repbar{3}_L}^{232\dot{2}}
\nonumber
\\&
-\frac{\alpha_{\rep{15}_R\otimes\repbar{3}_L}^{(2)}}{2\sqrt{2}}\cC_{\rep{15}_R \otimes \repbar{3}_L}^{332\dot{3}}
-\frac{\alpha_{\rep{3}_R\otimes\repbar{3}_L}^{(1)}}{2\sqrt{2}}\cC_{\rep{3}_R \otimes \repbar{3}_L}^{3\dot{2}}
-\frac{\alpha_{\rep{15}'_R\otimes\rep{6}_L}}{6\sqrt{2}}\cC_{\rep{15}'_R \otimes \rep{6}_L}^{1123\dot{3}\dot{3}}
+\frac{\alpha_{\rep{15}'_R\otimes\rep{6}_L}}{12\sqrt{2}}\cC_{\rep{15}'_R \otimes \rep{6}_L}^{1133\dot{2}\dot{3}}
\nonumber
\\&
+\frac{\alpha_{\rep{15}'_R\otimes\rep{6}_L}}{8\sqrt{2}}\cC_{\rep{15}'_R \otimes \rep{6}_L}^{1233\dot{1}\dot{3}}
-\frac{\alpha_{\rep{15}'_R\otimes\rep{6}_L}}{8\sqrt{2}}\cC_{\rep{15}'_R \otimes \rep{6}_L}^{1333\dot{1}\dot{2}}
-\frac{\alpha_{\rep{15}'_R\otimes\rep{6}_L}}{4\sqrt{2}}\cC_{\rep{15}'_R \otimes \rep{6}_L}^{2333\dot{1}\dot{1}}
+\frac{\alpha_{\repbar{6}_R\otimes\rep{6}_L}^{(1)}}{4\sqrt{2}}\cC_{\repbar{6}_R \otimes \rep{6}_L}^{12\dot{1}\dot{3}}
\nonumber
\\&
-\frac{\alpha_{\repbar{6}_R\otimes\rep{6}_L}^{(1)}}{2\sqrt{2}}\cC_{\repbar{6}_R \otimes \rep{6}_L}^{22\dot{2}\dot{3}}
-\frac{\alpha_{\repbar{6}_R\otimes\rep{6}_L}^{(2)}}{4\sqrt{2}}\cC_{\repbar{6}_R \otimes \rep{6}_L}^{23\dot{3}\dot{3}}
-\frac{\alpha_{\rep{3}_R\otimes\repbar{3}_L}^{(2)}}{4\sqrt{2}}\cC_{\rep{3}_R \otimes \repbar{3}_L}^{3\dot{2}}
+\frac{\alpha_{\rep{3}_R\otimes\repbar{3}_L}^{(3)}}{8\sqrt{2}}\cC_{\rep{3}_R \otimes \repbar{3}_L}^{3\dot{2}}\,,
\end{align}

\begin{align}
\cC_{p\Xi^{\eminus}}&=
\frac{3\,\alpha_{\rep{15}_R\otimes\repbar{3}_L}^{(1)}}{16}\cC_{\rep{15}_R \otimes \repbar{3}_L}^{131\dot{3}}
-\frac{\alpha_{\rep{15}_R\otimes\repbar{3}_L}^{(1)}}{8}\cC_{\rep{15}_R \otimes \repbar{3}_L}^{111\dot{1}}
-\frac{\alpha_{\rep{15}_R\otimes\repbar{3}_L}^{(1)}}{16}\cC_{\rep{15}_R \otimes \repbar{3}_L}^{122\dot{1}}
-\frac{\alpha_{\rep{15}_R\otimes\repbar{3}_L}^{(1)}}{16}\cC_{\rep{15}_R \otimes \repbar{3}_L}^{232\dot{3}}
\nonumber
\\&
+\alpha_{\rep{1}_R\otimes\rep{1}_L}\,\cC_{\rep{1}_R \otimes \rep{1}_L}
+\frac{\alpha_{\rep{15}'_R\otimes\rep{6}_L}}{24}\cC_{\rep{15}'_R \otimes \rep{6}_L}^{1123\dot{2}\dot{3}}
-\frac{\alpha_{\rep{15}'_R\otimes\rep{6}_L}}{6}\cC_{\rep{15}'_R \otimes \rep{6}_L}^{1133\dot{2}\dot{2}}
-\frac{\alpha_{\rep{15}'_R\otimes\rep{6}_L}}{24}\cC_{\rep{15}'_R \otimes \rep{6}_L}^{1223\dot{1}\dot{3}}
\nonumber
\\&
+\frac{\alpha_{\rep{15}'_R\otimes\rep{6}_L}}{24}\cC_{\rep{15}'_R \otimes \rep{6}_L}^{1233\dot{1}\dot{2}}
+\frac{\alpha_{\repbar{6}_R\otimes\rep{6}_L}^{(1)}}{4}\cC_{\repbar{6}_R \otimes \rep{6}_L}^{13\dot{1}\dot{3}}
+\frac{\alpha_{\rep{3}_R\otimes\repbar{3}_L}^{(3)}}{8}\cC_{\rep{3}_R \otimes \repbar{3}_L}^{1\dot{1}}
+\frac{\alpha_{\rep{3}_R\otimes\repbar{3}_L}^{(3)}}{4}\cC_{\rep{3}_R \otimes \repbar{3}_L}^{2\dot{2}}
\nonumber
\\&
+\frac{\alpha_{\rep{3}_R\otimes\repbar{3}_L}^{(3)}}{8}\cC_{\rep{3}_R \otimes \repbar{3}_L}^{3\dot{3}}
+\frac{\alpha_{\rep{3}_R\otimes\repbar{3}_L}^{(1)}}{2}\cC_{\rep{3}_R \otimes \repbar{3}_L}^{1\dot{1}}
+\frac{\alpha_{\rep{3}_R\otimes\repbar{3}_L}^{(1)}}{2}\cC_{\rep{3}_R \otimes \repbar{3}_L}^{3\dot{3}}
+\frac{3\,\alpha_{\rep{15}_R\otimes\repbar{3}_L}^{(2)}}{8}\cC_{\rep{15}_R \otimes \repbar{3}_L}^{111\dot{1}}
\nonumber
\\&
+\frac{\alpha_{\rep{15}_R\otimes\repbar{3}_L}^{(2)}}{4}\cC_{\rep{15}_R \otimes \repbar{3}_L}^{121\dot{2}}
-\frac{\alpha_{\rep{15}_R\otimes\repbar{3}_L}^{(2)}}{16}\cC_{\rep{15}_R \otimes \repbar{3}_L}^{122\dot{1}}
+\frac{3\,\alpha_{\rep{15}_R\otimes\repbar{3}_L}^{(2)}}{16}\cC_{\rep{15}_R \otimes \repbar{3}_L}^{131\dot{3}}
-\frac{\alpha_{\rep{15}_R\otimes\repbar{3}_L}^{(2)}}{16}\cC_{\rep{15}_R \otimes \repbar{3}_L}^{232\dot{3}}
\nonumber
\\&
-\frac{3\,\alpha_{\rep{27}_R\otimes\rep{1}_L}}{20}\cC_{\rep{27}_R \otimes \rep{1}_L}^{1111}
+\frac{\alpha_{\rep{27}_R\otimes\rep{1}_L}}{20}\cC_{\rep{27}_R \otimes \rep{1}_L}^{2222}
-\frac{3\,\alpha_{\rep{27}_R\otimes\rep{1}_L}}{20}\cC_{\rep{27}_R \otimes \rep{1}_L}^{3333}
-\frac{\alpha_{\rep{3}_R\otimes\repbar{3}_L}^{(2)}}{4}\cC_{\rep{3}_R \otimes \repbar{3}_L}^{2\dot{2}}
\nonumber
\\&
+\frac{\alpha_{\repbar{6}_R\otimes\rep{6}_L}^{(2)}}{2}\cC_{\repbar{6}_R \otimes \rep{6}_L}^{11\dot{1}\dot{1}}
+\frac{\alpha_{\repbar{6}_R\otimes\rep{6}_L}^{(2)}}{4}\cC_{\repbar{6}_R \otimes \rep{6}_L}^{12\dot{1}\dot{2}}
+\frac{\alpha_{\repbar{6}_R\otimes\rep{6}_L}^{(2)}}{4}\cC_{\repbar{6}_R \otimes \rep{6}_L}^{13\dot{1}\dot{3}}
+\frac{\alpha_{\repbar{6}_R\otimes\rep{6}_L}^{(2)}}{2}\cC_{\repbar{6}_R \otimes \rep{6}_L}^{33\dot{3}\dot{3}}
\,,
\\[0.5cm]
\cC_{\Sigma^{+}\Sigma^{\eminus}}&
=
\frac{3\,\alpha_{\rep{15}_R\otimes\repbar{3}_L}^{(2)}}{8}\cC_{\rep{15}_R \otimes \repbar{3}_L}^{111\dot{1}}
+\frac{3\,\alpha_{\rep{15}_R\otimes\repbar{3}_L}^{(2)}}{16}\cC_{\rep{15}_R \otimes \repbar{3}_L}^{121\dot{2}}
+\frac{3\,\alpha_{\rep{15}_R\otimes\repbar{3}_L}^{(2)}}{16}\cC_{\rep{15}_R \otimes \repbar{3}_L}^{122\dot{1}}
+\frac{\alpha_{\rep{15}_R\otimes\repbar{3}_L}^{(2)}}{4}\cC_{\rep{15}_R \otimes \repbar{3}_L}^{131\dot{3}}
\nonumber
\\&
+\frac{3\,\alpha_{\rep{15}_R\otimes\repbar{3}_L}^{(2)}}{8}\cC_{\rep{15}_R \otimes \repbar{3}_L}^{222\dot{2}}
+\frac{\alpha_{\rep{15}_R\otimes\repbar{3}_L}^{(2)}}{4}\cC_{\rep{15}_R \otimes \repbar{3}_L}^{232\dot{3}}
-\frac{\alpha_{\rep{15}_R\otimes\repbar{3}_L}^{(1)}}{8}\cC_{\rep{15}_R \otimes \repbar{3}_L}^{111\dot{1}}
+\frac{3\,\alpha_{\rep{15}_R\otimes\repbar{3}_L}^{(1)}}{16}\cC_{\rep{15}_R \otimes \repbar{3}_L}^{121\dot{2}}
\nonumber
\\&
+\frac{3\,\alpha_{\rep{15}_R\otimes\repbar{3}_L}^{(1)}}{16}\cC_{\rep{15}_R \otimes \repbar{3}_L}^{122\dot{1}}
-\frac{\alpha_{\rep{15}_R\otimes\repbar{3}_L}^{(1)}}{8}\cC_{\rep{15}_R \otimes \repbar{3}_L}^{222\dot{2}}
-\frac{\alpha_{\rep{3}_R\otimes\repbar{3}_L}^{(2)}}{4}\cC_{\rep{3}_R \otimes \repbar{3}_L}^{3\dot{3}}
+\alpha_{\rep{1}_R\otimes\rep{1}_L}\,\cC_{\rep{1}_R \otimes \rep{1}_L}
\nonumber
\\&
+\frac{\alpha_{\rep{3}_R\otimes\repbar{3}_L}^{(1)}}{2}\cC_{\rep{3}_R \otimes \repbar{3}_L}^{1\dot{1}}
+\frac{\alpha_{\rep{3}_R\otimes\repbar{3}_L}^{(1)}}{2}\cC_{\rep{3}_R \otimes \repbar{3}_L}^{2\dot{2}}
-\frac{\alpha_{\rep{15}'_R\otimes\rep{6}_L}}{6}\cC_{\rep{15}'_R \otimes \rep{6}_L}^{1122\dot{3}\dot{3}}
+\frac{\alpha_{\rep{15}'_R\otimes\rep{6}_L}}{24}\cC_{\rep{15}'_R \otimes \rep{6}_L}^{1123\dot{2}\dot{3}}
\nonumber
\\&
+\frac{\alpha_{\rep{15}'_R\otimes\rep{6}_L}}{24}\cC_{\rep{15}'_R \otimes \rep{6}_L}^{1223\dot{1}\dot{3}}
-\frac{\alpha_{\rep{15}'_R\otimes\rep{6}_L}}{24}\cC_{\rep{15}'_R \otimes \rep{6}_L}^{1233\dot{1}\dot{2}}
+\frac{\alpha_{\rep{3}_R\otimes\repbar{3}_L}^{(3)}}{8}\cC_{\rep{3}_R \otimes \repbar{3}_L}^{1\dot{1}}
+\frac{\alpha_{\rep{3}_R\otimes\repbar{3}_L}^{(3)}}{8}\cC_{\rep{3}_R \otimes \repbar{3}_L}^{2\dot{2}}
\nonumber
\\&
+\frac{\alpha_{\rep{3}_R\otimes\repbar{3}_L}^{(3)}}{4}\cC_{\rep{3}_R \otimes \repbar{3}_L}^{3\dot{3}}
+\frac{\alpha_{\repbar{6}_R\otimes\rep{6}_L}^{(1)}}{4}\cC_{\repbar{6}_R \otimes \rep{6}_L}^{12\dot{1}\dot{2}}
-\frac{3\,\alpha_{\rep{27}_R\otimes\rep{1}_L}}{20}\cC_{\rep{27}_R \otimes \rep{1}_L}^{1111}
-\frac{3\,\alpha_{\rep{27}_R\otimes\rep{1}_L}}{20}\cC_{\rep{27}_R \otimes \rep{1}_L}^{2222}
\nonumber
\\&
+\frac{\alpha_{\rep{27}_R\otimes\rep{1}_L}}{20}\cC_{\rep{27}_R \otimes \rep{1}_L}^{3333}
+\frac{\alpha_{\repbar{6}_R\otimes\rep{6}_L}^{(2)}}{2}\cC_{\repbar{6}_R \otimes \rep{6}_L}^{11\dot{1}\dot{1}}
+\frac{\alpha_{\repbar{6}_R\otimes\rep{6}_L}^{(2)}}{4}\cC_{\repbar{6}_R \otimes \rep{6}_L}^{12\dot{1}\dot{2}}
+\frac{\alpha_{\repbar{6}_R\otimes\rep{6}_L}^{(2)}}{4}\cC_{\repbar{6}_R \otimes \rep{6}_L}^{13\dot{1}\dot{3}}
\nonumber
\\&
+\frac{\alpha_{\repbar{6}_R\otimes\rep{6}_L}^{(2)}}{2}\cC_{\repbar{6}_R \otimes \rep{6}_L}^{22\dot{2}\dot{2}}
+\frac{\alpha_{\repbar{6}_R\otimes\rep{6}_L}^{(2)}}{4}\cC_{\repbar{6}_R \otimes \rep{6}_L}^{23\dot{2}\dot{3}}
\,,
\\[0.5cm]
\cC_{\Xi^0\Xi^0}&
=\frac{\alpha_{\rep{15}_R\otimes\repbar{3}_L}^{(1)}}{2}\cC_{\rep{15}_R \otimes \repbar{3}_L}^{332\dot{2}}
+\frac{\alpha_{\rep{15}'_R\otimes\rep{6}_L}}{12}\cC_{\rep{15}'_R \otimes \rep{6}_L}^{1133\dot{3}\dot{3}}
-\frac{\alpha_{\rep{15}'_R\otimes\rep{6}_L}}{8}\cC_{\rep{15}'_R \otimes \rep{6}_L}^{1333\dot{1}\dot{3}}
+\frac{\alpha_{\rep{15}'_R\otimes\rep{6}_L}}{2}\cC_{\rep{15}'_R \otimes \rep{6}_L}^{3333\dot{1}\dot{1}}
\nonumber
\\&
+\frac{\alpha_{\repbar{6}_R\otimes\rep{6}_L}^{(1)}}{2}\cC_{\repbar{6}_R \otimes \rep{6}_L}^{22\dot{3}\dot{3}}
+\frac{\alpha_{\rep{27}_R\otimes\rep{1}_L}}{2}\cC_{\rep{27}_R \otimes \rep{1}_L}^{3322}
\,,
\end{align}

\begin{align}
\cC_{\Lambda\Sigma^0}&
=
\frac{\sqrt{3}\,\alpha_{\rep{15}_R\otimes\repbar{3}_L}^{(2)}}{8}\cC_{\rep{15}_R \otimes \repbar{3}_L}^{111\dot{1}}
+\frac{5\,\alpha_{\rep{15}_R\otimes\repbar{3}_L}^{(2)}}{16\sqrt{3}}\cC_{\rep{15}_R \otimes \repbar{3}_L}^{121\dot{2}}
-\frac{5\,\alpha_{\rep{15}_R\otimes\repbar{3}_L}^{(2)}}{16\sqrt{3}}\cC_{\rep{15}_R \otimes \repbar{3}_L}^{122\dot{1}}
+\frac{\alpha_{\rep{15}_R\otimes\repbar{3}_L}^{(2)}}{4\sqrt{3}}\cC_{\rep{15}_R \otimes \repbar{3}_L}^{131\dot{3}}
\nonumber
\\&
-\frac{\sqrt{3}\,\alpha_{\rep{15}_R\otimes\repbar{3}_L}^{(2)}}{8}\cC_{\rep{15}_R \otimes \repbar{3}_L}^{222\dot{2}}
-\frac{\alpha_{\rep{15}_R\otimes\repbar{3}_L}^{(2)}}{4\sqrt{3}}\cC_{\rep{15}_R \otimes \repbar{3}_L}^{232\dot{3}}
+\frac{\sqrt{3}\,\alpha_{\rep{15}_R\otimes\repbar{3}_L}^{(1)}}{8}\cC_{\rep{15}_R \otimes \repbar{3}_L}^{111\dot{1}}
+\frac{\alpha_{\rep{15}_R\otimes\repbar{3}_L}^{(1)}}{16\sqrt{3}}\cC_{\rep{15}_R \otimes \repbar{3}_L}^{121\dot{2}}
\nonumber
\\&
-\frac{\alpha_{\rep{15}_R\otimes\repbar{3}_L}^{(1)}}{16\sqrt{3}}\cC_{\rep{15}_R \otimes \repbar{3}_L}^{122\dot{1}}
-\frac{\alpha_{\rep{15}_R\otimes\repbar{3}_L}^{(1)}}{4\sqrt{3}}\cC_{\rep{15}_R \otimes \repbar{3}_L}^{131\dot{3}}
-\frac{\sqrt{3}\,\alpha_{\rep{15}_R\otimes\repbar{3}_L}^{(1)}}{8}\cC_{\rep{15}_R \otimes \repbar{3}_L}^{222\dot{2}}
+\frac{\alpha_{\rep{15}_R\otimes\repbar{3}_L}^{(1)}}{4\sqrt{3}}\cC_{\rep{15}_R \otimes \repbar{3}_L}^{232\dot{3}}
\nonumber
\\&
+\frac{\alpha_{\rep{3}_R\otimes\repbar{3}_L}^{(1)}}{2\sqrt{3}}\cC_{\rep{3}_R \otimes \repbar{3}_L}^{1\dot{1}}
-\frac{\alpha_{\rep{3}_R\otimes\repbar{3}_L}^{(1)}}{2\sqrt{3}}\cC_{\rep{3}_R \otimes \repbar{3}_L}^{2\dot{2}}
+\frac{\sqrt{3}\,\alpha_{\rep{27}_R\otimes\rep{1}_L}}{10}\cC_{\rep{27}_R \otimes \rep{1}_L}^{1111}
-\frac{\sqrt{3}\,\alpha_{\rep{27}_R\otimes\rep{1}_L}}{10}\cC_{\rep{27}_R \otimes \rep{1}_L}^{2222}
\nonumber
\\&
-\frac{\alpha_{\rep{15}'_R\otimes\rep{6}_L}}{8\sqrt{3}}\cC_{\rep{15}'_R \otimes \rep{6}_L}^{1123\dot{2}\dot{3}}
+\frac{\alpha_{\rep{15}'_R\otimes\rep{6}_L}}{4\sqrt{3}}\cC_{\rep{15}'_R \otimes \rep{6}_L}^{1133\dot{2}\dot{2}}
+\frac{\alpha_{\rep{15}'_R\otimes\rep{6}_L}}{8\sqrt{3}}\cC_{\rep{15}'_R \otimes \rep{6}_L}^{1223\dot{1}\dot{3}}
-\frac{\alpha_{\rep{15}'_R\otimes\rep{6}_L}}{4\sqrt{3}}\cC_{\rep{15}'_R \otimes \rep{6}_L}^{2233\dot{1}\dot{1}}
\nonumber
\\&
+\frac{\alpha_{\rep{3}_R\otimes\repbar{3}_L}^{(2)}}{4\sqrt{3}}\cC_{\rep{3}_R \otimes \repbar{3}_L}^{1\dot{1}}
-\frac{\alpha_{\rep{3}_R\otimes\repbar{3}_L}^{(2)}}{4\sqrt{3}}\cC_{\rep{3}_R \otimes \repbar{3}_L}^{2\dot{2}}
-\frac{\alpha_{\rep{3}_R\otimes\repbar{3}_L}^{(3)}}{8\sqrt{3}}\cC_{\rep{3}_R \otimes \repbar{3}_L}^{1\dot{1}}
+\frac{\alpha_{\rep{3}_R\otimes\repbar{3}_L}^{(3)}}{8\sqrt{3}}\cC_{\rep{3}_R \otimes \repbar{3}_L}^{2\dot{2}}
\nonumber
\\&
+\frac{\alpha_{\repbar{6}_R\otimes\rep{6}_L}^{(2)}}{2\sqrt{3}}\cC_{\repbar{6}_R \otimes \rep{6}_L}^{11\dot{1}\dot{1}}
+\frac{\alpha_{\repbar{6}_R\otimes\rep{6}_L}^{(2)}}{4\sqrt{3}}\cC_{\repbar{6}_R \otimes \rep{6}_L}^{12\dot{1}\dot{2}}
+\frac{\alpha_{\repbar{6}_R\otimes\rep{6}_L}^{(2)}}{4\sqrt{3}}\cC_{\repbar{6}_R \otimes \rep{6}_L}^{13\dot{1}\dot{3}}
-\frac{\alpha_{\repbar{6}_R\otimes\rep{6}_L}^{(2)}}{2\sqrt{3}}\cC_{\repbar{6}_R \otimes \rep{6}_L}^{22\dot{2}\dot{2}}
\nonumber
\\&
-\frac{\alpha_{\repbar{6}_R\otimes\rep{6}_L}^{(2)}}{4\sqrt{3}}\cC_{\repbar{6}_R \otimes \rep{6}_L}^{23\dot{2}\dot{3}}
+\frac{\alpha_{\repbar{6}_R\otimes\rep{6}_L}^{(1)}}{2\sqrt{3}}\cC_{\repbar{6}_R \otimes \rep{6}_L}^{11\dot{1}\dot{1}}
-\frac{\alpha_{\repbar{6}_R\otimes\rep{6}_L}^{(1)}}{4\sqrt{3}}\cC_{\repbar{6}_R \otimes \rep{6}_L}^{13\dot{1}\dot{3}}
-\frac{\alpha_{\repbar{6}_R\otimes\rep{6}_L}^{(1)}}{2\sqrt{3}}\cC_{\repbar{6}_R \otimes \rep{6}_L}^{22\dot{2}\dot{2}}
\nonumber
\\&
+\frac{\alpha_{\repbar{6}_R\otimes\rep{6}_L}^{(1)}}{4\sqrt{3}}\cC_{\repbar{6}_R \otimes \rep{6}_L}^{23\dot{2}\dot{3}}\,,
\\[0.5cm]
\cC_{\Lambda\Xi^0}&
= 
\frac{5\,\alpha_{\rep{15}_R\otimes\repbar{3}_L}^{(1)}}{16\sqrt{6}}\cC_{\rep{15}_R \otimes \repbar{3}_L}^{131\dot{2}}
+\frac{\alpha_{\rep{15}_R\otimes\repbar{3}_L}^{(1)}}{4\sqrt{6}}\cC_{\rep{15}_R \otimes \repbar{3}_L}^{132\dot{1}}
+\frac{3\sqrt{3}\,\alpha_{\rep{15}_R\otimes\repbar{3}_L}^{(1)}}{16\sqrt2}\cC_{\rep{15}_R \otimes \repbar{3}_L}^{232\dot{2}}
-\frac{\alpha_{\rep{15}_R\otimes\repbar{3}_L}^{(1)}}{\sqrt{6}}\cC_{\rep{15}_R \otimes \repbar{3}_L}^{332\dot{3}}
\nonumber
\\&
-\frac{\alpha_{\rep{3}_R\otimes\repbar{3}_L}^{(1)}}{2\sqrt{6}}\cC_{\rep{3}_R \otimes \repbar{3}_L}^{3\dot{2}}
+\frac{\alpha_{\rep{15}_R\otimes\repbar{3}_L}^{(2)}}{16\sqrt{6}}\cC_{\rep{15}_R \otimes \repbar{3}_L}^{131\dot{2}}
-\frac{\alpha_{\rep{15}_R\otimes\repbar{3}_L}^{(2)}}{4\sqrt{6}}\cC_{\rep{15}_R \otimes \repbar{3}_L}^{132\dot{1}}
-\frac{\sqrt{3}\,\alpha_{\rep{15}_R\otimes\repbar{3}_L}^{(2)}}{16\sqrt2}\cC_{\rep{15}_R \otimes \repbar{3}_L}^{232\dot{2}}
\nonumber
\\&
-\frac{\alpha_{\rep{15}_R\otimes\repbar{3}_L}^{(2)}}{2\sqrt{6}}\cC_{\rep{15}_R \otimes \repbar{3}_L}^{332\dot{3}}
-\frac{\alpha_{\rep{3}_R\otimes\repbar{3}_L}^{(2)}}{4\sqrt{6}}\cC_{\rep{3}_R \otimes \repbar{3}_L}^{3\dot{2}}
+\frac{\alpha_{\rep{15}'_R\otimes\rep{6}_L}}{4\sqrt{6}}\cC_{\rep{15}'_R \otimes \rep{6}_L}^{1133\dot{2}\dot{3}}
-\frac{\alpha_{\rep{15}'_R\otimes\rep{6}_L}}{8\sqrt{6}}\cC_{\rep{15}'_R \otimes \rep{6}_L}^{1233\dot{1}\dot{3}}
\nonumber
\\&
-\frac{\sqrt{3}\,\alpha_{\rep{15}'_R\otimes\rep{6}_L}}{8\sqrt2}\cC_{\rep{15}'_R \otimes \rep{6}_L}^{1333\dot{1}\dot{2}}
+\frac{\sqrt{3}\,\alpha_{\rep{15}'_R\otimes\rep{6}_L}}{4\sqrt2}\cC_{\rep{15}'_R \otimes \rep{6}_L}^{2333\dot{1}\dot{1}}
+\frac{\alpha_{\rep{3}_R\otimes\repbar{3}_L}^{(3)}}{8\sqrt{6}}\cC_{\rep{3}_R \otimes \repbar{3}_L}^{3\dot{2}}
\nonumber
\\&
+\frac{\sqrt{3}\,\alpha_{\rep{27}_R\otimes\rep{1}_L}}{10\sqrt2}\cC_{\rep{27}_R \otimes \rep{1}_L}^{1312}
+\frac{\sqrt{3}\,\alpha_{\rep{27}_R\otimes\rep{1}_L}}{5\sqrt2}\cC_{\rep{27}_R \otimes \rep{1}_L}^{2322}
+\frac{\alpha_{\repbar{6}_R\otimes\rep{6}_L}^{(1)}}{4\sqrt{6}}\cC_{\repbar{6}_R \otimes \rep{6}_L}^{12\dot{1}\dot{3}}
+\frac{\alpha_{\repbar{6}_R\otimes\rep{6}_L}^{(1)}}{2\sqrt{6}}\cC_{\repbar{6}_R \otimes \rep{6}_L}^{22\dot{2}\dot{3}}
\nonumber
\\&
-\frac{\alpha_{\repbar{6}_R\otimes\rep{6}_L}^{(1)}}{\sqrt{6}}\cC_{\repbar{6}_R \otimes \rep{6}_L}^{23\dot{3}\dot{3}}
-\frac{\alpha_{\repbar{6}_R\otimes\rep{6}_L}^{(2)}}{4\sqrt{6}}\cC_{\repbar{6}_R \otimes \rep{6}_L}^{23\dot{3}\dot{3}}\,,
\end{align}

\begin{align}
\cC_{\Lambda\Lambda}&
=
\frac{\alpha_{\rep{15}_R\otimes\repbar{3}_L}^{(2)}}{16}\cC_{\rep{15}_R \otimes \repbar{3}_L}^{111\dot{1}}
+\frac{\alpha_{\rep{15}_R\otimes\repbar{3}_L}^{(2)}}{32}\cC_{\rep{15}_R \otimes \repbar{3}_L}^{121\dot{2}}
+\frac{\alpha_{\rep{15}_R\otimes\repbar{3}_L}^{(2)}}{32}\cC_{\rep{15}_R \otimes \repbar{3}_L}^{122\dot{1}}
+\frac{\alpha_{\rep{15}_R\otimes\repbar{3}_L}^{(2)}}{16}\cC_{\rep{15}_R \otimes \repbar{3}_L}^{222\dot{2}}
\nonumber
\\&
+\frac{\alpha_{\rep{1}_R\otimes\rep{1}_L}}{2}\cC_{\rep{1}_R \otimes \rep{1}_L}
-\frac{\alpha_{\rep{3}_R\otimes\repbar{3}_L}^{(2)}}{12}\cC_{\rep{3}_R \otimes \repbar{3}_L}^{1\dot{1}}
-\frac{\alpha_{\rep{3}_R\otimes\repbar{3}_L}^{(2)}}{12}\cC_{\rep{3}_R \otimes \repbar{3}_L}^{2\dot{2}}
+\frac{\alpha_{\rep{3}_R\otimes\repbar{3}_L}^{(2)}}{24}\cC_{\rep{3}_R \otimes \repbar{3}_L}^{3\dot{3}}
\nonumber
\\&
+\frac{\alpha_{\rep{3}_R\otimes\repbar{3}_L}^{(1)}}{12}\cC_{\rep{3}_R \otimes \repbar{3}_L}^{1\dot{1}}
+\frac{\alpha_{\rep{3}_R\otimes\repbar{3}_L}^{(1)}}{12}\cC_{\rep{3}_R \otimes \repbar{3}_L}^{2\dot{2}}
+\frac{\alpha_{\rep{3}_R\otimes\repbar{3}_L}^{(1)}}{3}\cC_{\rep{3}_R \otimes \repbar{3}_L}^{3\dot{3}}
+\frac{\alpha_{\rep{15}'_R\otimes\rep{6}_L}}{8}\cC_{\rep{15}'_R \otimes \rep{6}_L}^{1133\dot{2}\dot{2}}
\nonumber
\\&
-\frac{\alpha_{\rep{15}'_R\otimes\rep{6}_L}}{16}\cC_{\rep{15}'_R \otimes \rep{6}_L}^{1233\dot{1}\dot{2}}
+\frac{\alpha_{\rep{15}'_R\otimes\rep{6}_L}}{8}\cC_{\rep{15}'_R \otimes \rep{6}_L}^{2233\dot{1}\dot{1}}
+\frac{5\,\alpha_{\rep{3}_R\otimes\repbar{3}_L}^{(3)}}{48}\cC_{\rep{3}_R \otimes \repbar{3}_L}^{1\dot{1}}
+\frac{5\,\alpha_{\rep{3}_R\otimes\repbar{3}_L}^{(3)}}{48}\cC_{\rep{3}_R \otimes \repbar{3}_L}^{2\dot{2}}
\nonumber
\\&
+\frac{\alpha_{\rep{3}_R\otimes\repbar{3}_L}^{(3)}}{24}\cC_{\rep{3}_R \otimes \repbar{3}_L}^{3\dot{3}}
+\frac{\alpha_{\rep{15}_R\otimes\repbar{3}_L}^{(1)}}{16}\cC_{\rep{15}_R \otimes \repbar{3}_L}^{111\dot{1}}
+\frac{\alpha_{\rep{15}_R\otimes\repbar{3}_L}^{(1)}}{32}\cC_{\rep{15}_R \otimes \repbar{3}_L}^{121\dot{2}}
+\frac{\alpha_{\rep{15}_R\otimes\repbar{3}_L}^{(1)}}{32}\cC_{\rep{15}_R \otimes \repbar{3}_L}^{122\dot{1}}
\nonumber
\\&
-\frac{\alpha_{\rep{15}_R\otimes\repbar{3}_L}^{(1)}}{8}\cC_{\rep{15}_R \otimes \repbar{3}_L}^{131\dot{3}}
+\frac{\alpha_{\rep{15}_R\otimes\repbar{3}_L}^{(1)}}{16}\cC_{\rep{15}_R \otimes \repbar{3}_L}^{222\dot{2}}
-\frac{\alpha_{\rep{15}_R\otimes\repbar{3}_L}^{(1)}}{8}\cC_{\rep{15}_R \otimes \repbar{3}_L}^{232\dot{3}}
+\frac{\alpha_{\repbar{6}_R\otimes\rep{6}_L}^{(1)}}{12}\cC_{\repbar{6}_R \otimes \rep{6}_L}^{11\dot{1}\dot{1}}
\nonumber
\\&
+\frac{\alpha_{\repbar{6}_R\otimes\rep{6}_L}^{(1)}}{24}\cC_{\repbar{6}_R \otimes \rep{6}_L}^{12\dot{1}\dot{2}}
-\frac{\alpha_{\repbar{6}_R\otimes\rep{6}_L}^{(1)}}{12}\cC_{\repbar{6}_R \otimes \rep{6}_L}^{13\dot{1}\dot{3}}
+\frac{\alpha_{\repbar{6}_R\otimes\rep{6}_L}^{(1)}}{12}\cC_{\repbar{6}_R \otimes \rep{6}_L}^{22\dot{2}\dot{2}}
-\frac{\alpha_{\repbar{6}_R\otimes\rep{6}_L}^{(1)}}{12}\cC_{\repbar{6}_R \otimes \rep{6}_L}^{23\dot{2}\dot{3}}
\nonumber
\\
&
+\frac{\alpha_{\repbar{6}_R\otimes\rep{6}_L}^{(1)}}{3}\cC_{\repbar{6}_R \otimes \rep{6}_L}^{33\dot{3}\dot{3}}
+\frac{\alpha_{\repbar{6}_R\otimes\rep{6}_L}^{(2)}}{12}\cC_{\repbar{6}_R \otimes \rep{6}_L}^{11\dot{1}\dot{1}}
+\frac{\alpha_{\repbar{6}_R\otimes\rep{6}_L}^{(2)}}{24}\cC_{\repbar{6}_R \otimes \rep{6}_L}^{12\dot{1}\dot{2}}
+\frac{\alpha_{\repbar{6}_R\otimes\rep{6}_L}^{(2)}}{24}\cC_{\repbar{6}_R \otimes \rep{6}_L}^{13\dot{1}\dot{3}}
\nonumber
\\&
+\frac{\alpha_{\repbar{6}_R\otimes\rep{6}_L}^{(2)}}{12}\cC_{\repbar{6}_R \otimes \rep{6}_L}^{22\dot{2}\dot{2}}
+\frac{\alpha_{\repbar{6}_R\otimes\rep{6}_L}^{(2)}}{24}\cC_{\repbar{6}_R \otimes \rep{6}_L}^{23\dot{2}\dot{3}}
+\frac{\alpha_{\repbar{6}_R\otimes\rep{6}_L}^{(2)}}{3}\cC_{\repbar{6}_R \otimes \rep{6}_L}^{33\dot{3}\dot{3}}
+\frac{3\,\alpha_{\rep{27}_R\otimes\rep{1}_L}}{40}\cC_{\rep{27}_R \otimes \rep{1}_L}^{1111}
\nonumber
\\
&
+\frac{3\,\alpha_{\rep{27}_R\otimes\rep{1}_L}}{40}\cC_{\rep{27}_R \otimes \rep{1}_L}^{2222}
+\frac{9\,\alpha_{\rep{27}_R\otimes\rep{1}_L}}{40}\cC_{\rep{27}_R \otimes \rep{1}_L}^{3333}\,,
\\[0.5cm]
\cC_{\Sigma^0\Sigma^0}
&=
\frac{3\,\alpha_{\rep{15}_R\otimes\repbar{3}_L}^{(2)}}{16}\cC_{\rep{15}_R \otimes \repbar{3}_L}^{111\dot{1}}
+\frac{3\,\alpha_{\rep{15}_R\otimes\repbar{3}_L}^{(2)}}{32}\cC_{\rep{15}_R \otimes \repbar{3}_L}^{121\dot{2}}
+\frac{3\,\alpha_{\rep{15}_R\otimes\repbar{3}_L}^{(2)}}{32}\cC_{\rep{15}_R \otimes \repbar{3}_L}^{122\dot{1}}
+\frac{\alpha_{\rep{15}_R\otimes\repbar{3}_L}^{(2)}}{8}\cC_{\rep{15}_R \otimes \repbar{3}_L}^{131\dot{3}}
\nonumber
\\&
+\frac{3\,\alpha_{\rep{15}_R\otimes\repbar{3}_L}^{(2)}}{16}\cC_{\rep{15}_R \otimes \repbar{3}_L}^{222\dot{2}}
+\frac{\alpha_{\rep{15}_R\otimes\repbar{3}_L}^{(2)}}{8}\cC_{\rep{15}_R \otimes \repbar{3}_L}^{232\dot{3}}
+\frac{3\,\alpha_{\rep{15}_R\otimes\repbar{3}_L}^{(1)}}{16}\cC_{\rep{15}_R \otimes \repbar{3}_L}^{111\dot{1}}
-\frac{5\,\alpha_{\rep{15}_R\otimes\repbar{3}_L}^{(1)}}{32}\cC_{\rep{15}_R \otimes \repbar{3}_L}^{121\dot{2}}
\nonumber
\\
&-\frac{5\,\alpha_{\rep{15}_R\otimes\repbar{3}_L}^{(1)}}{32}\cC_{\rep{15}_R \otimes \repbar{3}_L}^{122\dot{1}}
+\frac{3\,\alpha_{\rep{15}_R\otimes\repbar{3}_L}^{(1)}}{16}\cC_{\rep{15}_R \otimes \repbar{3}_L}^{222\dot{2}}
+\frac{\alpha_{\rep{1}_R\otimes\rep{1}_L}}{2}\cC_{\rep{1}_R \otimes \rep{1}_L}
+\frac{\alpha_{\rep{3}_R\otimes\repbar{3}_L}^{(1)}}{4}\cC_{\rep{3}_R \otimes \repbar{3}_L}^{1\dot{1}}
\nonumber
\\&
+\frac{\alpha_{\rep{3}_R\otimes\repbar{3}_L}^{(1)}}{4}\cC_{\rep{3}_R \otimes \repbar{3}_L}^{2\dot{2}}
+\frac{7\alpha_{\rep{27}_R\otimes\rep{1}_L}}{40}\cC_{\rep{27}_R \otimes \rep{1}_L}^{1111}
+\frac{7\alpha_{\rep{27}_R\otimes\rep{1}_L}}{40}\cC_{\rep{27}_R \otimes \rep{1}_L}^{2222}
+\frac{\alpha_{\rep{27}_R\otimes\rep{1}_L}}{40}\cC_{\rep{27}_R \otimes \rep{1}_L}^{3333}
\nonumber
\\&
+\frac{\alpha_{\rep{3}_R\otimes\repbar{3}_L}^{(3)}}{16}\cC_{\rep{3}_R \otimes \repbar{3}_L}^{1\dot{1}}
+\frac{\alpha_{\rep{3}_R\otimes\repbar{3}_L}^{(3)}}{16}\cC_{\rep{3}_R \otimes \repbar{3}_L}^{2\dot{2}}
+\frac{\alpha_{\rep{3}_R\otimes\repbar{3}_L}^{(3)}}{8}\cC_{\rep{3}_R \otimes \repbar{3}_L}^{3\dot{3}}
+\frac{\alpha_{\repbar{6}_R\otimes\rep{6}_L}^{(1)}}{4}\cC_{\repbar{6}_R \otimes \rep{6}_L}^{11\dot{1}\dot{1}}
\nonumber
\\&
-\frac{\alpha_{\repbar{6}_R\otimes\rep{6}_L}^{(1)}}{8}\cC_{\repbar{6}_R \otimes \rep{6}_L}^{12\dot{1}\dot{2}}
+\frac{\alpha_{\repbar{6}_R\otimes\rep{6}_L}^{(1)}}{4}\cC_{\repbar{6}_R \otimes \rep{6}_L}^{22\dot{2}\dot{2}}
+\frac{\alpha_{\rep{15}'_R\otimes\rep{6}_L}}{6}\cC_{\rep{15}'_R \otimes \rep{6}_L}^{1122\dot{3}\dot{3}}
-\frac{\alpha_{\rep{15}'_R\otimes\rep{6}_L}}{24}\cC_{\rep{15}'_R \otimes \rep{6}_L}^{1123\dot{2}\dot{3}}
\nonumber
\\
&
+\frac{\alpha_{\rep{15}'_R\otimes\rep{6}_L}}{24}\cC_{\rep{15}'_R \otimes \rep{6}_L}^{1133\dot{2}\dot{2}}
-\frac{\alpha_{\rep{15}'_R\otimes\rep{6}_L}}{24}\cC_{\rep{15}'_R \otimes \rep{6}_L}^{1223\dot{1}\dot{3}}
+\frac{\alpha_{\rep{15}'_R\otimes\rep{6}_L}}{48}\cC_{\rep{15}'_R \otimes \rep{6}_L}^{1233\dot{1}\dot{2}}
+\frac{\alpha_{\rep{15}'_R\otimes\rep{6}_L}}{24}\cC_{\rep{15}'_R \otimes \rep{6}_L}^{2233\dot{1}\dot{1}}
\nonumber
\\
&
+\frac{\alpha_{\repbar{6}_R\otimes\rep{6}_L}^{(2)}}{4}\cC_{\repbar{6}_R \otimes \rep{6}_L}^{11\dot{1}\dot{1}}
+\frac{\alpha_{\repbar{6}_R\otimes\rep{6}_L}^{(2)}}{8}\cC_{\repbar{6}_R \otimes \rep{6}_L}^{12\dot{1}\dot{2}}
+\frac{\alpha_{\repbar{6}_R\otimes\rep{6}_L}^{(2)}}{8}\cC_{\repbar{6}_R \otimes \rep{6}_L}^{13\dot{1}\dot{3}}
+\frac{\alpha_{\repbar{6}_R\otimes\rep{6}_L}^{(2)}}{4}\cC_{\repbar{6}_R \otimes \rep{6}_L}^{22\dot{2}\dot{2}}
\nonumber
\\&
+\frac{\alpha_{\repbar{6}_R\otimes\rep{6}_L}^{(2)}}{8}\cC_{\repbar{6}_R \otimes \rep{6}_L}^{23\dot{2}\dot{3}}
-\frac{\alpha_{\rep{3}_R\otimes\repbar{3}_L}^{(2)}}{8}\cC_{\rep{3}_R \otimes \repbar{3}_L}^{3\dot{3}}
\,,
\end{align}

\begin{align}
\cC_{\Xi^{\eminus}\Sigma^+}
&=
\frac{3\,\alpha_{\rep{15}_R\otimes\repbar{3}_L}^{(1)}}{16}\cC_{\rep{15}_R \otimes \repbar{3}_L}^{131\dot{2}}
+\frac{\alpha_{\rep{15}_R\otimes\repbar{3}_L}^{(1)}}{4}\cC_{\rep{15}_R \otimes \repbar{3}_L}^{132\dot{1}}
-\frac{\alpha_{\rep{15}_R\otimes\repbar{3}_L}^{(1)}}{16}\cC_{\rep{15}_R \otimes \repbar{3}_L}^{232\dot{2}}
+\frac{\alpha_{\rep{27}_R\otimes\rep{1}_L}}{5}\cC_{\rep{27}_R \otimes \rep{1}_L}^{1312}
\nonumber
\\&
-\frac{\alpha_{\rep{27}_R\otimes\rep{1}_L}}{10}\cC_{\rep{27}_R \otimes \rep{1}_L}^{2322}
-\frac{\alpha_{\rep{15}_R\otimes\repbar{3}_L}^{(2)}}{16}\cC_{\rep{15}_R \otimes \repbar{3}_L}^{131\dot{2}}
+\frac{\alpha_{\rep{15}_R\otimes\repbar{3}_L}^{(2)}}{4}\cC_{\rep{15}_R \otimes \repbar{3}_L}^{132\dot{1}}
+\frac{3\,\alpha_{\rep{15}_R\otimes\repbar{3}_L}^{(2)}}{16}\cC_{\rep{15}_R \otimes \repbar{3}_L}^{232\dot{2}}
\nonumber
\\&
+\frac{\alpha_{\rep{15}_R\otimes\repbar{3}_L}^{(2)}}{2}\cC_{\rep{15}_R \otimes \repbar{3}_L}^{332\dot{3}}
+\frac{\alpha_{\rep{3}_R\otimes\repbar{3}_L}^{(1)}}{2}\cC_{\rep{3}_R \otimes \repbar{3}_L}^{3\dot{2}}
-\frac{\alpha_{\rep{15}'_R\otimes\rep{6}_L}}{12}\cC_{\rep{15}'_R \otimes \rep{6}_L}^{1123\dot{3}\dot{3}}
+\frac{\alpha_{\rep{15}'_R\otimes\rep{6}_L}}{12}\cC_{\rep{15}'_R \otimes \rep{6}_L}^{1133\dot{2}\dot{3}}
\nonumber
\\
&
+\frac{\alpha_{\rep{15}'_R\otimes\rep{6}_L}}{24}\cC_{\rep{15}'_R \otimes \rep{6}_L}^{1233\dot{1}\dot{3}}
-\frac{\alpha_{\rep{15}'_R\otimes\rep{6}_L}}{8}\cC_{\rep{15}'_R \otimes \rep{6}_L}^{1333\dot{1}\dot{2}}
+\frac{\alpha_{\rep{3}_R\otimes\repbar{3}_L}^{(2)}}{4}\cC_{\rep{3}_R \otimes \repbar{3}_L}^{3\dot{2}}
-\frac{\alpha_{\rep{3}_R\otimes\repbar{3}_L}^{(3)}}{8}\cC_{\rep{3}_R \otimes \repbar{3}_L}^{3\dot{2}}
\nonumber
\\
&
+\frac{\alpha_{\repbar{6}_R\otimes\rep{6}_L}^{(1)}}{4}\cC_{\repbar{6}_R \otimes \rep{6}_L}^{12\dot{1}\dot{3}}
+\frac{\alpha_{\repbar{6}_R\otimes\rep{6}_L}^{(2)}}{4}\cC_{\repbar{6}_R \otimes \rep{6}_L}^{23\dot{3}\dot{3}}
\,.\label{eq:C14}
\end{align}

\vspace{+0.3cm}
\noindent
In Eqs.~\eqref{eq:Cffhh_eq15}--\eqref{eq:Cffhh_eq20}, we report the six coefficients $\cC_{f_1f_2}^{h_1h_2}$ that mediate the contact interactions and enter the dinucleon decay channels discussed in Section~\ref{sec:pheno_num_constrs_exps}.
\begin{align}
\cC_{pp}^{\pi^{\eminus}\pi^{\eminus}}&=
\frac{5\,\alpha_{\rep{15}'_R\otimes\rep{6}_L}}{8}\cC_{\rep{15}'_R\otimes\rep{6}_L}^{1222\dot1\dot2}
-\frac{\alpha_{\rep{15}_R\otimes\repbar{3}_L}^{(1)}}{2}\cC_{\rep{15}_R\otimes\repbar{3}_L}^{223\dot3}
-\frac{13\,\alpha_{\rep{15}'_R\otimes\rep{6}_L}}{12}\cC_{\rep{15}'_R\otimes\rep{6}_L}^{1122\dot2\dot2}
\nonumber
\\&
-\frac{\alpha_{\rep{15}'_R\otimes\rep{6}_L}}{2}\cC_{\rep{15}'_R\otimes\rep{6}_L}^{2222\dot1\dot1}
-\frac{\alpha_{\rep{27}_R\otimes\rep{1}_L}}{2}\cC_{\rep{27}_R\otimes\rep{1}_L}^{2233}
-\frac{\alpha_{\repbar{6}_R\otimes\rep{6}_L}^{(1)}}{2}\cC_{\repbar{6}_R\otimes\rep{6}_L}^{33\dot2\dot2}
\,,\label{eq:Cffhh_eq15}
\\[0.5cm]
\cC_{pp}^{K^{\eminus} K^{\eminus}}&=
\frac{\alpha_{\rep{15}_R\otimes\repbar{3}_L}^{(1)}}{2}\cC_{\rep{15}_R\otimes\repbar{3}_L}^{111\dot 1}
+\frac{\alpha_{\rep{15}_R\otimes\repbar{3}_L}^{(1)}}{2}\cC_{\rep{15}_R\otimes\repbar{3}_L}^{131\dot 3}
-\frac{\alpha_{\rep{15}'_R\otimes\rep{6}_L}}{12}\cC_{\rep{15}'_R\otimes\rep{6}_L}^{1122\dot3\dot3}
-\frac{\alpha_{\rep{15}'_R\otimes\rep{6}_L}}{8}\cC_{\rep{15}'_R\otimes\rep{6}_L}^{1123\dot2\dot3}
\nonumber
\\&
-\frac{\alpha_{\rep{15}'_R\otimes\rep{6}_L}}{2}\cC_{\rep{15}'_R\otimes\rep{6}_L}^{1133\dot2\dot2}
+\frac{\alpha_{\rep{15}'_R\otimes\rep{6}_L}}{12}\cC_{\rep{15}'_R\otimes\rep{6}_L}^{1223\dot1\dot3}
+\frac{\alpha_{\rep{15}'_R\otimes\rep{6}_L}}{8}\cC_{\rep{15}'_R\otimes\rep{6}_L}^{1233\dot1\dot2}
-\frac{\alpha_{\rep{15}'_R\otimes\rep{6}_L}}{12}\cC_{\rep{15}'_R\otimes\rep{6}_L}^{2233\dot1\dot1}
\nonumber
\\&
-\frac{\alpha_{\rep{27}_R\otimes\rep{1}_L}}{2}\cC_{\rep{27}_R\otimes\rep{1}_L}^{1111}
-\frac{\alpha_{\rep{27}_R\otimes\rep{1}_L}}{2}\cC_{\rep{27}_R\otimes\rep{1}_L}^{3333}
-\frac{\alpha_{\repbar{6}_R\otimes\rep{6}_L}^{(1)}}{2}\cC_{\repbar{6}_R\otimes\rep{6}_L}^{11\dot1\dot1}
-\frac{\alpha_{\repbar{6}_R\otimes\rep{6}_L}^{(1)}}{2}\cC_{\repbar{6}_R\otimes\rep{6}_L}^{13\dot1\dot3}
\nonumber
\\&
-\frac{\alpha_{\repbar{6}_R\otimes\rep{6}_L}^{(1)}}{2}\cC_{\repbar{6}_R\otimes\rep{6}_L}^{33\dot3\dot3}
\,,
\\[0.5cm]
\cC_{pn}^{\pi^{\eminus} \pi^0}&=
\frac{\alpha_{\rep{15}_R\otimes\repbar{3}_L}^{(1)}}{\sqrt{2}}\cC_{\rep{15}_R\otimes\repbar{3}_L}^{223\dot3}
-\frac{11\,\alpha_{\rep{15}'_R\otimes\rep{6}_L}}{6\sqrt{2}}\cC_{\rep{15}'_R\otimes\rep{6}_L}^{1122\dot2\dot{2}}
-\frac{5\,\alpha_{\rep{15}'_R\otimes\rep{6}_L}}{4\sqrt{2}}\cC_{\rep{15}'_R\otimes\rep{6}_L}^{1222\dot1\dot{2}}
+\frac{\alpha_{\repbar{6}_R\otimes\rep{6}_L}^{(1)}}{\sqrt{2}}\cC_{\repbar{6}_R\otimes\rep{6}_L}^{33\dot2\dot2}
\nonumber
\\&
+\frac{5\,\alpha_{\rep{15}'_R\otimes\rep{6}_L}}{\sqrt{2}}\cC_{\rep{15}'_R\otimes\rep{6}_L}^{2222\dot1\dot{1}}
+\frac{\alpha_{\rep{27}_R\otimes\rep{1}_L}}{\sqrt{2}}\cC_{\rep{27}_R\otimes\rep{1}_L}^{2233}
\,,
\\[0.5cm]
\cC_{nn}^{\pi^0\pi^0}&=
\frac{\alpha_{\rep{15}'_R\otimes\rep{6}_L}}{8}\cC_{\rep{15}'_R\otimes\rep{6}_L}^{1222\dot1\dot2}
-\frac{\alpha_{\rep{15}_R\otimes\repbar{3}_L}^{(1)}}{2}\cC_{\rep{15}_R\otimes\repbar{3}_L}^{223\dot3}
-\frac{\alpha_{\rep{15}'_R\otimes\rep{6}_L}}{12}\cC_{\rep{15}'_R\otimes\rep{6}_L}^{1122\dot2\dot2}
-\frac{\alpha_{\repbar{6}_R\otimes\rep{6}_L}^{(1)}}{2}\cC_{\repbar{6}_R\otimes\rep{6}_L}^{33\dot2\dot2}
\nonumber
\\&
-\frac{\alpha_{\rep{27}_R\otimes\rep{1}_L}}{2}\cC_{\rep{27}_R\otimes\rep{1}_L}^{2233}
-\frac{9\,\alpha_{\rep{15}'_R\otimes\rep{6}_L}}{2}\cC_{\rep{15}'_R\otimes\rep{6}_L}^{2222\dot1\dot1}
\,,
\end{align}

\begin{align}
\cC_{nn}^{K^{\eminus} K^+}&=
\frac{\alpha_{\rep{15}'_R\otimes\rep{6}_L}}{8}\cC_{\rep{15}'_R\otimes\rep{6}_L}^{1222\dot1\dot2}
-\frac{\alpha_{\rep{15}_R\otimes\repbar{3}_L}^{(1)}}{2}\cC_{\rep{15}_R\otimes\repbar{3}_L}^{223\dot3}
-\frac{\alpha_{\rep{15}'_R\otimes\rep{6}_L}}{12}\cC_{\rep{15}'_R\otimes\rep{6}_L}^{1122\dot2\dot2}
-\frac{\alpha_{\rep{15}'_R\otimes\rep{6}_L}}{2}\cC_{\rep{15}'_R\otimes\rep{6}_L}^{2222\dot1\dot1}
\nonumber
\\&
-\frac{\alpha_{\rep{27}_R\otimes\rep{1}_L}}{2}\cC_{\rep{27}_R\otimes\rep{1}_L}^{2233}
-\frac{\alpha_{\repbar{6}_R\otimes\rep{6}_L}^{(1)}}{2}\cC_{\repbar{6}_R\otimes\rep{6}_L}^{33\dot2\dot2}
\,,
\\[0.5cm]
\cC_{nn}^{\pi^+\pi^{\eminus}}&=
\frac{13\,\alpha_{\rep{15}'_R\otimes\rep{6}_L}}{8}\cC_{\rep{15}'_R\otimes\rep{6}_L}^{1222\dot 1\dot 2}
-\frac{\alpha_{\rep{15}_R\otimes\repbar{3}_L}^{(1)}}{2}\cC_{\rep{15}_R\otimes\repbar{3}_L}^{223\dot 3}
-\frac{\alpha_{\repbar{6}_R\otimes\rep{6}_L}^{(1)}}{2}\cC_{\repbar{6}_R\otimes\rep{6}_L}^{33\dot 2\dot 2}
-\frac{\alpha_{\rep{27}_R\otimes\rep{1}_L}}{2}\cC_{\rep{27}_R\otimes\rep{1}_L}^{2233}
\nonumber
\\&
-\frac{5\,\alpha_{\rep{15}'_R\otimes\rep{6}_L}}{2}\cC_{\rep{15}'_R\otimes\rep{6}_L}^{2222\dot 1\dot 1}
-\frac{13\,\alpha_{\rep{15}'_R\otimes\rep{6}_L}}{12}\cC_{\rep{15}'_R\otimes\rep{6}_L}^{1122\dot 2\dot 2}
\,.\label{eq:Cffhh_eq20}
\end{align}

\vspace{+0.3cm}
\noindent
In Eq.~\eqref{eq:Cffhh_tilde}, we report the six coefficients $\widetilde\cC_{f_1f_2}^{h_1h_2}$ that mediate derivative contact interactions and enter the dinucleon decay channels discussed in Section~\ref{sec:pheno_num_constrs_exps}.
\begin{equation}\label{eq:Cffhh_tilde}
\begin{alignedat}{8}
&\widetilde\cC_{pp}^{\pi^{\eminus}\pi^{\eminus}}
&&=-\frac{2\,\alpha_{\rep{28}_R\otimes\rep1_L}}{15}\cC_{\rep{28}_R\otimes\rep1_L}^{112222}\,,
&&\qquad
\widetilde\cC_{pn}^{\pi^{\eminus} \pi^0}
&&=-\frac{8\sqrt2\,\alpha_{\rep{28}_R\otimes\rep1_L}}{15}\cC_{\rep{28}_R\otimes\rep1_L}^{112222}\,,
\\[0.5cm]
&\widetilde\cC_{pp}^{K^{\eminus} K^{\eminus}}
&&=-\frac{\alpha_{\rep{28}_R\otimes\rep1_L}}{45}\cC_{\rep{28}_R\otimes\rep1_L}^{112233}\,,
&&\qquad
\widetilde\cC_{nn}^{\pi^0\pi^0}
&&=-\frac{4\,\alpha_{\rep{28}_R\otimes\rep1_L}}{15}\cC_{\rep{28}_R\otimes\rep1_L}^{112222}\,,
\\[0.5cm]
&\widetilde\cC_{nn}^{K^{\eminus} K^+}&&=0\,,
&&\qquad
\widetilde\cC_{nn}^{\pi^+\pi^{\eminus}}&&
=\frac{4\,\alpha_{\rep{28}_R\otimes\rep1_L}}{15}\cC_{\rep{28}_R\otimes\rep1_L}^{112222}\,.
\end{alignedat}
\end{equation}

\section{Details on Computation of Dinucleon Decay Amplitudes}
\subsection{Computation of Exchange Diagrams}
\label{app:QFT_comp_matrix_elem}
In this subsection, we outline the key steps entering the evaluation of the exchange contributions to the amplitude, corresponding to diagrams $(b)$ and $(c)$ in Figure~\ref{fig:tree_lev_feyn_diags_dinuc}. We consider the process $f_1(p_1) f_2(p_2) \to h_1(k_1) h_2(k_2)$, where $p_{1,2}$ and $k_{1,2}$ denote the incoming and outgoing momenta, respectively. The computation can be performed either using standard Feynman rules or within the framework of the QFT time-ordered product. In both approaches, a key feature of the exchange diagrams is the structure of the fermionic propagator, which contains a quadratic term in the fields corresponding to a mixing between fermionic states.

From a perturbative perspective, this mixing can be treated as an insertion, as it does not modify the mass eigenstates at leading order. Nevertheless, it induces a non-trivial structure in the fermionic propagator. To make this explicit, we introduce the mass matrix $M$, taken to be diagonal in flavor space, and a matrix $\Delta$ encoding the off-diagonal mixing effects between fermions $f_i$ and $f_j$, with $\Delta_{ij} \ll M_{ii}, M_{jj}$. The fermionic propagator, viewed as a matrix in flavor space, then takes the form
\begin{equation}
    \dfrac{1}{\slashed p - M-\Delta}=\dfrac{1}{\slashed p-M}\sum_{n=0}^{\infty}\left(\dfrac{\Delta}{\slashed p-M}\right)^n=\dfrac{1}{\slashed p-M}+\dfrac{1}{\slashed p-M}\Delta\dfrac{1}{\slashed p -M}+\cO(\Delta^2)\,.
\end{equation}
This structure allows the mixing between $f_3$ and $f_4$ to be treated as an effective interaction insertion between two fermionic propagators. The matrix element is then evaluated following the standard fermion flow prescription, which fixes the ordering of the propagators and insertions. For the $t$-channel contribution, we get
\begin{equation}
\begin{alignedat}{2}
{\scalebox{0.96}{
$    i\cM_b=\frac{1}{f_\pi^2}i\cC^\sscript{LO}_{f_3f_1h_1}i\cC^\sscript{LO}_{f_4f_2h_2}\left[\dfrac{i}{\slashed p_t-m_{f_3}}i\slashed k_1\gamma^5u(p_1)\right]_{\alpha}i\cC_{f_3f_4}\cS_{f_3f_4} C_{\alpha\beta}\left[\dfrac{i}{-\slashed p_t-m_{f_4}}i\slashed k_2\gamma^5u(p_2)\right]_{\beta}\,,$
}}
\end{alignedat}
\end{equation}
where the propagator carries momentum $p_t = p_1 - k_1$, and $\alpha$ and $\beta$ are spinor indices. The spinor contractions can be rewritten using
\begin{equation}
    \Big[\Gamma u(p)\Big]_\alpha C_{\alpha\beta} \Big[\Gamma' u(q)\Big]_\beta=u^\intercal(p) \Gamma^\intercal C \Gamma' u(q)\,,
\end{equation}
where $\Gamma$ and $\Gamma'$ are combinations of $\gamma$-matrices. Using this relation, together with the identities of the charge-conjugation matrix
\begin{equation}\label{eq: C properties}
    C^{\eminus1}=C^\dag=C^\intercal=-C\,,
    \qquad 
    C(\gamma^\mu)^\intercal C=\gamma^\mu\,\qquad C(\gamma^5)^\intercal C=-\gamma^5\,,
\end{equation}
the amplitude $\cM_b$ can be cast into the form given in Eq.~\eqref{eq:t_and_u_channels_dinuc}.

The same matrix element can be obtained directly within the QFT $S$-matrix formulation. The resulting expression trivially matches the one derived using Feynman rules, while rendering the structure of the fermionic propagator, including the role of charge conjugation, more transparent. Neglecting overall couplings and normalization factors, the relevant transition can be written as
\begin{equation}
{\scalebox{0.95}{
$\begin{alignedat}{2}
    \mel{h_1 h_2}{S}{f_1 f_2}\propto \int d^4x\, d^4y\, d^4z\,
    \wick{\big[\partial_\mu h_1(x)\bar f_1^c(x)\gamma^\mu\gamma^5 \c1 f_3^c(x)\big] \big[\c1{\bar{f}}_3^c(y) \c2 f_4(y)\big]\big[\c2{\bar{f}}_4(z) \gamma^\nu \gamma^5f_2(z) \partial_\nu h_1(z)\big]}\,, 
\end{alignedat}$
}}
\end{equation}
The Wick contraction involving $f_4$ can be expressed as
\begin{equation}
    \langle f_4(y) \bar f_4(z)\rangle =iS_F(y-z)=\int \dfrac{d^4p}{(2\pi)^4}e^{-ip\cdot (y-z)}\dfrac{i(\slashed p+m_{f_3})}{p^2-m_{f_3}^2}\,,
\end{equation}
while the contraction involving charge-conjugated fields takes the form
\begin{equation}
    \langle f_3^c(x) \bar f_3^c(y)\rangle= -C\Big[f_3(y) \bar f_3(x)\Big]^\intercal C= - C iS_F^\intercal(y-x) C
    =\int \dfrac{d^4q}{(2\pi)^4}e^{-iq\cdot (y-x)}\dfrac{i(-\slashed q+m_{f_4})}{q^2-m_{f_4}^2}\,.
\end{equation}
Expanding the fields in plane waves and performing the integration over the space-time coordinates, one obtains
\begin{equation}
\begin{split}
    \mel{h_1 h_2}{S}{f_1 f_2}\propto \int d^4p\,d^4q& \,\bar u^c(p_1) \slashed k_1 \gamma^5 (\slashed q-m_{f_3})(\slashed p+m_{f_4})\slashed k_2\gamma^5 u(p_2)\\
    &\times \delta^{4}(p+q)\delta^{4}(p_1-q-k_1)\delta^{4}(p_2-k_2-p)\,.
\end{split}
\end{equation}
Upon integrating over the internal momenta and simplifying the $\gamma^5$ structure, this expression reproduces the numerator of $\cM_b$ given in Eq.~\eqref{eq:t_and_u_channels_dinuc}.

\subsection{Analytical Structure of Kinematic Functions}
\label{app:F_and_G_func_matr_elem}
The evaluation of the squared amplitude requires summing over the polarizations of the external fermionic states. Starting from 
\begin{equation}
    \cM= u^\intercal(p_1) C \Gamma u(p_2)\,,
    \qquad
    \cM^*=-\bar u(p_2)\widetilde{\Gamma}C\bar u^\intercal(p_1)\,,\qquad \widetilde{\Gamma}=\gamma^0\Gamma^\dag \gamma^0\,,
\end{equation}
and noting that
\begin{equation}
    \sum_{\mathrm{spin}}C\Big[\bar u^\intercal(p_1)u^\intercal(p_1)\Big] C=C\big[\slashed p_1+m_{f_1}\big]^\intercal C =-\slashed p_1+m_{f_1}\,,
\end{equation}
the amplitude can be expressed as
\begin{equation}
    |\cM|^2=\frac{1}{4}\sum_{\mathrm{spin}}\cM \cM^*=\frac{1}{4}\Tr\Big[(\slashed p_2+m_{f_2})\widetilde{\Gamma}(\slashed p_1-m_{f_1})\Gamma\Big]\,.
\end{equation}
This form makes explicit the Dirac structure of the squared amplitude and provides a convenient starting point for its decomposition in terms of kinematic functions $\mathcal F_j$, $\mathcal G_j$, $\mathcal H_j$, $\mathcal J_j$ and $\mathcal K_j$ introduced in Eq.~\eqref{eq:amp_squared_fully_general_FG}. These functions encode the spin-summed Dirac structures and their dependence on the external momenta, and can be evaluated numerically for a given process and choice of kinematical configuration:
\begin{equation}
{\scalebox{0.945}{
$    
\begin{alignedat}{2}
        [\mathcal F_t]_{f_1f_2h_1h_2}^{f_3f_4f_3'f_4'}&=\frac{\Tr\lzs (\slashed p_1-m_{f_1})\slashed k_1(\slashed p_t+m_{f_3})(\slashed p_t+m_{f_4})\slashed k_2(\slashed p_2+m_{f_2})\slashed k_2(\slashed p_t+m_{f_4'})(\slashed p_t+m_{f_3'})\slashed k_1 \dzs}{(t-m_{f_3}^2)(t-m_{f_4}^2)(t-m_{f_3'}^2)(t-m_{f_4'}^2)}\,,
        \\[0.15cm]
        [\mathcal F_u]_{f_1f_2h_1h_2}^{f_3f_4f_3'f_4'}&=\frac{\Tr\lzs (\slashed p_1-m_{f_1})\slashed k_2 (\slashed p_u+m_{f_3})(\slashed p_u+m_{f_4})\slashed k_1(\slashed p_2+m_{f_2})\slashed k_1(\slashed p_u+m_{f_4'})(\slashed p_u+m_{f_3'})\slashed k_2 \dzs}{(u-m_{f_3}^2)(u-m_{f_4}^2)(u-m_{f_3'}^2)(u-m_{f_4'}^2)}\,,
        \\[0.15cm]
        [\mathcal F_{tu}]_{f_1f_2h_1h_2}^{f_3f_4f_3'f_4'}&=\frac{\Tr\lzs (\slashed p_1-m_{f_1})\slashed k_1(\slashed p_t+m_{f_3})(\slashed p_t+m_{f_4})\slashed k_2 (\slashed p_2+m_{f_2})\slashed k_1 (\slashed p_u+m_{f_4'})(\slashed p_u+m_{f_3'})\slashed k_2 \dzs}{(t-m_{f_3}^2)(t-m_{f_4}^2)(u-m_{f_3'}^2)(u-m_{f_4'}^2)}\,,
        \\[0.15cm]
        [\mathcal G_t]_{f_1f_2h_1h_2}^{f_3f_4}&=\frac{\Tr\lzs (\slashed p_1-m_{f_1})(\slashed p_2+m_{f_2})\slashed k_2 (\slashed p_t+m_{f_4})(\slashed p_t+m_{f_3})\slashed k_1 \dzs}{(t-m_{f_3}^2)(t-m_{f_4}^2)}\,,
        \\[0.15cm]
        [\mathcal G_u]_{f_1f_2h_1h_2}^{f_3f_4}&=\frac{\Tr\lzs (\slashed p_1-m_{f_1})(\slashed p_2+m_{f_2})\slashed k_1 (\slashed p_u+m_{f_4})(\slashed p_u+m_{f_3})\slashed k_2 \dzs}{(u-m_{f_3}^2)(u-m_{f_4}^2)}\,,
        \\[0.15cm]
        [\mathcal H_t]^{ff'}_{f_1f_2h_1h_2}&=\frac{\Tr\lzs (\slashed p_1-m_{f_1})(\slashed p_t-m_f)\slashed k_2(\slashed p_2-m_{f_2})\slashed k_2(\slashed p_t-m_{f'}) \dzs}{(t-m_f^2)(t-m_{f'}^2)}\,,
        \\[0.15cm]
        [\mathcal H_u]^{ff'}_{f_1f_2h_1h_2}&=\frac{\Tr\lzs (\slashed p_1-m_{f_1})(\slashed p_u-m_f)\slashed k_1(\slashed p_2-m_{f_2})\slashed k_1(\slashed p_u-m_{f'}) \dzs}{(u-m_f^2)(u-m_{f'}^2)}\,,
        \\[0.15cm]
        [\mathcal H_{tu}]^{ff'}_{f_1f_2h_1h_2}&=\frac{\Tr\lzs (\slashed p_1-m_{f_1})(\slashed p_t-m_f)\slashed k_2(\slashed p_2-m_{f_2})\slashed k_1(\slashed p_u-m_{f'}) \dzs}{(t-m_f^2)(u-m_{f'}^2)}\,,
\end{alignedat}
$
}}
\end{equation}
The explicit computation of these functions has been performed using \texttt{FeynCalc}~\cite{Mertig:1990an,Shtabovenko:2016sxi,Shtabovenko:2020gxv,Shtabovenko:2023idz}.

\clearpage
\begin{table}[t]
\centering
\scalebox{0.85}{
\begin{tabular}{cc@{\hspace{0.6cm}}c@{\hspace{0.6cm}}c@{\hspace{0.6cm}}cc}
\toprule
\multirow{1}{*}{\textbf{Chiral irrep}}
&\multirow{1}{*}{\textbf{Component}}
&\multirow{1}{*}{\textbf{Coefficient}} 
&\textbf{Operator}
\\
\midrule
\addlinespace[0.2cm]
\multirow{19}{*}{$\begin{array}{cc}\repbar6_R\otimes \rep6_L\\[2pt]\bm{(a,b,\dot a,\dot b)}\end{array}$}
&$(3,3,\dot 2,\dot 2)$&$\cC_{\repbar6_R\otimes \rep6_L}^{33\dot2\dot2}$&$\cO_{RRL}^{121222}$
\\[0.2cm]
\noalign{\vskip 0.0cm}
\cdashline{2-5}[.4pt/2pt]
\noalign{\vskip 0.2cm}
&$(1,3,\dot 1,\dot 2)$&$\cC_{\repbar6_R\otimes \rep6_L}^{13\dot1\dot2}$&$\cO_{RRL}^{122312}$
\\[0.2cm]
&$(2,3,\dot 2,\dot 2)$&$\cC_{\repbar6_R\otimes \rep6_L}^{23\dot2\dot2}$&$-\cO_{RRL}^{121322}$
\\[0.2cm]
&$(3,3,\dot 2,\dot 3)$&$\cC_{\repbar6_R\otimes \rep6_L}^{33\dot2\dot3}$&$\cO_{RRL}^{121223}$
\\[0.2cm]
\noalign{\vskip 0.0cm}
\cdashline{2-5}[.4pt/2pt]
\noalign{\vskip 0.2cm}
&$(1,1,\dot 1,\dot 1)$&$\cC_{\repbar6_R\otimes \rep6_L}^{11\dot 1\dot 1}$&$\cO_{RRL}^{232311}$ 
\\[0.2cm]
&$(1,2,\dot 1,\dot 2)$ & $\cC_{\repbar6_R\otimes \rep6_L}^{12\dot 1\dot 2}$&$-\cO_{RRL}^{132312}$
\\[0.2cm]
&$(1,3,\dot 1,\dot 3)$ & $\cC_{\repbar6_R\otimes \rep6_L}^{13\dot 1\dot 3}$&$\cO_{RRL}^{122313}$
\\[0.2cm]
&$(2,2,\dot 2,\dot 2)$ & $\cC_{\repbar6_R\otimes \rep6_L}^{22\dot 2\dot 2}$&$\cO_{RRL}^{131322}$
\\[0.2cm]
&$(2,3,\dot 2,\dot 3)$ & $\cC_{\repbar6_R\otimes \rep6_L}^{23\dot 2\dot 3}$&$-\cO_{RRL}^{121323}$
\\[0.2cm]
&$(3,3,\dot 3,\dot 3)$ & $\cC_{\repbar6_R\otimes \rep6_L}^{33\dot 3\dot 3}$&$\cO_{RRL}^{121233}$
\\[0.2cm]
\noalign{\vskip 0.0cm}
\cdashline{2-5}[.4pt/2pt]
\noalign{\vskip 0.2cm}
&$(1,2,\dot 1,\dot 3)$&$\cC_{\repbar6_R\otimes \rep6_L}^{12\dot 1\dot 3}$&$-\cO_{RRL}^{132313}$
\\[0.2cm]
&$(2,2,\dot 2,\dot 3)$&$\cC_{\repbar6_R\otimes \rep6_L}^{22\dot 2\dot 3}$&$\cO_{RRL}^{131323}$
\\[0.2cm]
&$(2,3,\dot 3,\dot 3)$ & $\cC_{\repbar6_R\otimes \rep6_L}^{23\dot 3\dot 3}$&$-\cO_{RRL}^{121333}$
\\[0.2cm]
\noalign{\vskip 0.0cm}
\cdashline{2-5}[.4pt/2pt]
\noalign{\vskip 0.2cm}
&$(2,2,\dot 3,\dot 3)$&$\cC_{\repbar6_R\otimes \rep6_L}^{22\dot 3\dot 3}$&$\cO_{RRL}^{131333}$
\\[0.2cm]
\midrule
\addlinespace[0.2cm]
\multirow{22}{*}{$\begin{array}{cc}\rep{15}_{\bm R}\bm\otimes\repbar3_{\bm L}\\[2pt]\bm{(a,b,c,\dot a)}\end{array}$}
&$(2,2,3,\dot 3)$
&$\cC_{\rep{15}_R\otimes\repbar3_L}^{223\dot 3}$&$\cO_{LRR}^{121222}$
\\[0.2cm]
\noalign{\vskip 0.0cm}
\cdashline{2-5}[.4pt/2pt]
\noalign{\vskip 0.2cm}
&$(1,2,3,\dot 1)$&$\cC_{\rep{15}_R\otimes\repbar3_L}^{123\dot 1}$&$\cO_{LRR}^{231212}$
\\[0.2cm]
&$(2,2,3,\dot 2)$&$\cC_{\rep{15}_R\otimes\repbar3_L}^{223\dot 2}$&$-\cO_{LRR}^{131222}$
\\[0.2cm]
&$(1,2,1,\dot 3)$&$\cC_{\rep{15}_R\otimes\repbar3_L}^{121\dot 3}$&
$-\frac{1}{4} \cO_{LRR}^{121223} + \frac{1}{4} \cO_{LRR}^{121322} + \frac{3}{4} \cO_{LRR}^{122312}$
\\[0.2cm]
&$(2,2,2,\dot 3)$&$\cC_{\rep{15}_R\otimes\repbar3_L}^{222\dot 3}$&
$-\frac{1}{2} \cO_{LRR}^{121223} - \frac{1}{2} \cO_{LRR}^{121322} - \frac{1}{2} \cO_{LRR}^{122312}$
\\[0.2cm]
\noalign{\vskip 0.0cm}
\cdashline{2-5}[.4pt/2pt]
\noalign{\vskip 0.2cm}
&$(1,1,1,\dot 1)$&$\cC_{\rep{15}_R\otimes\repbar3_L}^{111\dot 1}$&$-\frac{1}{2} \cO_{LRR}^{231213} + \frac{1}{2} \cO_{LRR}^{231312} + \frac{1}{2} \cO_{LRR}^{232311}$
\\[0.2cm]
&$(1,2,2,\dot 1)$&$\cC_{\rep{15}_R\otimes\repbar3_L}^{122\dot 1}$&$-\frac{1}{4} \cO_{LRR}^{231213} - \frac{3}{4} \cO_{LRR}^{231312} - \frac{1}{4} \cO_{LRR}^{232311}$
\\[0.2cm]
&$(1,2,1,\dot 2)$&$\cC_{\rep{15}_R\otimes\repbar3_L}^{121\dot 2}$&$\frac{1}{4} \cO_{LRR}^{131223} - \frac{1}{4} \cO_{LRR}^{131322} - \frac{3}{4} \cO_{LRR}^{132312}$
\\[0.2cm]
&$(2,2,2,\dot 2)$&$\cC_{\rep{15}_R\otimes\repbar3_L}^{222\dot 2}$&$\frac{1}{2} \cO_{LRR}^{131223} + \frac{1}{2} \cO_{LRR}^{131322} + \frac{1}{2} \cO_{LRR}^{132312}$
\\[0.2cm]
&$(1,3,1,\dot 3)$&$\cC_{\rep{15}_R\otimes\repbar3_L}^{131\dot 3}$&$-\frac{1}{4} \cO_{LRR}^{121233} + \frac{1}{4} \cO_{LRR}^{121323} + \frac{3}{4} \cO_{LRR}^{122313}$
\\[0.2cm]
&$(2,3,2,\dot 3)$&$\cC_{\rep{15}_R\otimes\repbar3_L}^{232\dot 3}$&$-\frac{1}{4} \cO_{LRR}^{121233} - \frac{3}{4} \cO_{LRR}^{121323} - \frac{1}{4} \cO_{LRR}^{122313}$
\\[0.2cm]
\noalign{\vskip 0.0cm}
\cdashline{2-5}[.4pt/2pt]
\noalign{\vskip 0.2cm}
&$(1,3,2,\dot 1)$&$\cC_{\rep{15}_R\otimes\repbar3_L}^{132\dot 1}$&$-\cO_{LRR}^{231313}$
\\[0.2cm]
&$(1,3,1,\dot 2)$&$\cC_{\rep{15}_R\otimes\repbar3_L}^{131\dot 2}$&$\frac{1}{4} \cO_{LRR}^{131233} - \frac{1}{4} \cO_{LRR}^{131323} - \frac{3}{4} \cO_{LRR}^{132313}$
\\[0.2cm]
&$(2,3,2,\dot 2)$&$\cC_{\rep{15}_R\otimes\repbar3_L}^{232\dot 2}$&$\frac{1}{4} \cO_{LRR}^{131233} + \frac{3}{4} \cO_{LRR}^{131323} + \frac{1}{4} \cO_{LRR}^{132313}$
\\[0.2cm]
&$(3,3,2,\dot 3)$&$\cC_{\rep{15}_R\otimes\repbar3_L}^{332\dot 3}$&$-\cO_{LRR}^{121333}$
\\[0.2cm]
\noalign{\vskip 0.0cm}
\cdashline{2-5}[.4pt/2pt]
\noalign{\vskip 0.2cm}
&$(3,3,2,\dot 2)$&$\cC_{\rep{15}_R\otimes\repbar3_L}^{332\dot 2}$&$\cO_{LRR}^{131333}$
\\[0.2cm]
\bottomrule
\end{tabular}
}
\caption{Part I of the complete chiral operator basis. The dashed line separates operators according to the number of strange quarks.}
\label{tab:chiral_basis_P1}
\end{table}
\begin{table}[t]
\centering
\scalebox{0.88}{
\begin{tabular}{cc@{\hspace{0.6cm}}c@{\hspace{0.6cm}}c@{\hspace{0.6cm}}cc}
\toprule
\multirow{1}{*}{\textbf{Chiral irrep}}
&\multirow{1}{*}{\textbf{Component}}
&\multirow{1}{*}{\textbf{Coefficient}} 
&\textbf{Operator}
\\
\midrule
\addlinespace[0.2cm]
\multirow{7}{*}{\vspace{-0.2cm}$\begin{array}{c}\rep3_{\bm R}\bm\otimes\repbar3_{\bm L}\\[2pt]\bm{(a,\dot b)}\end{array}$}
&$(2,\dot 3)$
&$\cC_{\rep3_R\otimes\repbar3_L}^{2\dot3}$&$\frac{1}{2}\cO_{LRR}^{121223}-\frac{1}{2}\cO_{LRR}^{121322}+\frac{1}{2}\cO_{LRR}^{122312}$
\\[0.2cm]
\noalign{\vskip 0.0cm}
\cdashline{2-5}[.4pt/2pt]
\noalign{\vskip 0.2cm}
&$(1,\dot 1)$&$\cC_{\rep3_R\otimes\repbar3_L}^{1\dot 1}$&$\frac{1}{2}\cO_{LRR}^{231213}-\frac{1}{2}\cO_{LRR}^{231312}+\frac{1}{2}\cO_{LRR}^{232311}$
\\[0.2cm]
&$(2,\dot 2)$&$\cC_{\rep3_R\otimes\repbar3_L}^{2\dot 2}$&$-\frac{1}{2}\cO_{LRR}^{131223}+\frac{1}{2}\cO_{LRR}^{131322}-\frac{1}{2}\cO_{LRR}^{132312}$
\\[0.2cm]
&$(3,\dot 3)$&$\cC_{\rep3_R\otimes\repbar3_L}^{3\dot 3}$&$\frac{1}{2}\cO_{LRR}^{121233}-\frac{1}{2}\cO_{LRR}^{121323}+\frac{1}{2}\cO_{LRR}^{122313}$
\\[0.2cm]
\noalign{\vskip 0.0cm}
\cdashline{2-5}[.4pt/2pt]
\noalign{\vskip 0.2cm}
&$(3,\dot 2)$&$\cC_{\rep3_R\otimes\repbar3_L}^{3\dot 2}$&$-\frac{1}{2}\cO_{LRR}^{131233}+\frac{1}{2}\cO_{LRR}^{131323}-\frac{1}{2}\cO_{LRR}^{132313}$
\\[0.2cm]
\midrule
\addlinespace[0.2cm]
\multirow{30}{*}{\vspace{-0.2cm}$\begin{array}{c}\rep{15}_{\bm R}'\bm\otimes \rep6_{\bm L}\\[2pt]\bm{(a,b,c,d,\dot a,\dot b)}\end{array}$}
&$(1,1,2,2,\dot2,\dot2)$&$\cC_{\rep{15}_R'\otimes \rep6_L}^{1122\dot 2\dot 2}$&$\frac{1}{3}\widetilde\cO_{RRL}^{112222}+\frac{2}{3}\widetilde\cO_{RRL}^{121222}$
\\[0.2cm]
&$(1,2,2,2,\dot1,\dot2)$&$\cC_{\rep{15}_R'\otimes \rep6_L}^{1222\dot 1\dot 2}$&$\widetilde\cO_{RRL}^{122212}$
\\[0.2cm]
&$(2,2,2,2,\dot1,\dot1)$&$\cC_{\rep{15}_R'\otimes \rep6_L}^{1122\dot 2\dot 2}$&$\widetilde\cO_{RRL}^{222211}$
\\[0.2cm]
\noalign{\vskip 0.0cm}
\cdashline{2-5}[.4pt/2pt]
\noalign{\vskip 0.2cm}
&$(1,1,2,2,\dot2,\dot3)$&$\cC_{\rep{15}_R'\otimes \rep6_L}^{1122\dot 2\dot 3}$&$\frac{1}{3} \widetilde\cO_{RRL}^{112223} + \frac{2}{3} \widetilde\cO_{RRL}^{121223}$
\\[0.2cm]
&$(1,1,2,3,\dot2,\dot2)$&$\cC_{\rep{15}_R'\otimes \rep6_L}^{1123\dot 2\dot 2}$&$\frac{1}{3} \widetilde\cO_{RRL}^{112322} + \frac{2}{3} \widetilde\cO_{RRL}^{121322}$
\\[0.2cm]
&$(1,2,2,2,\dot1,\dot3)$&$\cC_{\rep{15}_R'\otimes \rep6_L}^{1222\dot 1\dot 3}$&$\widetilde\cO_{RRL}^{122213}$
\\[0.2cm]
&$(1,2,2,3,\dot1,\dot2)$&$\cC_{\rep{15}_R'\otimes \rep6_L}^{1223\dot 1\dot 2}$&$\frac{2}{3} \widetilde\cO_{RRL}^{122312} + \frac{1}{3} \widetilde\cO_{RRL}^{132212}$
\\[0.2cm]
&$(2,2,2,3,\dot1,\dot1)$&$\cC_{\rep{15}_R'\otimes \rep6_L}^{2223\dot 1\dot 1}$&$\widetilde\cO_{RRL}^{222311}$
\\[0.2cm]
\noalign{\vskip 0.0cm}
\cdashline{2-5}[.4pt/2pt]
\noalign{\vskip 0.2cm}
&$(1,1,2,2,\dot3,\dot3)$ &$\cC_{\rep{15}_R'\otimes \rep6_L}^{1122\dot 3\dot 3}$ &$\frac{1}{3} \widetilde\cO_{RRL}^{112233} + \frac{2}{3} \widetilde\cO_{RRL}^{121233}$
\\[0.2cm]
&$(1,1,2,3,\dot2,\dot3)$ &$\cC_{\rep{15}_R'\otimes \rep6_L}^{1123\dot 2\dot 3}$ &$\frac{1}{3} \widetilde\cO_{RRL}^{112323} + \frac{2}{3} \widetilde\cO_{RRL}^{121323}$
\\[0.2cm]
&$(1,1,3,3,\dot2,\dot2)$ & $\cC_{\rep{15}_R'\otimes \rep6_L}^{1133\dot 2\dot 2}$&$\frac{1}{3} \widetilde\cO_{RRL}^{113322} + \frac{2}{3} \widetilde\cO_{RRL}^{131322}$
\\[0.2cm]
&$(1,2,2,3,\dot1,\dot3)$ &$\cC_{\rep{15}_R'\otimes \rep6_L}^{1223\dot 1\dot 3}$ &$\frac{2}{3} \widetilde\cO_{RRL}^{122313} + \frac{1}{3} \widetilde\cO_{RRL}^{132213}$
\\[0.2cm]
&$(1,2,3,3,\dot1,\dot2)$ &$\cC_{\rep{15}_R'\otimes \rep6_L}^{1233\dot 1\dot 2}$ &$\frac{1}{3} \widetilde\cO_{RRL}^{123312} + \frac{2}{3} \widetilde\cO_{RRL}^{132312}$
\\[0.2cm]
&$(2,2,3,3,\dot1,\dot1)$ &$\cC_{\rep{15}_R'\otimes \rep6_L}^{2233\dot 1\dot 1}$ &$\frac{1}{3} \widetilde\cO_{RRL}^{223311} + \frac{2}{3} \widetilde\cO_{RRL}^{232311}$
\\[0.2cm]
\noalign{\vskip 0.0cm}
\cdashline{2-5}[.4pt/2pt]
\noalign{\vskip 0.2cm}
&$(1,1,2,3,\dot3,\dot3)$&$\cC_{\rep{15}_R'\otimes \rep6_L}^{1123\dot 3\dot 3}$ & $\frac{1}{3} \widetilde\cO_{RRL}^{112333} + \frac{2}{3} \widetilde\cO_{RRL}^{121333}$
\\[0.2cm]
&$(1,1,3,3,\dot2,\dot3)$&$\cC_{\rep{15}_R'\otimes \rep6_L}^{1133\dot 2\dot 3}$ & $\frac{1}{3} \widetilde\cO_{RRL}^{113323} + \frac{2}{3} \widetilde\cO_{RRL}^{131323}$
\\[0.2cm]
&$(1,2,3,3,\dot1,\dot3)$ &$\cC_{\rep{15}_R'\otimes \rep6_L}^{1233\dot 1\dot 3}$& $\frac{1}{3} \widetilde\cO_{RRL}^{123313} + \frac{2}{3} \widetilde\cO_{RRL}^{132313}$
\\[0.2cm]
&$(1,3,3,3,\dot1,\dot2)$&$\cC_{\rep{15}_R'\otimes \rep6_L}^{1333\dot 1\dot 2}$ & $\widetilde\cO_{RRL}^{133312}$
\\[0.2cm]
&$(2,3,3,3,\dot1,\dot1)$ &$\cC_{\rep{15}_R'\otimes \rep6_L}^{2333\dot 1\dot 1}$& $\widetilde\cO_{RRL}^{233311}$
\\[0.2cm]
\noalign{\vskip 0.0cm}
\cdashline{2-5}[.4pt/2pt]
\noalign{\vskip 0.2cm}
&$(1,1,3,3,\dot3,\dot3)$&$\cC_{\rep{15}_R'\otimes \rep6_L}^{1133\dot 3\dot 3}$ & $\frac{1}{3} \widetilde\cO_{RRL}^{113333} + \frac{2}{3} \widetilde\cO_{RRL}^{131333}$
\\[0.2cm]
&$(1,3,3,3,\dot1,\dot3)$&$\cC_{\rep{15}_R'\otimes \rep6_L}^{1333\dot 1\dot 3}$ & $\widetilde\cO_{RRL}^{133313}$
\\[0.2cm]
&$(3,3,3,3,\dot1,\dot1)$&$\cC_{\rep{15}_R'\otimes \rep6_L}^{3333\dot 1\dot 1}$ & $\widetilde\cO_{RRL}^{333311}$
\\[0.2cm]
\midrule
\addlinespace[0.2cm]
\multirow{2}{*}{\vspace{-0.2cm}$\rep1_{\bm R}\bm\otimes\rep1_{\bm L}$}&---
&$\cC_{\rep1_{R}\otimes\rep1_{L}}$
&$\frac{1}{6}\cO_{RRR}^{121233}+\frac{1}{6}\cO_{RRR}^{131322}+\frac{1}{6}\cO_{RRR}^{232311}$
\\[0.1cm]
&&&$-\frac{1}{3}\cO_{RRR}^{121323}+\frac{1}{3}\cO_{RRR}^{122313}-\frac{1}{3}\cO_{RRR}^{132312}$
\\[0.2cm]
\bottomrule
\end{tabular}
}
\caption{Part II of the complete chiral operator basis. The dashed line separates operators according to the number of strange quarks.}
\label{tab:chiral_basis_P2}
\end{table}
\begin{table}[t]
\centering
\scalebox{0.9}{
\begin{tabular}{cc@{\hspace{0.6cm}}c@{\hspace{0.6cm}}c@{\hspace{0.6cm}}cc}
\toprule
\multirow{1}{*}{\textbf{Chiral irrep}}
&\multirow{1}{*}{\textbf{Component}}
&\multirow{1}{*}{\textbf{Coefficient}} 
&\textbf{Operator}
\\
\midrule
\addlinespace[0.2cm]
\multirow{17}{*}{$\begin{array}{c}\rep{27}_{\bm R}\bm\otimes\rep1_{\bm L}\\[2pt]\bm{(a,b,c,d)}\end{array}$}
&$(2,2,3,3)$&$\cC_{\rep{27}_R\otimes\rep1_L}^{2233}$&$\cO_{RRR}^{121222}$
\\[0.2cm]
\noalign{\vskip 0.0cm}
\cdashline{2-5}[.4pt/2pt]
\noalign{\vskip 0.2cm}
&$(1,2,1,3)$&$\cC_{\rep{27}_R\otimes\rep1_L}^{1213}$&$-\frac{1}{5} \cO_{RRR}^{121223} + \frac{1}{5} \cO_{RRR}^{121322} + \frac{4}{5} \cO_{RRR}^{122312}$
\\[0.2cm]
&$(2,2,2,3)$&$\cC_{\rep{27}_R\otimes\rep1_L}^{2223}$&$-\frac{2}{5} \cO_{RRR}^{121223} - \frac{3}{5} \cO_{RRR}^{121322} - \frac{2}{5} \cO_{RRR}^{122312}$
\\[0.2cm]
\noalign{\vskip 0.0cm}
\cdashline{2-5}[.4pt/2pt]
\noalign{\vskip 0.2cm}
&$(1,1,1,1)$&$\cC_{\rep{27}_R\otimes\rep1_L}^{1111}$
&$\frac{1}{10} \cO_{RRR}^{121233} - \frac{1}{5} \cO_{RRR}^{121323} - \frac{3}{5} \cO_{RRR}^{122313} $
\\[0.2cm]
&&&$+ \frac{1}{10} \cO_{RRR}^{131322} + \frac{3}{5} \cO_{RRR}^{132312} + \frac{3}{10} \cO_{RRR}^{232311}$
\\[0.2cm]
\noalign{\vskip 0.0cm}
\cdashline{2-5}[.4pt/2pt]
\noalign{\vskip 0.2cm}
&$(2,2,2,2)$&$\cC_{\rep{27}_R\otimes\rep1_L}^{2222}$
&$\frac{1}{10} \cO_{RRR}^{121233} + \frac{3}{5} \cO_{RRR}^{121323} + \frac{1}{5} \cO_{RRR}^{122313}$
\\[0.2cm]
&&&$ + \frac{3}{10} \cO_{RRR}^{131322} + \frac{3}{5} \cO_{RRR}^{132312} + \frac{1}{10} \cO_{RRR}^{232311}$
\\[0.2cm]
\noalign{\vskip 0.0cm}
\cdashline{2-5}[.4pt/2pt]
\noalign{\vskip 0.2cm}
&$(3,3,3,3)$&$\cC_{\rep{27}_R\otimes\rep1_L}^{3333}$
&$\frac{3}{10} \cO_{RRR}^{121233} + \frac{3}{5} \cO_{RRR}^{121323} - \frac{3}{5} \cO_{RRR}^{122313}$
\\[0.2cm]
&&&$ + \frac{1}{10} \cO_{RRR}^{131322} - \frac{1}{5} \cO_{RRR}^{132312} + \frac{1}{10} \cO_{RRR}^{232311}$
\\[0.2cm]
\noalign{\vskip 0.0cm}
\cdashline{2-5}[.4pt/2pt]
\noalign{\vskip 0.2cm}
&$(1,3,1,2)$&$\cC_{\rep{27}_R\otimes\rep1_L}^{1312}$
&$\frac{1}{5} \cO_{RRR}^{121333} - \frac{1}{5} \cO_{RRR}^{131323} - \frac{4}{5} \cO_{RRR}^{132313}$
\\[0.2cm]
&$(2,3,2,2)$&$\cC_{\rep{27}_R\otimes\rep1_L}^{2322}$
&$\frac{2}{5} \cO_{RRR}^{121333} + \frac{3}{5} \cO_{RRR}^{131323} + \frac{2}{5} \cO_{RRR}^{132313}$
\\[0.2cm]
\noalign{\vskip 0.0cm}
\cdashline{2-5}[.4pt/2pt]
\noalign{\vskip 0.2cm}
&$(3,3,2,2)$&$\cC_{\rep{27}_R\otimes\rep1_L}^{3322}$&$\cO_{RRR}^{131333}$
\\[0.2cm]
\midrule
\addlinespace[0.2cm]
\multirow{9}{*}{$\begin{array}{c}\rep{28}_{\bm R}\bm\otimes\rep1_{\bm L}\\[2pt]\bm{(a,b,c,d,e,f)}\end{array}$}
&$(1,1,2,2,2,2)$&$\cC_{\rep{28}_R\otimes\rep1_L}^{112222}$
&$\frac{1}{5}\widetilde\cO_{RRR}^{112222}+\frac{4}{5}\widetilde\cO_{RRR}^{121222}$
\\[0.2cm]
\noalign{\vskip 0.0cm}
\cdashline{2-5}[.4pt/2pt]
\noalign{\vskip 0.2cm}
&$(1,1,2,2,2,3)$&$\cC_{\rep{28}_R\otimes\rep1_L}^{112223}$
&$\frac{1}{5} \widetilde\cO_{RRR}^{112223} + \frac{2}{5} \widetilde\cO_{RRR}^{121223} + \frac{2}{5} \widetilde\cO_{RRR}^{121322}$
\\[0.2cm]
\noalign{\vskip 0.0cm}
\cdashline{2-5}[.4pt/2pt]
\noalign{\vskip 0.2cm}
&$(1,1,2,2,3,3)$&$\cC_{\rep{28}_R\otimes\rep1_L}^{112233}$
&$\frac{1}{15} \widetilde\cO_{RRR}^{112233} + \frac{2}{15} \widetilde\cO_{RRR}^{112323} + \frac{2}{15} \widetilde\cO_{RRR}^{121233}$
\\[0.2cm]
&&&$+ \frac{8}{15} \widetilde\cO_{RRR}^{121323} + \frac{2}{15} \widetilde\cO_{RRR}^{131322}$
\\[0.2cm]
\noalign{\vskip 0.0cm}
\cdashline{2-5}[.4pt/2pt]
\noalign{\vskip 0.2cm}
&$(1,1,2,3,3,3)$&$\cC_{\rep{28}_R\otimes\rep1_L}^{112333}$
&$\frac{1}{5} \widetilde\cO_{RRR}^{112333} + \frac{2}{5} \widetilde\cO_{RRR}^{121333} + \frac{2}{5} \widetilde\cO_{RRR}^{131323}$
\\[0.2cm]
\noalign{\vskip 0.0cm}
\cdashline{2-5}[.4pt/2pt]
\noalign{\vskip 0.2cm}
&$(1,1,3,3,3,3)$&$\cC_{\rep{28}_R\otimes\rep1_L}^{113333}$
&$\frac{1}{5} \widetilde\cO_{RRR}^{113333} + \frac{4}{5} \widetilde\cO_{RRR}^{131333}$
\\[0.2cm]
\bottomrule
\end{tabular}
}

\caption{Part III of the complete chiral operator basis. The dashed line separates operators according to the number of strange quarks.}
\label{tab:chiral_basis_P3}
\end{table}

\begin{table}[t]
    \centering
\scalebox{0.85}{
\begin{tabular}{c@{\hspace{1.2cm}}c@{\hspace{0.8cm}}c}
\toprule
\multirow{1}{*}{\textbf{Chiral}}
&\multirow{2}{*}{\textbf{SMEFT mapping}}
\\
\multirow{1}{*}{\textbf{coefficient}}
\\
\midrule
\addlinespace[0.2cm]
$\cC_{\rep{27}_R\otimes\rep1_L}^{2233}$
&$\frac{1}{4}[\cC_{d^4u^2}^{(1)}-\cC_{d^4u^2}^{(2)}]_{111111}$
\\[0.2cm]
$\cC_{\rep{28}_R\otimes\rep1_L}^{112222}$
&$\frac{5}{16}[\cC_{d^4u^2}^{(1)}+\cC_{d^4u^2}^{(2)}]_{111111}$
\\[0.2cm]
\midrule
\addlinespace[0.2cm]
$\cC_{\rep{27}_R\otimes\rep1_L}^{1213}$
&$\frac{3}{4}[\cC_{d^4u^2}^{(2)}-\cC_{d^4u^2}^{(1)}]_{111112}+\frac{3}{4}[\cC_{d^4u^2}^{(2)}-\cC_{d^4u^2}^{(1)}]_{111121}+\frac{1}{2}[\cC_{d^4u^2}^{(1)}-\cC_{d^4u^2}^{(2)}]_{111211}$
\\[0.2cm]
&$+\frac{1}{2}[\cC_{d^4u^2}^{(1)}-\cC_{d^4u^2}^{(2)}]_{121111}$
&\\[0.2cm]
\noalign{\vskip 0.0cm}
\cdashline{1-3}[.4pt/2pt]
\noalign{\vskip 0.2cm}
$\cC_{\rep{27}_R\otimes\rep1_L}^{2223}$
&$\frac{1}{4}[\cC_{d^4u^2}^{(2)}-\cC_{d^4u^2}^{(1)}]_{111112}+\frac{1}{4}[\cC_{d^4u^2}^{(2)}-\cC_{d^4u^2}^{(1)}]_{111121}+\frac{1}{4}[\cC_{d^4u^2}^{(2)}-\cC_{d^4u^2}^{(1)}]_{111211}$
\\[0.2cm]
&$+\frac{1}{4}[\cC_{d^4u^2}^{(2)}-\cC_{d^4u^2}^{(1)}]_{121111}$
&\\[0.2cm]
\noalign{\vskip 0.0cm}
\cdashline{1-3}[.4pt/2pt]
\noalign{\vskip 0.2cm}
$\cC_{\rep{28}_R\otimes\rep1_L}^{112223}$
&$\frac{5}{8}[\cC_{d^4u^2}^{(1)}+\cC_{d^4u^2}^{(2)}]_{111112}+\frac{5}{8}[\cC_{d^4u^2}^{(1)}+\cC_{d^4u^2}^{(2)}]_{111121}$
\\[0.2cm]
\midrule
\addlinespace[0.2cm]
$\cC_{\rep{27}_R\otimes\rep1_L}^{1111}$
&$\phantom{-}\frac{3}{4}[\cC_{d^4u^2}^{(1)}-\cC_{d^4u^2}^{(2)}]_{111122}+\frac{1}{8}[\cC_{d^4u^2}^{(1)}-\cC_{d^4u^2}^{(2)}]_{111212}-\frac{1}{2}[\cC_{d^4u^2}^{(1)}-\cC_{d^4u^2}^{(2)}]_{111221}$
\\[0.2cm]
&$-\frac{1}{2}[\cC_{d^4u^2}^{(1)}-\cC_{d^4u^2}^{(2)}]_{121211}+\frac{1}{8}[\cC_{d^4u^2}^{(1)}]_{121112}+\frac{1}{2}[\cC_{d^4u^2}^{(2)}]_{121112}+\frac{1}{8}[\cC_{d^4u^2}^{(1)}]_{121121}$
&\\[0.2cm]
&$+\frac{1}{2}[\cC_{d^4u^2}^{(2)}]_{121121}$
\\[0.2cm]
\noalign{\vskip 0.0cm}
\cdashline{1-3}[.4pt/2pt]
\noalign{\vskip 0.2cm}
$\cC_{\rep{27}_R\otimes\rep1_L}^{2222}$
&$-\frac{1}{2}[\cC_{d^4u^2}^{(1)}-\cC_{d^4u^2}^{(2)}]_{111122}+\frac{1}{8}[\cC_{d^4u^2}^{(1)}-\cC_{d^4u^2}^{(2)}]_{111212}+\frac{1}{8}[\cC_{d^4u^2}^{(1)}-\cC_{d^4u^2}^{(2)}]_{111221}$
\\[0.2cm]
&$+\frac{1}{8}[\cC_{d^4u^2}^{(1)}-\cC_{d^4u^2}^{(2)}]_{121112}+\frac{1}{8}[\cC_{d^4u^2}^{(1)}-\cC_{d^4u^2}^{(2)}]_{121121}+\frac{3}{4}[\cC_{d^4u^2}^{(1)}-\cC_{d^4u^2}^{(2)}]_{121211}$
&\\[0.2cm]
\noalign{\vskip 0.0cm}
\cdashline{1-3}[.4pt/2pt]
\noalign{\vskip 0.2cm}
$\cC_{\rep{27}_R\otimes\rep1_L}^{3333}$
&$\phantom{+}\frac{3}{4}[\cC_{d^4u^2}^{(1)}-\cC_{d^4u^2}^{(2)}]_{111122}+\frac{1}{8}[\cC_{d^4u^2}^{(1)}-\cC_{d^4u^2}^{(2)}]_{111212}+\frac{1}{8}[\cC_{d^4u^2}^{(1)}-\cC_{d^4u^2}^{(2)}]_{111221}$
\\[0.2cm]
&$+\frac{1}{8}[\cC_{d^4u^2}^{(1)}-\cC_{d^4u^2}^{(2)}]_{121112}+\frac{1}{8}[\cC_{d^4u^2}^{(1)}-\cC_{d^4u^2}^{(2)}]_{121121}-\frac{1}{2}[\cC_{d^4u^2}^{(1)}-\cC_{d^4u^2}^{(2)}]_{121211}$
&\\[0.2cm]
\noalign{\vskip 0.0cm}
\cdashline{1-3}[.4pt/2pt]
\noalign{\vskip 0.2cm}
$\cC_{\rep{1}_R\otimes\rep1_L}$
&$\frac{3}{4}[\cC_{d^4u^2}^{(1)}-\cC_{d^4u^2}^{(2)}]_{121211}-\frac{3}{8}[\cC_{d^4u^2}^{(1)}-\cC_{d^4u^2}^{(2)}]_{111212}-\frac{3}{8}[\cC_{d^4u^2}^{(1)}]_{121112}+\frac{3}{8}[\cC_{d^4u^2}^{(2)}]_{121121}$
\\[0.2cm]
\noalign{\vskip 0.0cm}
\cdashline{1-3}[.4pt/2pt]
\noalign{\vskip 0.2cm}
$\cC_{\rep{28}_R\otimes\rep1_L}^{112233}$
&$\frac{15}{8}[\cC_{d^4u^2}^{(1)}-\cC_{d^4u^2}^{(2)}]_{111122}$
\\[0.2cm]
\midrule
\addlinespace[0.2cm]
$\cC_{\rep{27}_R\otimes\rep1_L}^{2322}$
&$\frac{1}{4}[\cC_{d^4u^2}^{(1)}-\cC_{d^4u^2}^{(2)}]_{111222}+\frac{1}{4}[\cC_{d^4u^2}^{(1)}-\cC_{d^4u^2}^{(2)}]_{121122}+\frac{1}{4}[\cC_{d^4u^2}^{(1)}-\cC_{d^4u^2}^{(2)}]_{121212}$
\\[0.2cm]
&$+\frac{1}{4}[\cC_{d^4u^2}^{(1)}-\cC_{d^4u^2}^{(2)}]_{121221}$
&\\[0.2cm]
\noalign{\vskip 0.0cm}
\cdashline{1-3}[.4pt/2pt]
\noalign{\vskip 0.2cm}
$\cC_{\rep{27}_R\otimes\rep1_L}^{1312}$
&$\frac{3}{4}[\cC_{d^4u^2}^{(1)}-\cC_{d^4u^2}^{(2)}]_{111222}+\frac{3}{4}[\cC_{d^4u^2}^{(1)}-\cC_{d^4u^2}^{(2)}]_{121122}-\frac{1}{2}[\cC_{d^4u^2}^{(1)}-\cC_{d^4u^2}^{(2)}]_{121212}$
\\[0.2cm]
&$-\frac{1}{2}[\cC_{d^4u^2}^{(1)}-\cC_{d^4u^2}^{(2)}]_{121221}$
&\\[0.2cm]
\noalign{\vskip 0.0cm}
\cdashline{1-3}[.4pt/2pt]
\noalign{\vskip 0.2cm}
$\cC_{\rep{28}_R\otimes\rep1_L}^{112333}$
&$\frac{5}{8}[\cC_{d^4u^2}^{(1)}+\cC_{d^4u^2}^{(2)}]_{111222}+\frac{5}{8}[\cC_{d^4u^2}^{(1)}+\cC_{d^4u^2}^{(2)}]_{121122}$
\\[0.2cm]
\midrule
\addlinespace[0.2cm]
$\cC_{\rep{27}_R\otimes\rep1_L}^{3322}$
&$\frac{1}{4}[\cC_{d^4u^2}^{(1)}-\cC_{d^4u^2}^{(2)}]_{121222}$
\\[0.2cm]
$\cC_{\rep{28}_R\otimes\rep1_L}^{113333}$
&$\frac{5}{16}[\cC_{d^4u^2}^{(1)}+\cC_{d^4u^2}^{(2)}]_{121222}$
\\[0.2cm]
\bottomrule
\end{tabular}
}
    \caption{Projection of the $\cO_{d^4u^2}^{(1,2)}$ SMEFT operators onto the chiral basis.}
    \label{tab:SMEFT_proj_Od4u2}
\end{table}
\begin{table}[t]
    \centering
\scalebox{0.9}{
\begin{tabular}{c@{\hspace{1.2cm}}c@{\hspace{0.6cm}}c}
\toprule
\multirow{1}{*}{\textbf{Chiral}}
&\multirow{2}{*}{\textbf{SMEFT mapping}}
\\
\multirow{1}{*}{\textbf{coefficient}}
\\
\midrule
\addlinespace[0.2cm]
$\cC_{\rep{15}_R\otimes\repbar3_L}^{223\dot 3}$&$\frac{1}{2}[\cC_{d^3Q^2u}]_{111111}$
\\[0.2cm]
\midrule
\addlinespace[0.2cm]
$\cC_{\rep{15}_R\otimes\repbar3_L}^{222\dot 3}$&$-\frac{1}{2}[\cC_{d^3Q^2u}]_{111211}-\frac{1}{2}[\cC_{d^3Q^2u}]_{112111}-\frac{1}{2}[\cC_{d^3Q^2u}]_{121111}$
\\[0.2cm]
$\cC_{\rep{15}_R\otimes\repbar3_L}^{121\dot 3}$&$-[\cC_{d^3Q^2u}]_{112111}$
\\[0.2cm]
$\cC_{\rep{15}_R\otimes\repbar3_L}^{223\dot 2}$&$-\frac{1}{4}[\cC_{d^3Q^2u}]_{111112}-\frac{1}{4}[\cC_{d^3Q^2u}]_{111121}$
\\[0.2cm]
$\cC_{\rep3_R\otimes\repbar3_L}^{2\dot 3}$&$-\frac{1}{2}[\cC_{d^3Q^2u}]_{121111}+\frac{1}{2}[\cC_{d^3Q^2u}]_{111211}$
\\[0.2cm]
$\cC_{\rep{15}_R'\otimes \rep6_L}^{1222\dot 1\dot 3}$&$-\frac{1}{4}[\cC_{d^3Q^2u}]_{111121}+\frac{1}{4}[\cC_{d^3Q^2u}]_{111112}$
\\[0.2cm]
\midrule
\addlinespace[0.2cm]
$\cC_{\rep{15}_R\otimes\repbar3_L}^{131\dot 3}$&$-\frac{1}{2}[\cC_{d^3Q^2u}]_{112211}+\frac{1}{2}[\cC_{d^3Q^2u}]_{121211}-\frac{1}{2}[\cC_{d^3Q^2u}]_{122111}$
\\[0.2cm]
$\cC_{\rep{15}_R\otimes\repbar3_L}^{232\dot 3}$&$-\frac{1}{2}[\cC_{d^3Q^2u}]_{112211}-\frac{1}{2}[\cC_{d^3Q^2u}]_{121211}-\frac{1}{2}[\cC_{d^3Q^2u}]_{122111}$
\\[0.2cm]
$\cC_{\rep{15}_R\otimes\repbar3_L}^{121\dot 2}$&$\frac{1}{2}[\cC_{d^3Q^2u}]_{112112}+\frac{1}{2}[\cC_{d^3Q^2u}]_{112121}$
\\[0.2cm]
\noalign{\vskip 0.0cm}
\cdashline{1-3}[.4pt/2pt]
\noalign{\vskip 0.2cm}
$\cC_{\rep{15}_R\otimes\repbar3_L}^{222\dot 2}$&$\frac{1}{4}[\cC_{d^3Q^2u}]_{111212}+\frac{1}{4}[\cC_{d^3Q^2u}]_{111221}+\frac{1}{4}[\cC_{d^3Q^2u}]_{112112}+\frac{1}{4}[\cC_{d^3Q^2u}]_{112121}$&
\\[0.2cm]
&$+\frac{1}{4}[\cC_{d^3Q^2u}]_{121112}+\frac{1}{4}[\cC_{d^3Q^2u}]_{121121}$
\\[0.2cm]
\noalign{\vskip 0.0cm}
\cdashline{1-3}[.4pt/2pt]
\noalign{\vskip 0.2cm}
$\cC_{\rep3_R\otimes\repbar3_L}^{2\dot 2}$&$-\frac{1}{4}[\cC_{d^3Q^2u}]_{111212}-\frac{1}{4}[\cC_{d^3Q^2u}]_{111221}+\frac{1}{4}[\cC_{d^3Q^2u}]_{121112}+\frac{1}{4}[\cC_{d^3Q^2u}]_{121121}$&
\\[0.2cm]
$\cC_{\rep3_R\otimes\repbar3_L}^{3\dot 3}$&$\frac{1}{2}[\cC_{d^3Q^2u}]_{112211}-\frac{1}{2}[\cC_{d^3Q^2u}]_{122111}$&
\\[0.2cm]
$\cC_{\repbar6_R\otimes\rep6_L}^{13\dot 1\dot 3}$&$\frac{1}{4}[\cC_{d^3Q^2u}]_{111212}-\frac{1}{4}[\cC_{d^3Q^2u}]_{111221}-\frac{1}{4}[\cC_{d^3Q^2u}]_{112112}+\frac{1}{4}[\cC_{d^3Q^2u}]_{112121}$&
\\[0.2cm]
$\cC_{\rep{15}_R'\otimes\rep6_L}^{1223\dot1\dot 3}$&$\frac{3}{8}[\cC_{d^3Q^2u}]_{111212}-\frac{3}{8}[\cC_{d^3Q^2u}]_{111221}+\frac{3}{8}[\cC_{d^3Q^2u}]_{112112}-\frac{3}{8}[\cC_{d^3Q^2u}]_{112121}$&
\\[0.2cm]
\midrule
\addlinespace[0.2cm]
$\cC_{\rep{15}_R\otimes\repbar3_L}^{332\dot 3}$&$-\frac{1}{2}[\cC_{d^3Q^2u}]_{122211}$
\\[0.2cm]
\noalign{\vskip 0.0cm}
\cdashline{1-3}[.4pt/2pt]
\noalign{\vskip 0.2cm}
$\cC_{\rep{15}_R\otimes\repbar3_L}^{131\dot 2}$&$\frac{1}{4}[\cC_{d^3Q^2u}]_{112212}+\frac{1}{4}[\cC_{d^3Q^2u}]_{112221}-\frac{1}{4}[\cC_{d^3Q^2u}]_{121212}-\frac{1}{4}[\cC_{d^3Q^2u}]_{121221}$
\\[0.2cm]
&$+\frac{1}{4}[\cC_{d^3Q^2u}]_{122112}+\frac{1}{4}[\cC_{d^3Q^2u}]_{122121}$
\\[0.2cm]
\noalign{\vskip 0.0cm}
\cdashline{1-3}[.4pt/2pt]
\noalign{\vskip 0.2cm}
$\cC_{\rep{15}_R\otimes\repbar3_L}^{232\dot 2}$&$\frac{1}{4}[\cC_{d^3Q^2u}]_{112212}+\frac{1}{4}[\cC_{d^3Q^2u}]_{112221}+\frac{1}{4}[\cC_{d^3Q^2u}]_{121212}+\frac{1}{4}[\cC_{d^3Q^2u}]_{121221}$
\\[0.2cm]
&$+\frac{1}{4}[\cC_{d^3Q^2u}]_{122112}+\frac{1}{4}[\cC_{d^3Q^2u}]_{122121}$
\\[0.2cm]
\noalign{\vskip 0.0cm}
\cdashline{1-3}[.4pt/2pt]
\noalign{\vskip 0.2cm}
$\cC_{\rep3_R\otimes\repbar3_L}^{3\dot2}$&$-\frac{1}{4}[\cC_{d^3Q^2u}]_{112212}-\frac{1}{4}[\cC_{d^3Q^2u}]_{112221}+\frac{1}{4}[\cC_{d^3Q^2u}]_{122112}+\frac{1}{4}[\cC_{d^3Q^2u}]_{122121}$
\\[0.2cm]
$\cC_{\repbar6_R\otimes\rep6_L}^{12\dot 1\dot 3}$&$-\frac{1}{4}[\cC_{d^3Q^2u}]_{121212}+\frac{1}{4}[\cC_{d^3Q^2u}]_{121221}+\frac{1}{4}[\cC_{d^3Q^2u}]_{122112}-\frac{1}{4}[\cC_{d^3Q^2u}]_{122121}$
\\[0.2cm]
$\cC_{\rep{15}_R'\otimes\rep6_L}^{1233\dot1\dot 3}$&$\frac{3}{4}[\cC_{d^3Q^2u}]_{112212}-\frac{3}{4}[\cC_{d^3Q^2u}]_{112221}$
\\[0.2cm]
\midrule
\addlinespace[0.2cm]
$\cC_{\rep{15}_R\otimes\repbar3_L}^{332\dot 2}$&$\frac{1}{4}[\cC_{d^3Q^2u}]_{122212}+\frac{1}{4}[\cC_{d^3Q^2u}]_{122221}$
\\[0.2cm]
\bottomrule
\end{tabular}
}
    \caption{Projection of the $\cO_{d^3Q^2u}$ SMEFT operator onto the chiral basis.}
    \label{tab:SMEFT_proj_Od3Q2u}
\end{table}
\begin{table}[t]
    \centering
\scalebox{0.92}{
\begin{tabular}{c@{\hspace{1.4cm}}c@{\hspace{0.8cm}}c}
\toprule
\multirow{1}{*}{\textbf{Chiral}}
&\multirow{2}{*}{\textbf{SMEFT mapping}}
\\
\multirow{1}{*}{\textbf{coefficient}}
\\
\midrule
\addlinespace[0.2cm]
$\cC_{\repbar6_L\otimes\rep6_R}^{\dot 3\dot 322}$&$\frac{1}{2}[\cC_{d^2Q^4}^{(1)}]_{111111}-\frac{1}{2}[\cC_{d^2Q^4}^{(2)}]_{111111}$
\\[0.2cm]
$\cC_{\rep{15}_L'\otimes\rep6_R}^{\dot 1\dot 1\dot 2\dot 222}$&$\frac{3}{2}[\cC_{d^2Q^4}^{(1)}]_{111111}+\frac{3}{2}[\cC_{d^2Q^4}^{(2)}]_{111111}$
\\[0.2cm]
\midrule
\addlinespace[0.2cm]
$\cC_{\repbar6_L\otimes\rep6_R}^{\dot 3\dot 323}$&$\frac{1}{2}[\cC_{d^2Q^4}^{(1)}]_{121111}+\frac{1}{2}[\cC_{d^2Q^4}^{(1)}]_{211111}-\frac{1}{2}[\cC_{d^2Q^4}^{(2)}]_{121111}-\frac{1}{2}[\cC_{d^2Q^4}^{(2)}]_{211111}$
\\[0.2cm]
\noalign{\vskip 0.0cm}
\cdashline{1-3}[.4pt/2pt]
\noalign{\vskip 0.2cm}
$\cC_{\repbar6_L\otimes\rep6_R}^{\dot 2\dot 322}$&$-\frac{1}{4}[\cC_{d^2Q^4}^{(1)}]_{111112}-\frac{1}{4}[\cC_{d^2Q^4}^{(1)}]_{111121}-\frac{1}{4}[\cC_{d^2Q^4}^{(1)}]_{111211}-\frac{1}{4}[\cC_{d^2Q^4}^{(1)}]_{112111}$
\\[0.2cm]
&$+\frac{1}{4}[\cC_{d^2Q^4}^{(2)}]_{111112}+\frac{1}{4}[\cC_{d^2Q^4}^{(2)}]_{111121}+\frac{1}{4}[\cC_{d^2Q^4}^{(2)}]_{111211}+\frac{1}{4}[\cC_{d^2Q^4}^{(2)}]_{112111}$
\\[0.2cm]
\noalign{\vskip 0.0cm}
\cdashline{1-3}[.4pt/2pt]
\noalign{\vskip 0.2cm}
$\cC_{\rep{15}'_L\otimes\rep6_R}^{\dot1\dot1\dot2\dot223}$&$\frac{3}{2}[\cC_{d^2Q^4}^{(1)}]_{121111}+\frac{3}{2}[\cC_{d^2Q^4}^{(1)}]_{211111}+\frac{3}{2}[\cC_{d^2Q^4}^{(2)}]_{121111}+\frac{3}{2}[\cC_{d^2Q^4}^{(2)}]_{211111}$
\\[0.2cm]
\noalign{\vskip 0.0cm}
\cdashline{1-3}[.4pt/2pt]
\noalign{\vskip 0.2cm}
$\cC_{\rep{15}'_L\otimes\rep6_R}^{\dot1\dot1\dot2\dot322}$&$\frac{3}{4}[\cC_{d^2Q^4}^{(1)}]_{111112}+\frac{3}{4}[\cC_{d^2Q^4}^{(1)}]_{111121}+\frac{3}{4}[\cC_{d^2Q^4}^{(1)}]_{111211}+\frac{3}{4}[\cC_{d^2Q^4}^{(1)}]_{112111}$
\\[0.2cm]
&$+\frac{3}{4}[\cC_{d^2Q^4}^{(2)}]_{111112}+\frac{3}{4}[\cC_{d^2Q^4}^{(2)}]_{111121}+\frac{3}{4}[\cC_{d^2Q^4}^{(2)}]_{111211}+\frac{3}{4}[\cC_{d^2Q^4}^{(2)}]_{112111}$
\\[0.2cm]
\noalign{\vskip 0.0cm}
\cdashline{1-3}[.4pt/2pt]
\noalign{\vskip 0.2cm}
$\cC_{\repbar3_R\otimes\rep{15}_L}^{1\dot1\dot2\dot3}$&$[\cC_{d^2Q^4}^{(2)}]_{211111}-[\cC_{d^2Q^4}^{(2)}]_{121111}$
\\[0.2cm]
\midrule
\addlinespace[0.2cm]
$\cC_{\repbar6_L\otimes\rep6_R}^{\dot3\dot333}$&$\frac{1}{2}[\cC_{d^2Q^4}^{(1)}]_{221111}-\frac{1}{2}[\cC_{d^2Q^4}^{(2)}]_{221111}$
\\[0.2cm]
$\cC_{\repbar6_L\otimes\rep6_R}^{\dot2\dot222}$&$\frac{1}{4}[\cC_{d^2Q^4}^{(1)}]_{111221}+\frac{1}{4}[\cC_{d^2Q^4}^{(1)}]_{112112}-\frac{1}{4}[\cC_{d^2Q^4}^{(2)}]_{111221}-\frac{1}{4}[\cC_{d^2Q^4}^{(2)}]_{112112}$
\\[0.2cm]
\noalign{\vskip 0.0cm}
\cdashline{1-3}[.4pt/2pt]
\noalign{\vskip 0.2cm}
$\cC_{\repbar6_L\otimes\rep6_R}^{\dot2\dot323}$&$-\frac{1}{4} [\cC_{d^2Q^4}^{(1)}]_{121112}-\frac{1}{4} [\cC_{d^2Q^4}^{(1)}]_{121121}-\frac{1}{4} [\cC_{d^2Q^4}^{(1)}]_{121211}-\frac{1}{4} [\cC_{d^2Q^4}^{(1)}]_{122111}$
\\[0.2cm]
&$-\frac{1}{4} [\cC_{d^2Q^4}^{(1)}]_{211112}-\frac{1}{4} [\cC_{d^2Q^4}^{(1)}]_{211121}-\frac{1}{4} [\cC_{d^2Q^4}^{(1)}]_{211211}-\frac{1}{4} [\cC_{d^2Q^4}^{(1)}]_{212111}$
\\[0.2cm]
&$+\frac{1}{4} [\cC_{d^2Q^4}^{(2)}]_{121112}+\frac{1}{4} [\cC_{d^2Q^4}^{(2)}]_{121121}+\frac{1}{4} [\cC_{d^2Q^4}^{(2)}]_{121211}+\frac{1}{4} [\cC_{d^2Q^4}^{(2)}]_{122111}$
\\[0.2cm]
&$+\frac{1}{4} [\cC_{d^2Q^4}^{(2)}]_{211112}+\frac{1}{4} [\cC_{d^2Q^4}^{(2)}]_{211121}+\frac{1}{4} [\cC_{d^2Q^4}^{(2)}]_{211211}+\frac{1}{4} [\cC_{d^2Q^4}^{(2)}]_{212111}$
\\[0.2cm]
\noalign{\vskip 0.0cm}
\cdashline{1-3}[.4pt/2pt]
\noalign{\vskip 0.2cm}
$\cC_{\rep{15}'_L\otimes\rep6_R}^{\dot1\dot1\dot2\dot233}$&$\frac{3}{2}[\cC_{d^2Q^4}^{(1)}]_{221111}+\frac{3}{2}[\cC_{d^2Q^4}^{(2)}]_{221111}$&
\\[0.2cm]
$\cC_{\rep{15}'_L\otimes\rep6_R}^{\dot1\dot1\dot3\dot322}$&$\frac{3}{4}[\cC_{d^2Q^4}^{(1)}]_{111122}+\frac{3}{4}[\cC_{d^2Q^4}^{(1)}]_{112211}+\frac{3}{4}[\cC_{d^2Q^4}^{(2)}]_{111122}+\frac{3}{4}[\cC_{d^2Q^4}^{(2)}]_{112211}$&
\\[0.2cm]
\noalign{\vskip 0.0cm}
\cdashline{1-3}[.4pt/2pt]
\noalign{\vskip 0.2cm}
$\cC_{\rep{15}'_L\otimes\rep6_R}^{\dot1\dot1\dot2\dot323}$&$\frac{3}{4} [\cC_{d^2Q^4}^{(1)}]_{121112}+\frac{3}{4} [\cC_{d^2Q^4}^{(1)}]_{121121}+\frac{3}{4} [\cC_{d^2Q^4}^{(1)}]_{121211}+\frac{3}{4} [\cC_{d^2Q^4}^{(1)}]_{122111}$
\\[0.2cm]
&$+\frac{3}{4} [\cC_{d^2Q^4}^{(1)}]_{211112}+\frac{3}{4} [\cC_{d^2Q^4}^{(1)}]_{211121}+\frac{3}{4} [\cC_{d^2Q^4}^{(1)}]_{211211}+\frac{3}{4} [\cC_{d^2Q^4}^{(1)}]_{212111}$
\\[0.2cm]
&$+\frac{3}{4} [\cC_{d^2Q^4}^{(2)}]_{121112}+\frac{3}{4} [\cC_{d^2Q^4}^{(2)}]_{121121}+\frac{3}{4} [\cC_{d^2Q^4}^{(2)}]_{121211}+\frac{3}{4} [\cC_{d^2Q^4}^{(2)}]_{122111}$
\\[0.2cm]
&$+\frac{3}{4} [\cC_{d^2Q^4}^{(2)}]_{211112}+\frac{3}{4} [\cC_{d^2Q^4}^{(2)}]_{211121}+\frac{3}{4} [\cC_{d^2Q^4}^{(2)}]_{211211}+\frac{3}{4} [\cC_{d^2Q^4}^{(2)}]_{212111}$
\\[0.2cm]
\noalign{\vskip 0.0cm}
\cdashline{1-3}[.4pt/2pt]
\noalign{\vskip 0.2cm}
$\cC_{\repbar3_R\otimes\rep{15}_L}^{1\dot 1\dot 1\dot 1}$&$\frac{1}{2}[\cC_{d^2Q^4}^{(2)}]_{121121}+\frac{1}{2}[\cC_{d^2Q^4}^{(2)}]_{122111}-\frac{1}{2}[\cC_{d^2Q^4}^{(2)}]_{211121}-\frac{1}{2}[\cC_{d^2Q^4}^{(2)}]_{212111}$
\\[0.2cm]
\noalign{\vskip 0.0cm}
\cdashline{1-3}[.4pt/2pt]
\noalign{\vskip 0.2cm}
$\cC_{\repbar3_R\otimes\rep{15}_L}^{1\dot 1\dot 2\dot 2}$
&$\frac{1}{2} [\cC_{d^2Q^4}^{(2)}]_{121112}+\frac{1}{2} [\cC_{d^2Q^4}^{(2)}]_{121121}+\frac{1}{2} [\cC_{d^2Q^4}^{(2)}]_{121211}+\frac{1}{2} [\cC_{d^2Q^4}^{(2)}]_{122111}$
\\[0.2cm]
&$-\frac{1}{2} [\cC_{d^2Q^4}^{(2)}]_{211112}-\frac{1}{2} [\cC_{d^2Q^4}^{(2)}]_{211121}-\frac{1}{2} [\cC_{d^2Q^4}^{(2)}]_{211211}-\frac{1}{2} [\cC_{d^2Q^4}^{(2)}]_{212111}$
\\[0.2cm]
\bottomrule
\end{tabular}
}
    \caption{Part I of the projection of the $\cO_{d^2Q^4}^{(1,2)}$ SMEFT operators onto the chiral basis. In this case, the mapping involves parity-conjugated basis elements {\color{black}with respect to the ones explicitly written in Tables \ref{tab:chiral_basis_P1}--\ref{tab:chiral_basis_P3}}, which are obtained by exchanging $R \leftrightarrow L$.}
    \label{tab:SMEFT_proj_Od2Q4_part1}
\end{table}
\begin{table}[t]
    \centering
\scalebox{0.94}{
\begin{tabular}{c@{\hspace{1.4cm}}c@{\hspace{0.8cm}}c}
\toprule
\multirow{1}{*}{\textbf{Chiral}}
&\multirow{2}{*}{\textbf{SMEFT mapping}}
\\
\multirow{1}{*}{\textbf{coefficient}}
\\
\midrule
\addlinespace[0.2cm]
$\cC_{\repbar3_R\otimes\rep3_L}^{1\dot 1}$
&$\frac{1}{2} [\cC_{d^2Q^4}^{(1)}]_{121112}-\frac{1}{2} [\cC_{d^2Q^4}^{(1)}]_{121121}+\frac{1}{2} [\cC_{d^2Q^4}^{(1)}]_{121211}-\frac{1}{2} [\cC_{d^2Q^4}^{(1)}]_{122111}$
\\[0.2cm]
&$-\frac{1}{2} [\cC_{d^2Q^4}^{(1)}]_{211112}+\frac{1}{2} [\cC_{d^2Q^4}^{(1)}]_{211121}-\frac{1}{2} [\cC_{d^2Q^4}^{(1)}]_{211211}+\frac{1}{2} [\cC_{d^2Q^4}^{(1)}]_{212111}$
\\[0.2cm]
&$-\frac{1}{4} [\cC_{d^2Q^4}^{(2)}]_{121112}+\frac{1}{4} [\cC_{d^2Q^4}^{(2)}]_{121121}-\frac{1}{4} [\cC_{d^2Q^4}^{(2)}]_{121211}+\frac{1}{4} [\cC_{d^2Q^4}^{(2)}]_{122111}$
\\[0.2cm]
&$+\frac{1}{4} [\cC_{d^2Q^4}^{(2)}]_{211112}-\frac{1}{4} [\cC_{d^2Q^4}^{(2)}]_{211121}+\frac{1}{4} [\cC_{d^2Q^4}^{(2)}]_{211211}-\frac{1}{4} [\cC_{d^2Q^4}^{(2)}]_{212111}$
\\[0.2cm]
\midrule
\addlinespace[0.2cm]
$\cC_{\repbar6_L\otimes\rep6_R}^{\dot2\dot 333}$
&$-\frac{1}{4} [\cC_{d^2Q^4}^{(1)}]_{221112}-\frac{1}{4} [\cC_{d^2Q^4}^{(1)}]_{221121}-\frac{1}{4} [\cC_{d^2Q^4}^{(1)}]_{221211}-\frac{1}{4} [\cC_{d^2Q^4}^{(1)}]_{222111}$
\\[0.2cm]
&$+\frac{1}{4} [\cC_{d^2Q^4}^{(2)}]_{221112}+\frac{1}{4} [\cC_{d^2Q^4}^{(2)}]_{221121}+\frac{1}{4} [\cC_{d^2Q^4}^{(2)}]_{221211}+\frac{1}{4} [\cC_{d^2Q^4}^{(2)}]_{222111}$
\\[0.2cm]
\noalign{\vskip 0.0cm}
\cdashline{1-3}[.4pt/2pt]
\noalign{\vskip 0.2cm}
$\cC_{\repbar6_L\otimes\rep6_R}^{\dot2\dot 223}$
&$\frac{1}{4} [\cC_{d^2Q^4}^{(1)}]_{121221}+\frac{1}{4} [\cC_{d^2Q^4}^{(1)}]_{122112}+\frac{1}{4} [\cC_{d^2Q^4}^{(1)}]_{211221}+\frac{1}{4} [\cC_{d^2Q^4}^{(1)}]_{212112}$
\\[0.2cm]
&$-\frac{1}{4} [\cC_{d^2Q^4}^{(2)}]_{121221}-\frac{1}{4} [\cC_{d^2Q^4}^{(2)}]_{122112}-\frac{1}{4} [\cC_{d^2Q^4}^{(2)}]_{211221}-\frac{1}{4} [\cC_{d^2Q^4}^{(2)}]_{212112}$
\\[0.2cm]
\noalign{\vskip 0.0cm}
\cdashline{1-3}[.4pt/2pt]
\noalign{\vskip 0.2cm}
$\cC_{\rep{15}'_L\otimes\rep6_R}^{\dot1\dot1\dot2\dot333}$
&$\frac{3}{4} [\cC_{d^2Q^4}^{(1)}]_{221112}+\frac{3}{4} [\cC_{d^2Q^4}^{(1)}]_{221121}+\frac{3}{4} [\cC_{d^2Q^4}^{(1)}]_{221211}+\frac{3}{4} [\cC_{d^2Q^4}^{(1)}]_{222111}$
\\[0.2cm]
&$+\frac{3}{4} [\cC_{d^2Q^4}^{(2)}]_{221112}+\frac{3}{4} [\cC_{d^2Q^4}^{(2)}]_{221121}+\frac{3}{4} [\cC_{d^2Q^4}^{(2)}]_{221211}+\frac{3}{4} [\cC_{d^2Q^4}^{(2)}]_{222111}$
\\[0.2cm]
\noalign{\vskip 0.0cm}
\cdashline{1-3}[.4pt/2pt]
\noalign{\vskip 0.2cm}
$\cC_{\rep{15}'_L\otimes\rep6_R}^{\dot1\dot1\dot3\dot323}$
&$\frac{3}{4} [\cC_{d^2Q^4}^{(1)}]_{121122}+\frac{3}{4} [\cC_{d^2Q^4}^{(1)}]_{122211}+\frac{3}{4} [\cC_{d^2Q^4}^{(1)}]_{211122}+\frac{3}{4} [\cC_{d^2Q^4}^{(1)}]_{212211}$
\\[0.2cm]
&$+\frac{3}{4} [\cC_{d^2Q^4}^{(2)}]_{121122}+\frac{3}{4} [\cC_{d^2Q^4}^{(2)}]_{122211}+\frac{3}{4} [\cC_{d^2Q^4}^{(2)}]_{211122}+\frac{3}{4} [\cC_{d^2Q^4}^{(2)}]_{212211}$
\\[0.2cm]
\noalign{\vskip 0.0cm}
\cdashline{1-3}[.4pt/2pt]
\noalign{\vskip 0.2cm}
$\cC_{\repbar3_R\otimes\rep{15}_L}^{1\dot 1\dot 3\dot 2}$
&$\frac{1}{2} [\cC_{d^2Q^4}^{(2)}]_{121221}+\frac{1}{2} [\cC_{d^2Q^4}^{(2)}]_{122112}-\frac{1}{2} [\cC_{d^2Q^4}^{(2)}]_{211221}-\frac{1}{2} [\cC_{d^2Q^4}^{(2)}]_{212112}$
\\[0.2cm]
\midrule
\addlinespace[0.2cm]
$\cC_{\repbar6_L\otimes\rep6_R}^{\dot2\dot233}$
&$\frac{1}{4} [\cC_{d^2Q^4}^{(1)}]_{221221}+\frac{1}{4} [\cC_{d^2Q^4}^{(1)}]_{222112}-\frac{1}{4} [\cC_{d^2Q^4}^{(2)}]_{221221}-\frac{1}{4} [\cC_{d^2Q^4}^{(2)}]_{222112}$
\\[0.2cm]
$\cC_{\rep{15}_L'\otimes\rep6_R}^{\dot1\dot1\dot3\dot333}$
&$\frac{3}{4} [\cC_{d^2Q^4}^{(1)}]_{221122}+\frac{3}{4} [\cC_{d^2Q^4}^{(1)}]_{222211}+\frac{3}{4} [\cC_{d^2Q^4}^{(2)}]_{221122}+\frac{3}{4} [\cC_{d^2Q^4}^{(2)}]_{222211}$
\\[0.2cm]
\bottomrule
\end{tabular}
}
    \caption{Part II of the projection of the $\cO_{d^2Q^4}^{(1,2)}$ SMEFT operators onto the chiral basis. In this case, the mapping involves parity-conjugated basis elements {\color{black}with respect to the ones explicitly written in Tables \ref{tab:chiral_basis_P1}--\ref{tab:chiral_basis_P3}}, which are obtained by exchanging $R \leftrightarrow L$.}
    \label{tab:SMEFT_proj_Od2Q4_part2}
\end{table}

\clearpage
\bibliographystyle{JHEP}
\bibliography{References}

\end{document}